\RequirePackage{fix-cm}
\documentclass[twocolumn]{svjour3}          
\smartqed  
\smartqed  

\usepackage{amssymb}
\usepackage{amsmath}

\usepackage{graphicx}
\usepackage{epstopdf}
\usepackage{subfigure}
\usepackage{hyperref}
\usepackage{xcolor}
\usepackage{caption}
\newcommand\Partials[3]{\frac{\partial^{#3} #1}{\partial #2^{#3}}}
\newcommand\Partial[2]{\frac{\partial #1}{\partial #2}}

\journalname{Nonlinear Dynamics}
\begin{document}
\begin{sloppypar}

\title{Bifurcation analysis of quasi-periodic orbits of mechanical systems with 1:2 internal resonance via spectral submanifolds
}

\titlerunning{Bifurcation analysis of quasi-periodic orbits of systems with 1:2 internal resonance via SSMs}        

\author{Hongming Liang \and
        Shobhit Jain \and
        Mingwu Li
}


\institute{H. Liang\and M. Li\at
              Department of Mechanics and Aerospace Engineering, Southern University of Science and Technology, Shenzhen, 518055, China \\
              \email{limw@sustech.edu.cn (M. Li)}           
           \and
           S. Jain \at
              Delft Institute of Applied Mathematics, TU Delft, Mekelweg 4, 2628CD, Delft, The Netherlands
}

\date{Received: date / Accepted: date}

\maketitle

\begin{abstract}
A 1:2 internally resonant mechanical system can undergo secondary Hopf (Neimark-Sacker) bifurcations, resulting in a quasi-periodic response when the system is subject to harmonic excitation. While these quasi-periodic orbits have been observed in practice, their bifurcations are not well studied, especially in high-dimensional mechanical systems. This is mainly because of the challenges associated with the computation and bifurcation detection of these quasi-periodic motions. Here we present a computational framework to address these challenges via reductions on spectral submanifolds, which transforms quasi-periodic orbits of high-dimensional systems as limit cycles of four-dimensional reduced-order models. We apply the proposed framework to analyze bifurcations of quasi-periodic orbits in several mechanical systems exhibiting 1:2 internal resonance, including a finite element model of a shallow-curved shell. We uncover local bifurcations such as period-doubling and saddle-node, as well as global bifurcations such as homoclinic connections, isolas, and simple bifurcations of quasi-periodic orbits. We also observe cascades of period-doubling bifurcations of quasi-periodic orbits that eventually result in chaotic motions, as well as the coexistence of chaotic and quasi-periodic attractors. These findings elucidate the complex bifurcation mechanism of quasi-periodic orbits in 1:2 internally resonant systems.

\end{abstract}

\section{Introduction}  \label{sec:Introduction}

A system is in a 1:2 internal resonance if it admits a 1:2 relationship between two natural frequencies. Under appropriate nonlinear coupling, this internal resonance or even near internal resonance can result in energy transfer among the resonant modes~\cite{nayfeh1986ResponseTwodegreeoffreedomSystems,lacarbonara1999DIRECTTREATMENTDISCRETIZATIONS,nayfeh1987ParametricExcitationTwo}. Such 1:2 internal resonance phenomena are commonly found in nonlinear mechanical systems such as beams \cite{asadi2021StrongInternalResonance,oz2006TwotooneInternalResonances,xiong2014NonlinearForcedVibration,tien1994NonlinearDynamicsShallow}, cables\cite{srinil2007TwotooneResonantMultimodal,zheng2002SuperharmonicInternalResonances}, pipes \cite{alfosail2019TwotooneInternalResonance,zhang2016SupercriticalNonlinearVibration}, and shells \cite{li2022NonlinearAnalysisForceda}. A system with 1:2 internal resonance can display complex yet intriguing bifurcations. Thus, it is essential to understand the effects of 1:2 internal resonance if one aims to suppress the structural vibration amplitude or to utilize the internal resonance for design innovations that take nonlinearities into account. Indeed, in the field of mechatronics, 1:2 internal resonance phenomena in micro-electro-mechanical systems (MEMS) can be used to generate frequency combs (FCs) that are very important for precision measurements and synchronization applications \cite{alfosail2019TheoreticalExperimentalInvestigation,yu2023OnetotwoInternalResonance,bhosale2024MultiharmonicPhononicFrequency}. Similarly, a 1:2 internal resonance has also been used to enhance the performance of energy harvesters~\cite{chen2015internal,zhang2024internal}.

Consider a 1:2 internally resonant mechanical system subject to a harmonic excitation, the forced response curve (FRC) that depicts the nonlinear periodic response amplitude under the variations in excitation frequency is typically an M-shaped curve \cite{li2022NonlinearAnalysisForceda,yu2023OnetotwoInternalResonance,liNonlinearAnalysisForced2022c,gobat2021BackboneCurvesNeimarkSacker} (see Fig~\ref{fig:ep_x1_X2}), which reflects the coexistence of multiple-solutions and jumping characteristics of the system. In particular, secondary Hopf bifurcations (HB, also referred to as Neimark-Sacker bifurcations) and saddle-node (SN) bifurcations can be found along the FRC \cite{li2022NonlinearAnalysisForceda,liNonlinearAnalysisForced2022c} of such an M-shaped FRC. In fact, there are often two HB points on a given FRC, as shown in Fig.~\ref{fig:ep_x1_X2}, and the periodic orbits between these two HB points are unstable. 


A secondary HB point on an FRC indicates the birth of quasi-periodic orbits under the variations in excitation frequency. Therefore, one would expect the appearance of quasi-periodic orbits when the excitation frequency is within the frequency interval ended by the two HB points. Indeed, this has been confirmed by previous studies~\cite{gobat2021BackboneCurvesNeimarkSacker,li2019SingularityAnalysisResponse,liNonlinearAnalysisForced2022c}. In particular, the FRC of quasi-periodic orbits in a nonlinear two degrees-of-freedom system with 1:2 internal resonance was obtained via normal form analysis using SSMTool~\cite{liNonlinearAnalysisForced2022c}. Along the FRC of quasi-periodic orbits, both stable and unstable tori were computed, and multiple period-doubling (PD) and SN bifurcated tori were detected~\cite{liNonlinearAnalysisForced2022c}. These PD and SN bifurcations were further confirmed by a recent study where the associated invariant tori were obtained using a semi-analytical homotopy method~\cite{song2024StrongNonlinearMixing}. Moreover, a cascade of period-doubling bifurcations of quasi-periodic orbits leading to chaos was uncovered using long-time integrations and Fourier transforms~\cite{song2024StrongNonlinearMixing}.

Quasi-periodicity can be a transition state from periodicity to chaos~\cite{song2024StrongNonlinearMixing,yao2019TwoBifurcationRoutes,song2023MultipleSwitchingBifurcations}. A 1:2 internal resonance system can have the coexistence of quasi-periodic, periodic, and chaotic attractors~\cite{song2024StrongNonlinearMixing}. Therefore, it is important to study the bifurcations of quasi-periodic orbits under the variations in forcing frequency and amplitude. However, this is a challenging task for high-dimensional mechanical systems with 1:2 internal resonance. Various numerical methods for the computation of quasi-periodic orbits have been proposed in the past several decades, including harmonic balance~\cite{kim1996QUASIPERIODICRESPONSESTABILITY,guskov2012HarmonicBalanceBasedApproach,ju2017ModifiedTwoTimescaleIncremental,liao2020ContinuationStabilityAnalysis}, numerical integration~\cite{gobat2021BackboneCurvesNeimarkSacker,li2019SingularityAnalysisResponse}, and collocation~\cite{roose2007continuation,dankowicz2013recipes}. They are effective for low-dimensional systems but impractical for high-dimensional systems, e.g., finite element models. In addition, the stability analysis and bifurcation detection of quasi-periodic orbits are difficult and remain an active research area~\cite{breunung2022asymptotic,fiedler2024efficient}. Therefore, the few reported studies on the bifurcation analysis of quasi-periodic orbits of 1:2 internally resonant systems are restricted to systems with two degrees-of-freedom~\cite{liNonlinearAnalysisForced2022c,song2024StrongNonlinearMixing}, where forcing amplitudes are further fixed to simplify analysis.

Here, we aim to establish a unified analysis for the bifurcation of quasi-periodic orbits of 1:2 internally resonant mechanical systems, ranging from simple oscillators to complex finite element models of shallow shells with more than 1300 DOFs. We allow for variations in both forcing frequency and amplitude. Our analysis utilizes reduction on spectral submanifolds (SSMs), which have emerged as a powerful tool for constructing low-dimensional reduced-order models (ROMs) of high-dimensional nonlinear mechanical systems. Indeed, SSMs are low-dimensional attracting invariant manifolds, and hence, their reduced dynamics serve as a rigorous ROM for the full nonlinear system~\cite{haller2016nonlinear}. SSM-based model reduction has been successfully applied to extract backbone curves and FRCs~\cite{breunung2018explicit,li2024fast}, and to predict bifurcations of periodic and quasi-periodic orbits~\cite{liNonlinearAnalysisForced2022c}. Recent advances in SSM-based model reductions include treatments of constrained mechanical systems~\cite{li2023model}, parametric resonance~\cite{thurnher2024nonautonomous}, random vibration~\cite{xu2024nonlinear}, and general forcing~\cite{haller2024nonlinear}.

To the specific aim of this study, we will utilize the contributions in~\cite{liNonlinearAnalysisForced2022c} for analyzing bifurcation of quasi-periodic orbits in 1:2 internally resonant mechanical systems. As shown in~\cite{liNonlinearAnalysisForced2022c}, two-dimensional tori that contain the quasi-periodic orbits of the full system can be simply recovered as limit cycles of the SSM-based ROMs. Moreover, one can infer bifurcations of these quasi-periodic orbits from those of the associated limit cycles. In the current context, these SSM-based ROMs enable us to uncover the complex bifurcation mechanism of quasi-periodic orbits in 1:2 internally resonant mechanical systems, including homoclinic bifurcations, isolas of quasi-periodic orbits, and cascades of periodic doubling bifurcations leading to chaos. Such chaotic motions, for instance, have been observed in experiments~\cite{nayfeh1989ExperimentalInvestigationComplicated,nayfeh1990ExperimentalInvestigationResonantly}.

The rest of this paper is organized as follows. We present a computational framework via reductions on SSMs for the bifurcation analysis of quasi-periodic orbits of 1:2 internally resonant mechanical systems in Sect.~\ref{sec:comp-framework}. We then apply the computational framework to three representative examples of systems with 1:2 internal resonance, including two coupled nonlinear oscillators in Sect.~\ref{sec:coupled-nonlinear-oscillators}, a pinned-pinned shallow curved beam in Sect.~\ref{sec:beam}, and a finite element model with more than 1300 degrees-of-freedom for a shallow shell in Sect.~\ref{sec:shell}. In each example, we first compute the FRC of periodic orbits and then extract the FRC of quasi-periodic orbits born out of secondary Hopf bifurcations. We further investigate the local and global bifurcations of these quasi-periodic orbits under the variations in forcing frequency and amplitudes. In these examples, we also compare the SSM-based predictions and reference solutions of the original systems to demonstrate the accuracy and efficiency of the SSM-based reductions. As we will see, the SSM-based reductions make accurate predictions efficiently, paving the road for bifurcation analysis of quasi-periodic orbits for high-dimensional systems with 1:2 internal resonance. Finally, we draw conclusions in Sect.~\ref{sec:conclusion}.

\section{A computational framework via reductions on SSMs}
\label{sec:comp-framework}

\subsection{Problem statement}

We consider a periodically forced nonlinear mechanical system
\begin{equation}
\label{eq:eom-second-full}
\boldsymbol{M}\ddot{\boldsymbol{x}}+\boldsymbol{C}\dot{\boldsymbol{x}}+\boldsymbol{K}\boldsymbol{x}+\boldsymbol{f}(\boldsymbol{x},\dot{\boldsymbol{x}})=\epsilon \boldsymbol{f}^{\mathrm{ext}}(\Omega t),\quad 0<\epsilon\ll1
\end{equation}
where $\boldsymbol{x}\in\mathbb{R}^n$ is the generalized displacement vector; $\boldsymbol{M},\boldsymbol{C},\boldsymbol{K}\in\mathbb{R}^{n\times n}$ are the mass, damping and stiffness matrices; $\boldsymbol{f}(\boldsymbol{x},\dot{\boldsymbol{x}})$ is a $C^r$ smooth nonlinear function such that
$\boldsymbol{f}(\boldsymbol{x},\dot{\boldsymbol{x}})\sim \mathcal{O}(|\boldsymbol{x}|^2,|\boldsymbol{x}||\dot{\boldsymbol{x}}|,|\dot{\boldsymbol{x}}|^2)$; and $\epsilon \boldsymbol{f}^{\mathrm{ext}}(\Omega t)$ denotes external harmonic excitation. We note that we allow for asymmetric damping and stiffness matrices to account for gyroscopic and follower forces, as well as nonlinear damping.

Solving the linear part of~\eqref{eq:eom-second-full} leads to the generalized eigenvalue problem
\begin{equation}
\label{eq:eom-second-eig}
(\lambda_j^2\boldsymbol{M}+\lambda_j\boldsymbol{C}+\boldsymbol{K})\boldsymbol{\phi}_j=\boldsymbol{0},\quad \boldsymbol{\theta}_j^\ast(\lambda_j^2\boldsymbol{M}+\lambda_j\boldsymbol{C}+\boldsymbol{K})=\boldsymbol{0}
\end{equation}
where $\lambda_j$ is an eigenvalue, and $\boldsymbol{\phi}_j$ and $\boldsymbol{\theta}_j$ are associated right and left eigenvectors. We assume that the mechanical system~\eqref{eq:eom-second-full} is in a 1:2 internal resonance; namely, there exist two pairs of complex conjugate modes that are 1:2 internally resonant. Without loss of generality, we let $\lambda_1$ and $\bar{\lambda}_1$ be the eigenvalues of the first pair of resonant modes, and $\lambda_2$ and $\bar{\lambda}_2$ be the eigenvalues of the second pair of resonant modes, we have
\begin{equation}
    \lambda_2\approx2\lambda_1,\quad \bar{\lambda}_2\approx2\bar{\lambda}_1.
\end{equation}
In addition, we assume that there are no resonances between the spectrum $\{\lambda_1,\bar{\lambda}_1,\lambda_2,\bar{\lambda}_2\}$ and eigenvalues outside the spectrum.

We are interested in the forced dynamics when that the excitation frequency $\Omega$ is close to the natural frequency of the first pair of internally resonant modes, namely, $\mathrm{i}\Omega\approx\lambda_1$. Therefore, the first pair of modes is in external resonance. Due to the 1:2 internal resonance, energy transfer can occur between the two pairs of modes. We assume the origin of the system~\eqref{eq:eom-second-full} is asymptotically stable, namely, $\mathrm{Re}(\lambda_j)<0$ for $1\leq j\leq 2n$. Then the stable fixed point is perturbed as a stable small-amplitude periodic orbit when $\epsilon$ is sufficiently small~\cite{guckenheimer2013nonlinear}. However, when $\epsilon$ exceeds a critical value, the periodic orbit can become unstable via a secondary Hopf bifurcation (also referred to as Neimark-Sacker bifurcation), and quasi-periodic orbits appear. Our goal is to provide a thorough analysis of the bifurcations of the quasi-periodic orbits under the variations in $\epsilon$ and $\Omega$.

\subsection{Reduction on spectral submanifolds}

Solving quasi-periodic orbits of the mechanical system~\eqref{eq:eom-second-full} is challenging for $n\gg1$. We use reductions on spectral submanifolds (SSMs) to address this challenge. Next, we present a brief introduction to the theory of SSMs.

The second-order system~\eqref{eq:eom-second-full} can be transformed into a first-order system as below
\begin{equation}
\label{eq:full-first}
\boldsymbol{B}\dot{\boldsymbol{z}}	=\boldsymbol{A}\boldsymbol{z}+\boldsymbol{F}(\boldsymbol{z})+\epsilon\boldsymbol{F}^{\mathrm{ext}}({\Omega t}),
\end{equation}
where
\begin{gather}
\label{eq:zABF}
\boldsymbol{z}=\begin{pmatrix}\boldsymbol{x}\\\dot{\boldsymbol{x}}\end{pmatrix},\,\,
\boldsymbol{A}=\begin{pmatrix}-\boldsymbol{K} 
& \boldsymbol{0}\\\boldsymbol{0} & \boldsymbol{M}\end{pmatrix},\,\,
\boldsymbol{B}=\begin{pmatrix}\boldsymbol{C} 
& \boldsymbol{M}\\\boldsymbol{M} & \boldsymbol{0}\end{pmatrix},\nonumber\\
\boldsymbol{F}(\boldsymbol{z})=\begin{pmatrix}{-\boldsymbol{f}(\boldsymbol{x},\dot{\boldsymbol{x}})}\\\boldsymbol{0}\end{pmatrix},\,\,
\boldsymbol{F}^{\mathrm{ext}}(\Omega t) = \begin{pmatrix}\boldsymbol{f}^{\mathrm{ext}}(\Omega t)\\\boldsymbol{0}\end{pmatrix}.
\end{gather}
The associated generalized eigenvalue problem becomes (cf.~\eqref{eq:eom-second-eig})
\begin{equation}
\boldsymbol{A}\boldsymbol{v}_j=\lambda_j\boldsymbol{B}\boldsymbol{v}_j,\quad \boldsymbol{u}_j^\ast \boldsymbol{A}=\lambda_j \boldsymbol{u}_j^\ast \boldsymbol{B},
\end{equation}
where $\lambda_j$ is a generalized eigenvalue and $\boldsymbol{v}_j$ and $\boldsymbol{u}_j$ are the corresponding \emph{right} and \emph{left} eigenvectors, respectively. In particular, we have
\begin{equation}
    \boldsymbol{v}_j=\begin{pmatrix}\boldsymbol{\phi}_j\\\lambda_j\boldsymbol{\phi}_j\end{pmatrix},\quad \boldsymbol{u}_j=\begin{pmatrix}\boldsymbol{\theta}_j\\\bar{\lambda}_j\boldsymbol{\theta}_j\end{pmatrix}.
\end{equation}

In our setting, we take $\mathcal{E}=\mathrm{span}(\boldsymbol{v}_1,\bar{\boldsymbol{v}}_1,\boldsymbol{v}_2,\bar{\boldsymbol{v}}_2)$ as a master subspace for model reduction. It follows that we construct four-dimensional reduced-order models (ROMs) for the full system with a $2n$-dimensional phase space. A significant dimension reduction is obtained when $n$  is large. In the case of unforced system ($\epsilon=0$), the subspace $\mathcal{E}$ is invariant for the linear system $\boldsymbol{B}\dot{\boldsymbol{z}}=\boldsymbol{A}\boldsymbol{z}$, i.e., any trajectory of the linear system with an initial condition in $\mathcal{E}$ will stay in $\mathcal{E}$ for all times. 

Under the addition of the nonlinear term $\boldsymbol{F}(\boldsymbol{z})$ in system~\eqref{eq:full-first}, the master subspace $\mathcal{E}$ is perturbed into nonlinear normal modes (NNM), which are invariant manifolds tangent to $\mathcal{E}$ at the origin~\cite{shaw1993normal}. While there can be infinitely many NNMs associated with a given spectral subspace, there exists a unique, smoothest one among them under generically satisfied non-resonance conditions which is defined as the \emph{spectral submanifold} (SSM) associated with $\mathcal{E}$. We denote this SSM as $\mathcal{W}(\mathcal{E})$. Slow SSMs are associated with the master subspace with the lowest linear decay rates and attract nearby full system trajectories exponentially fast. Hence, the reduced dynamics on slow SSMs provides a rigorous ROM. The SSM $\mathcal{W}(\mathcal{E})$ can be viewed as an embedding of an open set in reduced coordinates $\boldsymbol{p}=(p_1,\bar{p}_1,p_2,\bar{p}_2)\in\mathbb{C}^4$ via a map $\boldsymbol{z}=\boldsymbol{W}(\boldsymbol{p}):\mathbb{C}^4\to\mathbb{R}^{2n}$. In addition, the reduced dynamics on the SSM can be expressed as $\dot{\boldsymbol{p}}=\boldsymbol{R}(\boldsymbol{p})$. 

Under further addition of the external periodic forcing $\epsilon\boldsymbol{F}^\mathrm{ext}(\Omega t)$, the autonomous SSM is perturbed as a unique, smoothest, periodic SSM, which we denote as $\mathcal{W}(\mathcal{E},\Omega t)$. Similarly to the autonomous SSM, the time-periodic SSM $\mathcal{W}(\mathcal{E},\Omega t)$ can be embedded via a map $\boldsymbol{W}_\epsilon(\boldsymbol{p},\Omega t):\mathbb{C}^4\times \mathbb{S}^1 \to\mathbb{R}^{2n}$. The reduced dynamics on this time-periodic SSM is non-autonomous and can be expressed as~\cite{jain2022HowComputeInvariant,li2022NonlinearAnalysisForceda}
\begin{equation}
\label{eq:red-reps}
\dot{\boldsymbol{p}}=\boldsymbol{R}_\epsilon(\boldsymbol{p},\Omega t)=\boldsymbol{R}(\boldsymbol{p})+\epsilon\boldsymbol{S}(\boldsymbol{p},\Omega t)+\mathcal{O}(\epsilon^2).
\end{equation}

We express the reduced coordinates in polar form as
\begin{equation}
\label{eq:transform}
    p_1=\rho_1e^{\mathrm{i}(\theta_1+\Omega t)},\quad p_2=\rho_2e^{\mathrm{i}(\theta_2+2\Omega t)}
\end{equation}
and adopt a leading-order approximation to the non-autonomous part $\boldsymbol{S}(\boldsymbol{p},\Omega t)$ ~\cite{li2022NonlinearAnalysisForceda}. Then the reduced dynamics is given as~\cite{li2022NonlinearAnalysisForceda}
\begin{equation}
\begin{aligned}
\dot{\rho}_1 = & \mathrm{Re}(\lambda_1)\rho_1+g_1(\boldsymbol{\rho},\boldsymbol{\theta})+\epsilon(\mathrm{Re}(f)\cos\theta_1 \\
&+\mathrm{Im}(f)\sin\theta_1),  \\
\dot{\theta}_1 = & \mathrm{Im}(\lambda_1)-\Omega+g_2(\boldsymbol{\rho},\boldsymbol{\theta})+{\epsilon}(\mathrm{Im}(f)\cos\theta_1 \\
&-\mathrm{Re}(f)\sin\theta_1)/{\rho_1},\\
\dot{\rho}_2 = & \mathrm{Re}(\lambda_2)\rho_2+g_3(\boldsymbol{\rho},\boldsymbol{\theta}), \\
\dot{\theta}_2 = & \mathrm{Im}(\lambda_2)-2\Omega+g_4(\boldsymbol{\rho},\boldsymbol{\theta}),\label{eq:red-nonauto-th}
\end{aligned}
\end{equation}
where $\boldsymbol{\rho}=(\rho_1,\rho_2)$, $\boldsymbol{\theta}=(\theta_1,\theta_2)$, $f$ is a modal force, and $g_i(\boldsymbol{\rho},\boldsymbol{\theta})$ for $1\leq i\leq4$ are nonlinear functions. Explicit expressions for $f$ and $g_i$ can be found in eqs. (37)-(38) of~\cite{li2022NonlinearAnalysisForceda}.

Due to the transformation~\eqref{eq:transform}, a fixed point of the vector field~\eqref{eq:red-nonauto-th} corresponds to a periodic orbit $\boldsymbol{p}_\mathrm{po}(t)$ on the SSM. Therefore, we can extract the forced response of periodic orbits from the zero-level contours of the vector field~\eqref{eq:red-nonauto-th}. We can further infer bifurcations of periodic orbits of the full system from bifurcations of fixed points of the vector field~\eqref{eq:red-nonauto-th}~\cite{liNonlinearAnalysisForced2022c}. In addition, a limit cycle of the SSM-based ROM~\eqref{eq:red-nonauto-th} corresponds to a two-dimensional invariant torus of the full system~\cite{liNonlinearAnalysisForced2022c}. Hence, we can predict bifurcations of quasi-periodic orbits that stay on tori via the bifurcations of the limit cycles of the ROM.

\begin{figure*}[!ht]
    \centering
    \includegraphics[width=0.6\textwidth]{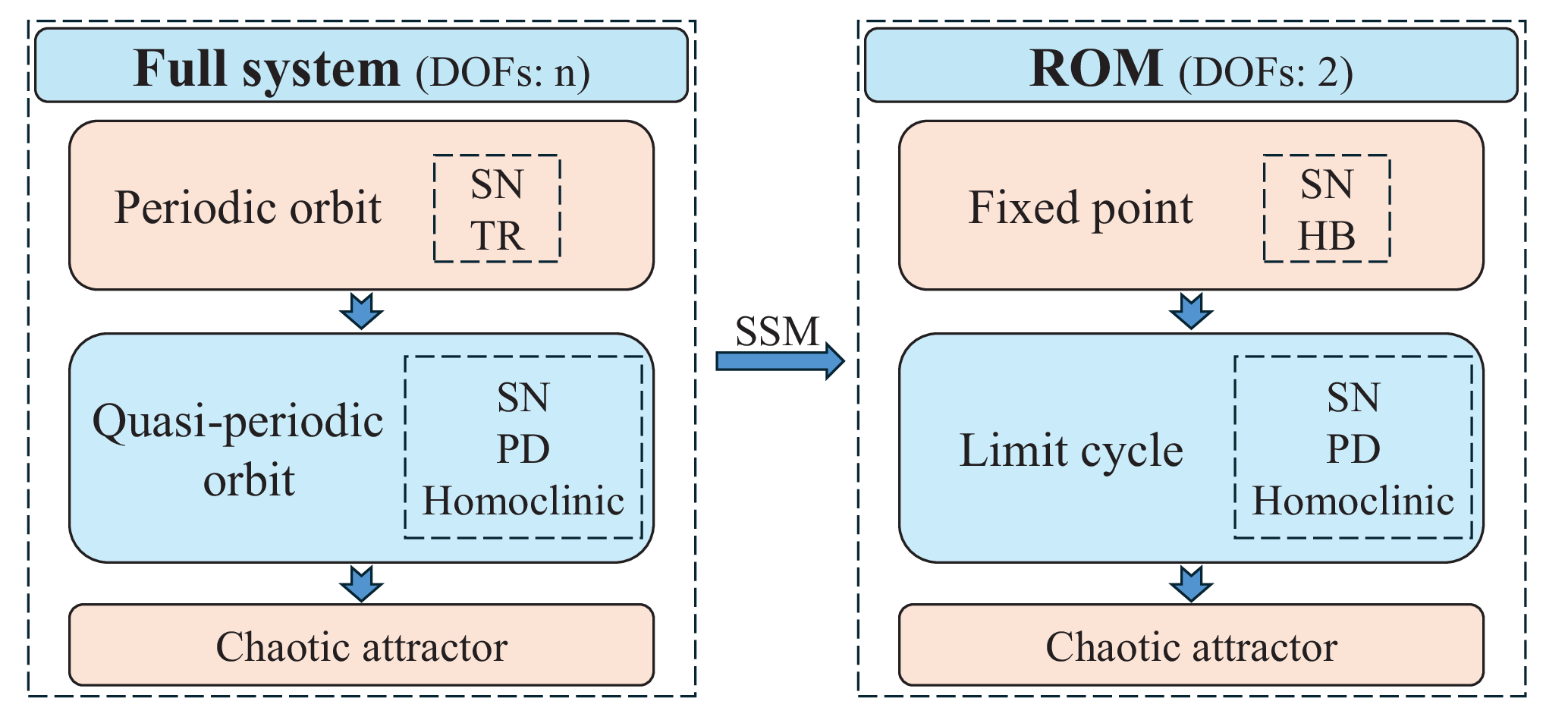}
    \caption{\small A computational framework for the nonlinear analysis of 1:2 internally resonant mechanical systems via reductions on spectral submanifolds (SSMs).}
    \label{fig:FulltoROM}
\end{figure*}

As summarized in Fig.~\ref{fig:FulltoROM}, SSM-based reductions enable effective bifurcation analysis of nonlinear dynamics of high-dimensional mechanical systems. Indeed, it achieves a significant dimension reduction because the $2n$-dimensional full system is reduced to a 4-dimensional ROM~\eqref{eq:red-nonauto-th}. Furthermore, thanks to the normal form parameterization style used in the polar ROM~\eqref{eq:red-nonauto-th}, a further simplification is obtained, i.e., quasi-periodic orbits of the full system are calculated as limit cycles of the ROM.

Inspired by the above discussion, we will present bifurcation analysis of quasi-periodic orbits in the Poincar{\'e} section of the full system~\eqref{eq:full-first}. In particular, we take $\{(\boldsymbol{z},t): \mathrm{mod}(t,2\pi/\Omega)=0\}$ as the Poincar{\'e} section. Let $T=2\pi/\Omega$, the associated Poincar{\'e} map maps the state $\boldsymbol{z}(t)$ from $t=kT$ to $t=(k+1)T$ under the flow generated by~\eqref{eq:full-first}. Therefore, we also call this map as period-$2\pi/\Omega$ map. We further define Poincar{\'e} intersection for an orbit/attractor as the intersection of the orbit/attrator with the Poincar{\'e} section.

We note that a chaotic attractor of the 4-dimensional SSM-based ROM~\eqref{eq:red-nonauto-th} also corresponds to a chaotic attractor of the full system~\eqref{eq:full-first}, as pointed out in Fig.~\ref{fig:FulltoROM}. We will use the aforementioned Poincar{\'e} intersection as a tool to distinguish chaotic attractors from quasi-periodic attractors. In particular, the Poincar{\'e} intersection of a quasi-periodic attractor will be a closed curve, while the Poincar{\'e} intersection of a chaotic attractor will be a strange attractor (cf. Fig.~\ref{fig:chaos&isola}). As an additional approach, we will also use the power spectral density (PSD) plot to distinguish them. The power spectral density (PSD) plot of a quasi-periodic attractor has isolated peaks (cf. Fig.~\ref{fig:PSD-qusia}), while the PSD plot of a chaotic attractor will have significant values for a broad frequency bandwidth  (cf. the middle panel of Fig.~\ref{fig:chaos_x2}).

\begin{remark}
    We have assumed the excitation frequency $\Omega$ is near the natural frequency of the first pair of internally resonant modes, namely, $\mathrm{i}\Omega\approx\lambda_1$. In this case, reduction via the associated four-dimensional SSM is locally exact because the SSM captures the inner resonance of the two pairs of modes and coupling between them and the rest high-order modes. This exact reduction enables us to uncover important information about the full system. However, such a four-dimensional reduction cannot capture all the dynamics of the full system. In particular, when the excitation frequency is near the natural frequency of the third pair of modes, i.e., $\mathrm{i}\Omega\approx\lambda_3$, we should perform model reduction on the two-dimensional SSM associated with the third pair of modes to capture the primary resonance of the third pair of modes. Therefore, one should select the master subspace for model reduction according to both external and internal resonances~\cite{li2022NonlinearAnalysisForceda}.
\end{remark}

\subsection{Computation of SSMs}

We still need to determine the map $\boldsymbol{W}_\epsilon(\boldsymbol{p},\Omega t)$ and the unknown coefficients for the nonlinear functions $g_i$ in~\eqref{eq:red-nonauto-th}. For the consistency with~\eqref{eq:red-reps}, we approximate $\boldsymbol{W}_\epsilon(\boldsymbol{p},\Omega t)$ as
\begin{equation}
    \boldsymbol{W}_\epsilon(\boldsymbol{p},\Omega t)=\boldsymbol{W}(\boldsymbol{p})+\epsilon\boldsymbol{X}(\boldsymbol{p},\Omega t)+\mathcal{O}(\epsilon^2).
\end{equation}
We further adopt a leading-order approximation to the non-autonomous part $\boldsymbol{X}(\boldsymbol{p},\Omega t)$, namely,
\begin{equation}
\label{eq:lead-trunc}
    \boldsymbol{X}(\boldsymbol{p},\Omega t)=\boldsymbol{X}_0(\Omega t)+\mathcal{O}(|\boldsymbol{p}|).
\end{equation}
Here $\boldsymbol{X}_0(\Omega t)=\boldsymbol{x}_0e^{\mathrm{i}\Omega t}+\bar{\boldsymbol{x}}_0e^{-\mathrm{i}\Omega t}$ and the coefficient vector $\boldsymbol{x}_0$ is obtained by solving a system of linear equations~\cite{jain2022HowComputeInvariant,li2022NonlinearAnalysisForceda}.

We use the computational procedure ~\cite{jain2022HowComputeInvariant} based on the parametrization method~\cite{CabreIII,Haro2016TheManifolds} to solve for the autonomous map $\boldsymbol{W}(\boldsymbol{p})$ and as well the unknown coefficients in $g_i$ (see eq.~\eqref{eq:red-nonauto-th}). The map $\boldsymbol{W}(\boldsymbol{p})$ is approximated by a truncated Taylor expansion in $\boldsymbol{p}$. Then the coefficients of the Taylor expansion, along with $g_i$, are determined from the invariance of SSM. In particular, they are solved from systems of linear equations in a recursive fashion, which allows us to automatically compute the SSM expansion up to arbitrary polynomial orders without change of coordinates using only the eigenvectors associated with the master subspace $\mathcal{E}$ ~\cite{jain2022HowComputeInvariant}. In practice, we truncate the Taylor expansion based on the convergence of forced response with increasing expansion orders~\cite{li2022NonlinearAnalysisForceda}. We use notation $\mathcal{O}(q)$ with $q\in\mathbb{N}$ to denote the Taylor expansion is truncated at the $q$-th order in this paper.

The above parameterization method has been implemented in SSMTool~\cite{ssmtool2}, an open-source package for the computation of invariant manifolds. We note that \textsc{coco}~\cite{COCO,dankowicz2013recipes,ahsan2022methods}, a continuation package, has been integrated into SSMTool~\cite{li2022NonlinearAnalysisForceda,liNonlinearAnalysisForced2022c}. This integration enables the detection of bifurcations of fixed points of~\eqref{eq:red-nonauto-th}, parameter continuation of these bifurcated fixed points, as well as the continuation of (bifurcated) limit cycles of the ROM~\eqref{eq:red-nonauto-th}. In particular, we use the po-toolbox of \textsc{coco} to obtain the limit cycles. In the po-toolbox, boundary-value problems are formulated for the limit cycles, and collocation methods are used to solve for the limit cycles and their associated monodromy matrices. The stability and bifurcation of these limit cycles are further analyzed using Floquet theory. Therefore, we can easily perform bifurcation analysis of quasi-periodic orbits using SSMTool. Indeed, we will use this package to study the bifurcation of quasi-periodic orbits of a suite of 1:2 internally resonant mechanical systems in the coming sections, ranging from simple oscillators to high-dimensional FE models.

\section{Two coupled nonlinear oscillators} 
\label{sec:coupled-nonlinear-oscillators}

As our first mechanical system with 1:2 internal resonance, we consider two coupled nonlinear oscillators. This is the simplest system with 1:2 internal resonance and acts as a benchmark study. Bifurcations of quasi-periodic orbits of such a two DOF system under variations of excitation frequency have been partially uncovered in recent studies~\cite{liNonlinearAnalysisForced2022c,song2024StrongNonlinearMixing}. Here, we revisit this simple system to present a thorough bifurcation analysis of quasi-periodic orbits under the variations in both excitation frequency and amplitude. As we will see, the system can display global bifurcations of quasi-periodic orbits such as homoclinic bifurcations. Moreover, with the help of parameter continuation, we clearly uncover a cascade of periodic doubling bifurcations, leading to chaotic attractors.

\subsection{Setup}

The governing equations of two coupled nonlinear oscillators are
\begin{equation}
\begin{split}
\ddot{x}_1+c_1\dot{x}_1+x_1+b_1x_1x_2=\epsilon f_1\cos \Omega t\\ 
\ddot{x}_2+c_2\dot{x}_2+4x_2+b_2x_{1}^{2}=\epsilon f_2\cos \Omega t. \label{eq:system1}
\end{split}
\end{equation}
The undamped natural frequencies of the linearized system are given as $\omega_1=1$ rad/s and $\omega_2=2$ rad/s. Thus, the system has a 1:2 internal resonance. In the following computations, system parameters are chosen as $c_1=0.005~\rm{N \cdot s/m}$, $c_2=0.01~\rm{N \cdot s/m}$, $b_1=b_2=1~\rm{N/m^2}$, $f_1=1~\rm{N}$, $f_2=0~\rm{N}$, and $\epsilon=0.01$, unless otherwise stated.

\subsection{Periodic and quasi-periodic orbits}

We first present the periodic and quasi-periodic orbits of the system under varying $\Omega$ obtained via reduction on SSM at $\mathcal{O}({3})$ truncation. We recall that these results for $x_1$ have been presented and also verified in~\cite{liNonlinearAnalysisForced2022c}. Here, we present the results for $x_1$ to make this paper self-contained. We will also present the forced response curve (FRC) of periodic and quasi-periodic orbits for $x_2$.

\begin{figure}[!ht]
    \centering
    \includegraphics[width=0.45\textwidth]{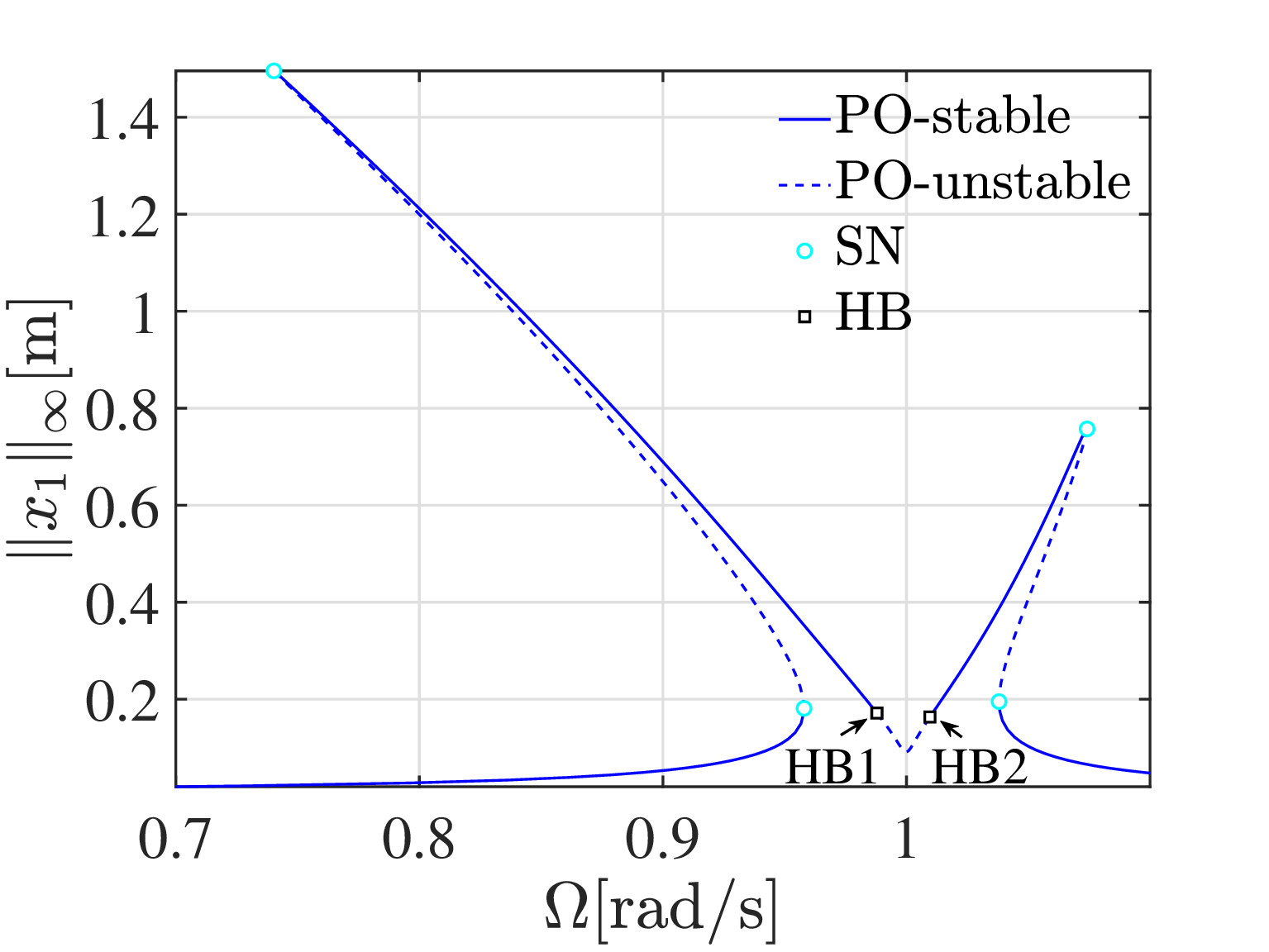}
    \includegraphics[width=0.45\textwidth]{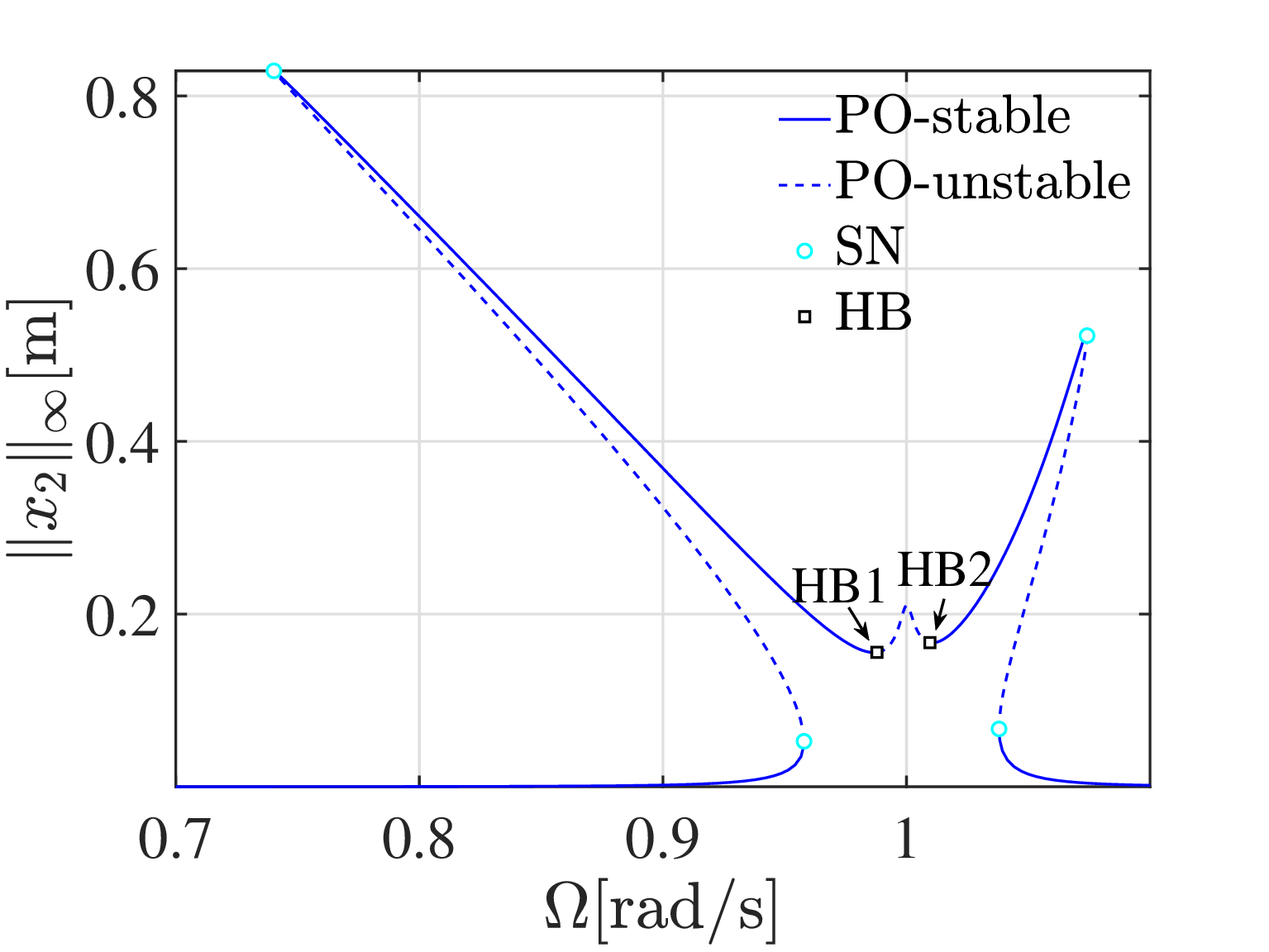}
    \caption{\small FRCs of periodic orbits of the coupled nonlinear oscillators in  (\ref{eq:system1}) with $\Omega \in$ [0.7,1.1]. In all figures throughout this paper, the solid and dashed lines denote stable and unstable solutions, unless otherwise stated. The circles and squares denote saddle-node (SN) and Hopf bifurcation (HB) points, respectively.}
    \label{fig:ep_x1_X2}
\end{figure}

The FRCs of periodic orbits (FRC-PO) of the two oscillators for $\Omega \in$ [0.7, 1.1] are shown in Fig.~\ref{fig:ep_x1_X2}. Here, $\Vert x_1 \Vert_\infty$ and $\Vert x_2 \Vert_\infty$ represent the infinite norm of $x_1(t)$ and $x_2(t)$, giving the amplitude of periodic or quasi-periodic response. There are four saddle-node (SN) bifurcation points and two secondary Hopf bifurcation (HB) points on the FRC-PO. In particular, the periodic orbits for $\Omega\in[\Omega_\mathrm{HB1},\Omega_\mathrm{HB2}]$ are unstable.

\begin{figure*}[!ht]
    \centering
    \includegraphics[width=0.45\textwidth]{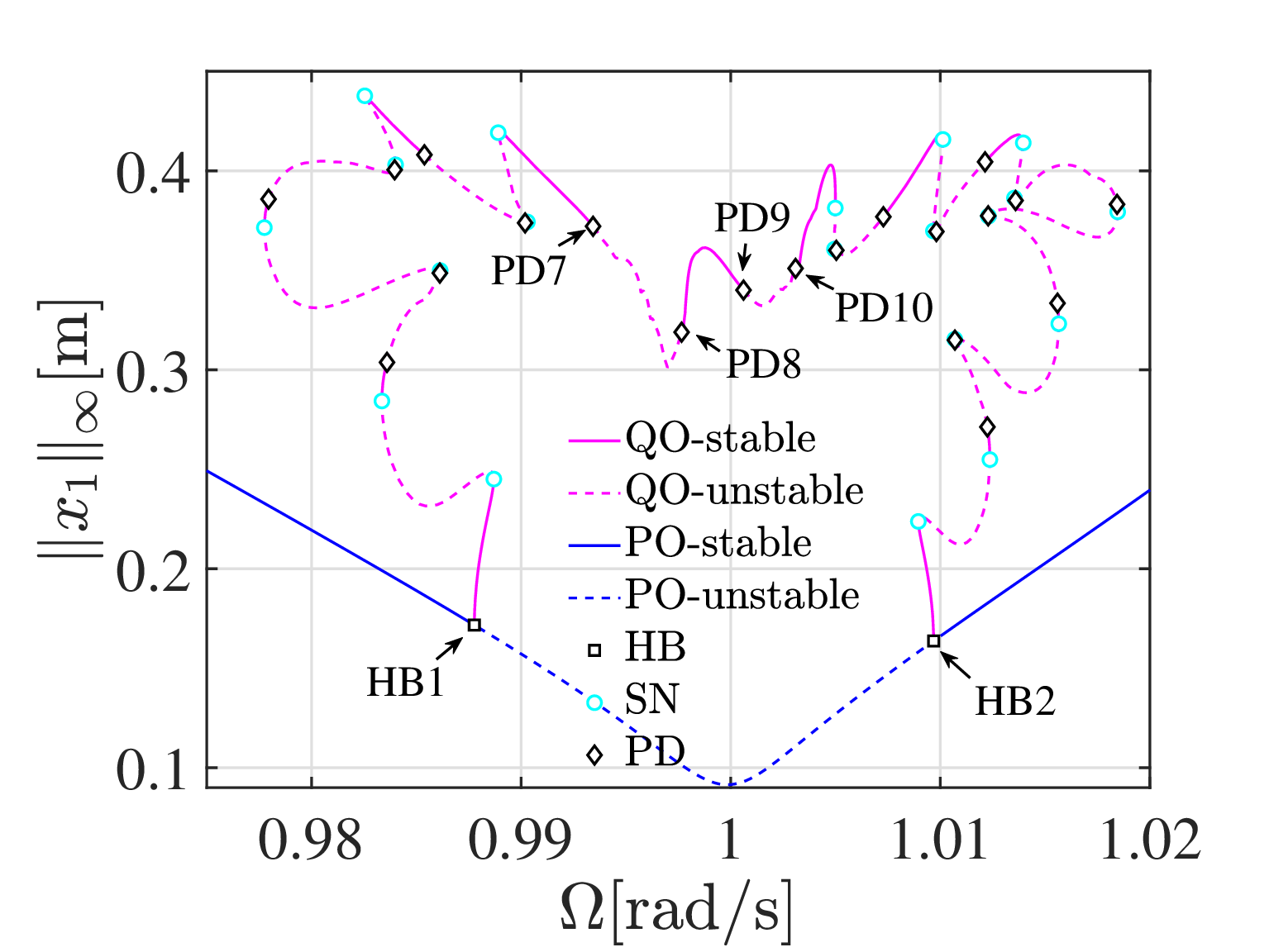}
    \includegraphics[width=0.45\textwidth]{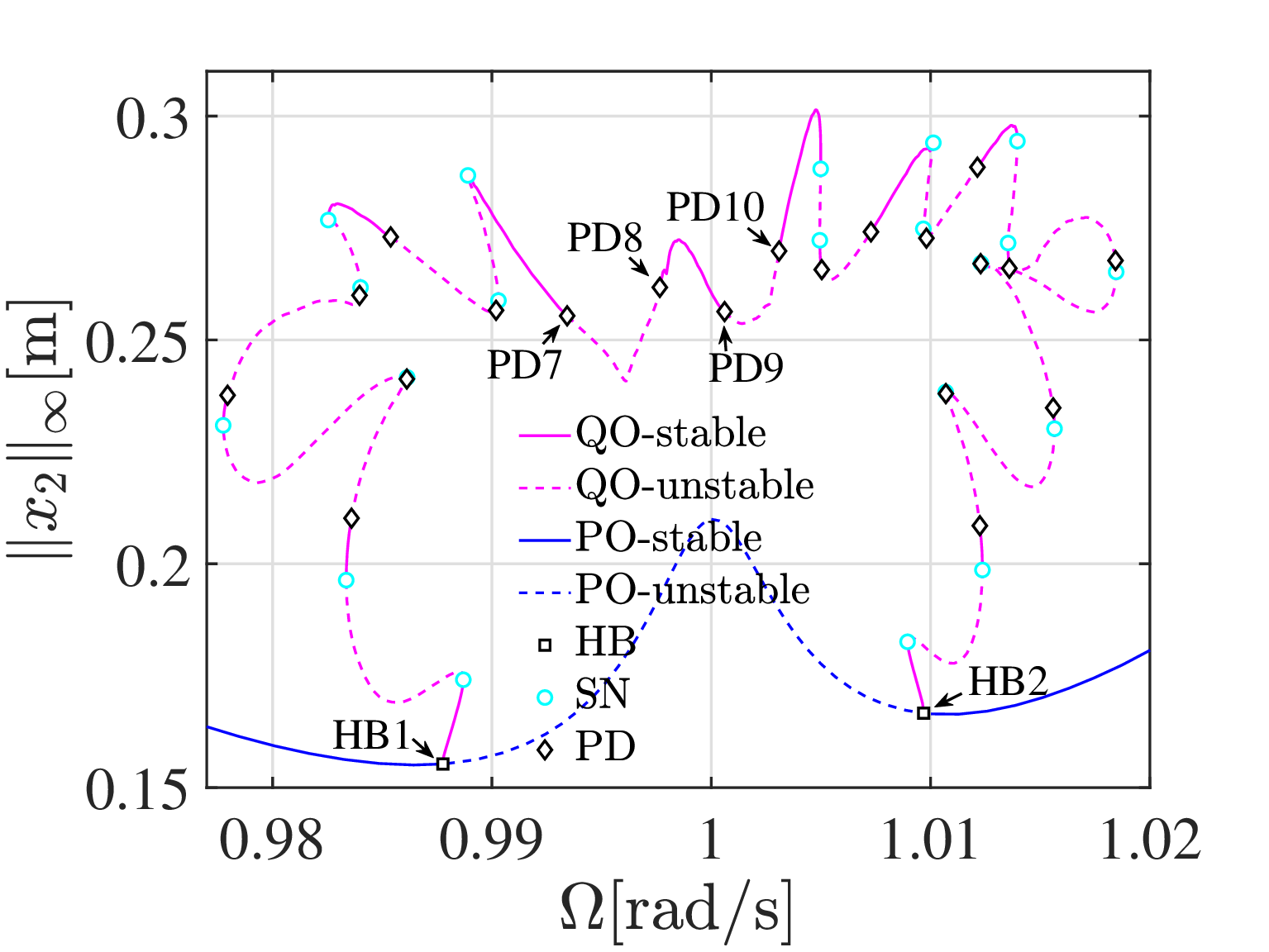}\\
    \caption{\small FRCs of quasi-periodic and periodic orbits of the system of two coupled oscillators in \eqref{eq:system1} with $\Omega \approx$ 1. The magenta solid/dashed lines denote the amplitudes of stable/unstable quasi-periodic orbits. The blue solid/dashed lines denote the amplitudes of stable/unstable periodic orbits. The circles, diamonds, and squares denote saddle-node (SN), period-double (PD), and Hopf bifurcation (HB) points, respectively.}
    \label{fig:HB2po_x1}
\end{figure*}

The secondary Hopf bifurcation marks the birth of quasi-periodic orbits. The FRCs of quasi-periodic orbits (FRC-QO) of the two oscillators under variations of $\Omega$ are shown in Fig.~\ref{fig:HB2po_x1}. We recall that these quasi-periodic orbits stay on some two-dimensional tori, which are computed as limit cycles of the ROM~\eqref{eq:red-nonauto-th} (cf. Fig.~\ref{fig:FulltoROM}). Therefore, the FRC-QO is obtained by parameter continuation of limit cycles that are born out of HB1. Along the FRC-QO, 20 period-doubling (PD) bifurcation points and 20 SN bifurcation points are observed, marking the complex bifurcations of quasi-periodic orbits. We note that the appearance of PD bifurcation points indicates that new families of quasi-periodic orbits with doubled periods can be obtained. We will analyze these new families of quasi-periodic orbits in Sect.~\ref{ssec:qusi2chaos}.

\subsection{PD and SN bifurcation curves}

Now we allow for the variations in the forcing amplitude $\epsilon$ to study how the PD and SN bifurcated quasi-periodic orbits in Fig.~\ref{fig:HB2po_x1} evolve with varying $\epsilon$. We expect that they will disappear if $\epsilon$ is sufficiently small. We note that both the PD and SN bifurcations are codimension-one bifurcation. Thus we will have a curve of PD or SN bifurcated quasi-periodic orbits under the variations in $(\Omega,\epsilon)$. Since the PD and SN bifurcations of quasi-periodic orbits correspond to PD and SN bifurcations of the limit cycles of the ROM~\eqref{eq:red-nonauto-th}, we use parameter continuation of PD or SN bifurcated limit cycles of the ROM to extract the bifurcation curve. In particular, we initialize our continuation runs with the bifurcation points on FRC-QO shown in Fig.~\ref{fig:HB2po_x1}.


\begin{figure*}[!ht]
    \centering
    \includegraphics[width=0.45\textwidth]{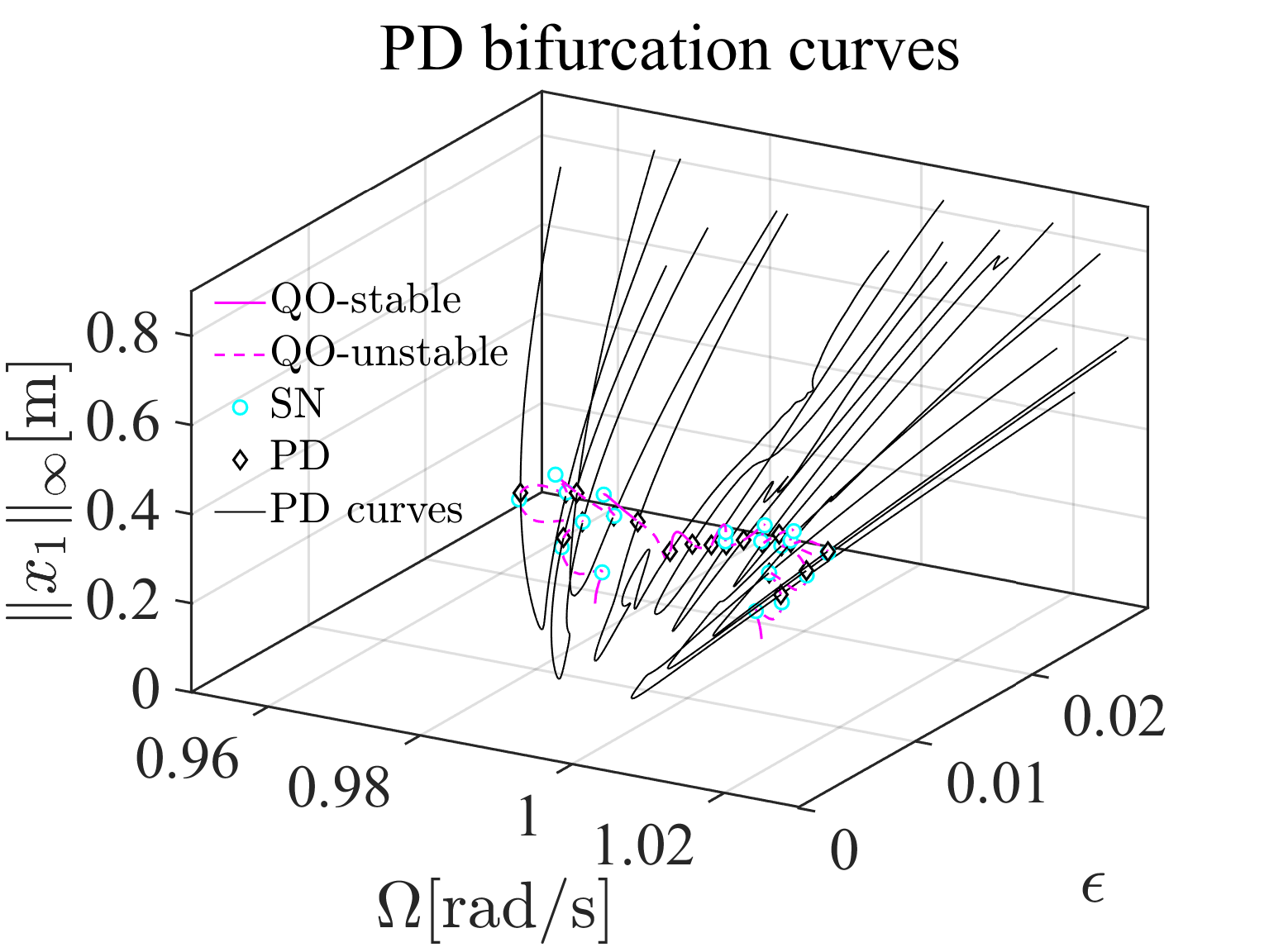} 
    \includegraphics[width=0.45\textwidth]{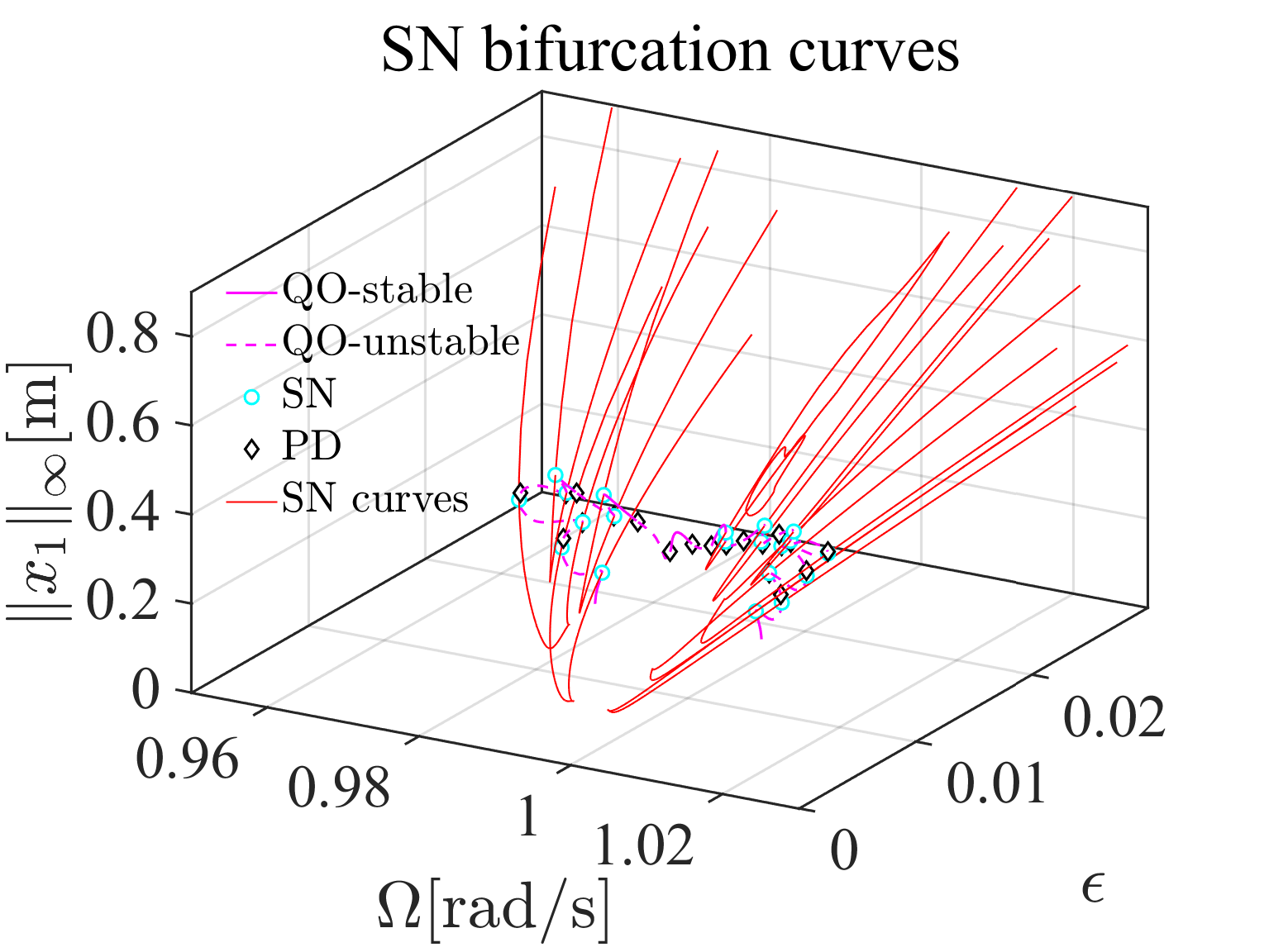} \\
    \includegraphics[width=0.45\textwidth]{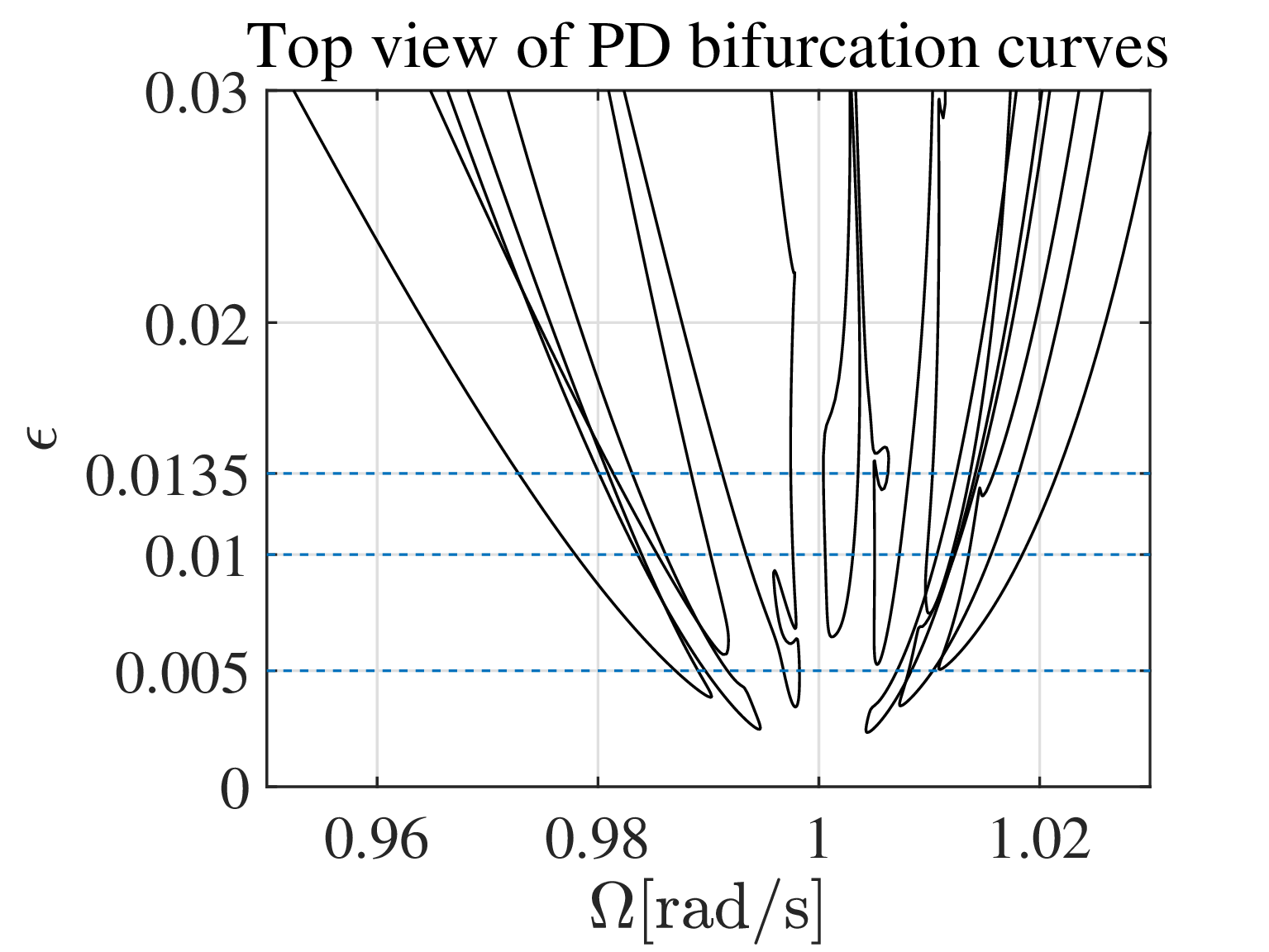}
    \includegraphics[width=0.45\textwidth]{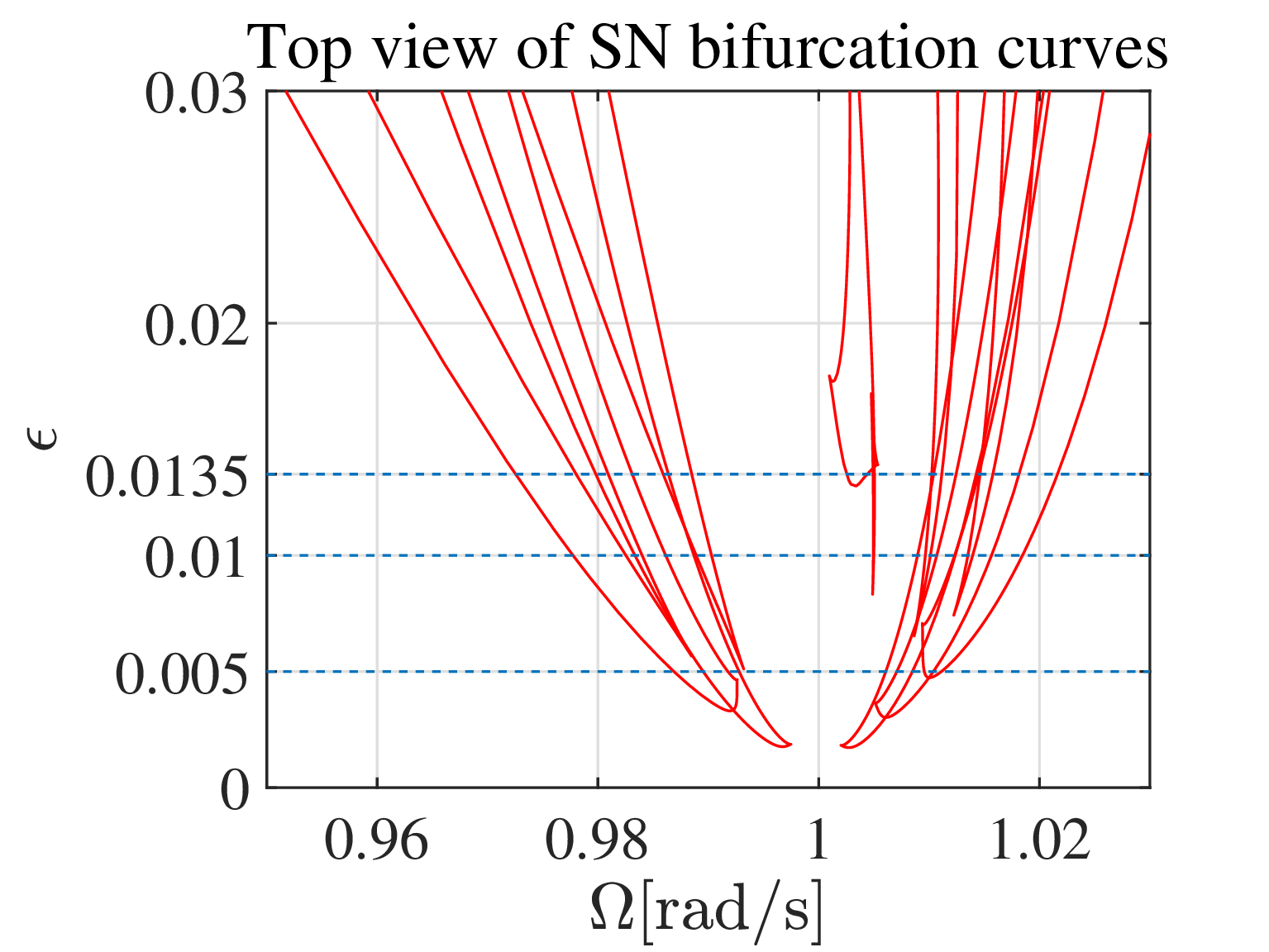}
    \caption{\small Bifurcation curves of quasi-periodic orbits of two coupled oscillators. The upper-left and upper-right panels present the projection of the continuation paths of period-doubling (PD) and saddle-node (SN) bifurcated quasi-periodic orbit onto $(\Omega,\epsilon,||x_1||_{\infty})$. The lower two panels give the top view of the corresponding upper panels. The black/red solid lines denote the continuation path of PD/SN bifurcated quasi-periodic orbits.}
    \label{fig:po2PD_x1}
\end{figure*}

The continuation paths for the PD and SN bifurcation points with $(\Omega,\epsilon) \in $ [0.7, 1.1] $\times$ [0.0001, 0.03] are shown in the left and right panels of Fig.~\ref{fig:po2PD_x1}, respectively. As the upper bound of $\epsilon$ is increased to 0.03, we use $\mathcal{O}(7)$ truncation for SSM reduction to ensure converged predictions. Here, the lower panels are the projection of the corresponding upper panels. We observe from the figure that multiple branches exist for the PD or SN continuation path. The amplitudes of SN and PD bifurcated quasi-periodic orbits decrease as $\epsilon$ decreases, as seen in the upper two panels. Moreover, some PD or SN branches merge and then disappear as $\epsilon$ decreases, resulting in a dynamic change in the number of PD or SN bifurcation points on FRC-QO. For example, when $\epsilon$ is equal to 0.01, there are 20 SN and 20 PD bifurcation points, while at $\epsilon$ = 0.005, there are only 10 SN and 10 PD bifurcation points, and at $\epsilon$ = 0.0135, both numbers are increased to 22.

\subsection{Homoclinic bifurcation of quasi-periodic orbits}

We observe from the right panel of Fig.~\ref{fig:HB2po_x1} that the amplitude of the quasi-periodic orbits is close to that of the unstable periodic orbits for $\Omega\approx1$. If the forcing amplitude $\epsilon$ is increased, it may occur that the quasi-periodic orbits intersect with the periodic orbits. In this subsection, we will discuss how the intersection of quasi-periodic and periodic orbits leads to the emergence of homoclinic bifurcations of quasi-periodic orbits. 

\begin{figure}[!ht]
    \centering
    \includegraphics[width=0.5\textwidth]{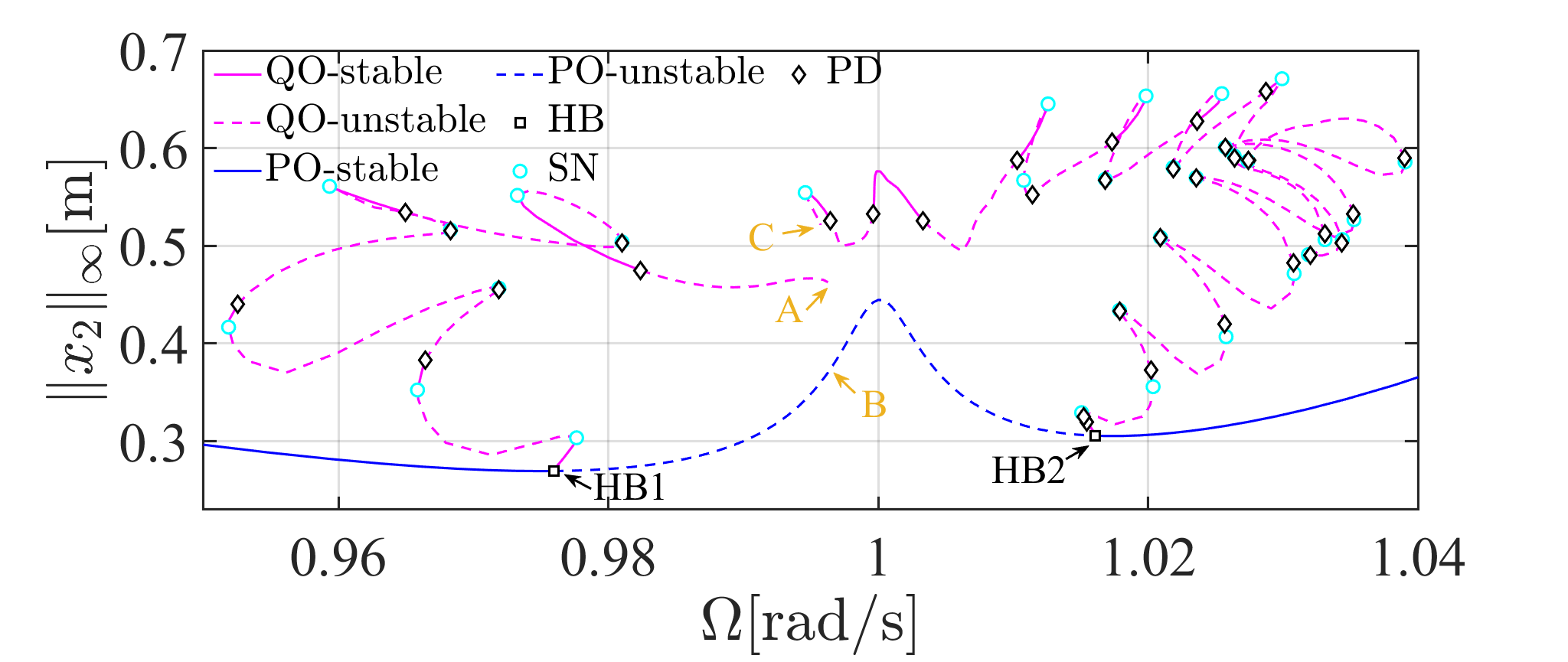}\\
    \includegraphics[width=0.5\textwidth]{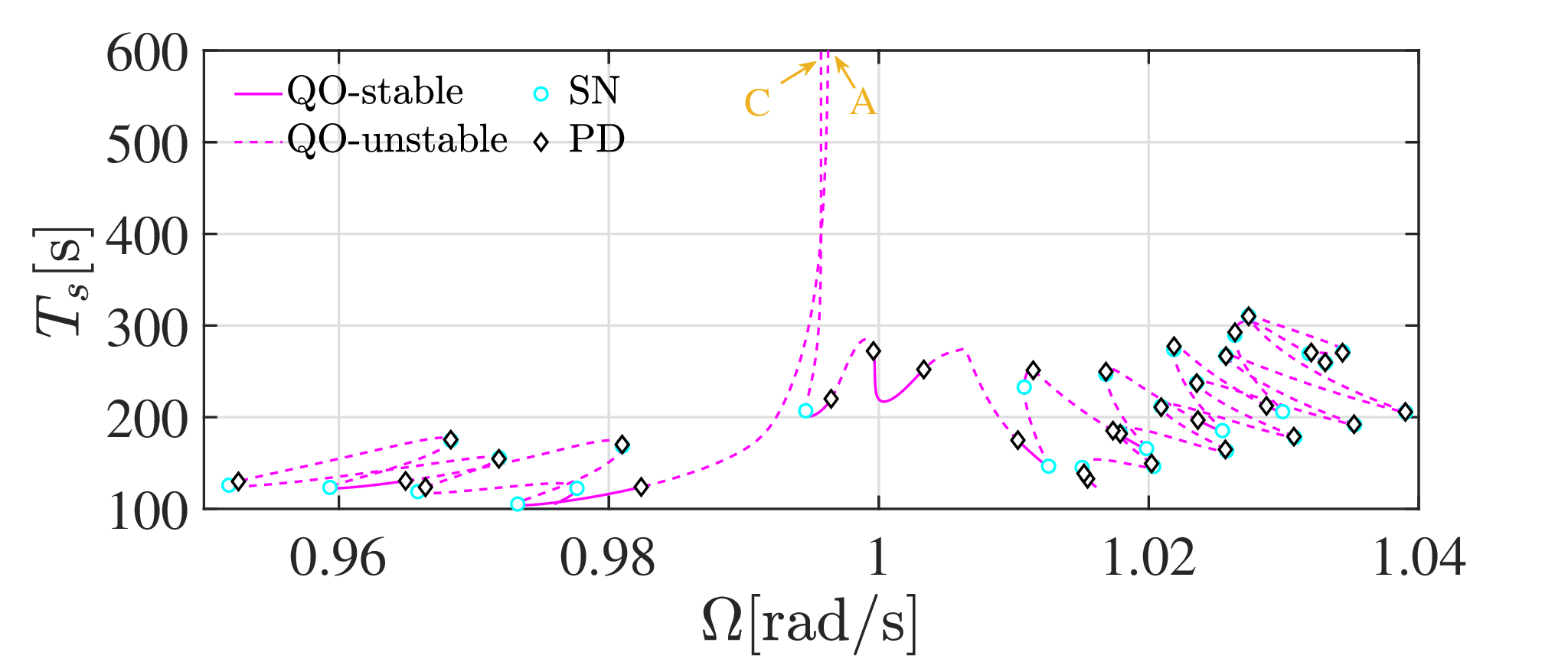}
    \caption{\small The upper panel presents the FRCs of quasi-periodic and periodic orbits of the second oscillator in  (\ref{eq:system1}) with $\epsilon$ = 0.03. Each of these quasi-periodic orbits corresponds to a limit cycle of the SSM-based ROM. The lower panel presents the period of these corresponding limit cycles. The magenta solid/dashed lines denote stable/unstable quasi-periodic orbits. The blue solid/dashed lines denote the amplitudes of stable/unstable periodic orbits. The circles, diamonds, and squares denote saddle-node (SN), period-double (PD), and Hopf bifurcation (HB) points, respectively.}
    \label{fig:homo_x1}
\end{figure}

We first compute the FRC-PO and FRC-QO with varying $\Omega$ but fixed $\epsilon=0.03$. The obtained results are plotted in Fig.~\ref{fig:homo_x1}. Here, the continuation run of quasi-periodic orbits starting at HB1 is terminated at point A of FRC-QO shown in Fig.~\ref{fig:homo_x1}, where the amplitude $||x_2||_\infty$ barely changes. We note that each quasi-periodic orbit on the FRC-QO consists of two incommensurable base frequencies: one is the excitation frequency $\Omega$, and the other one is much slower and given by $2\pi/T_s$, where $T_s$ is the period of the corresponding limit cycle of the SSM-based ROM~\eqref{eq:red-nonauto-th}~\cite{liNonlinearAnalysisForced2022c}. As seen in the lower panel of Fig.~\ref{fig:homo_x1}, $T_s\to\infty$ as the continuation approaches point A. Therefore, the quasi-periodic orbit undergoes a infinite-period bifurcation~\cite{Strogatz2018Nonlinear} when $\Omega\to\Omega_\mathrm{A}$. As mentioned in~\cite{Strogatz2018Nonlinear}, there are two kinds of infinite-period bifurcation: saddle-node on invariant circle bifurcation (SNIC) and homoclinic bifurcation. Next, we show that our observed infinite-period bifurcation is the second kind.

\begin{figure}[!ht]
    \centering
    \includegraphics[width=0.45\textwidth]{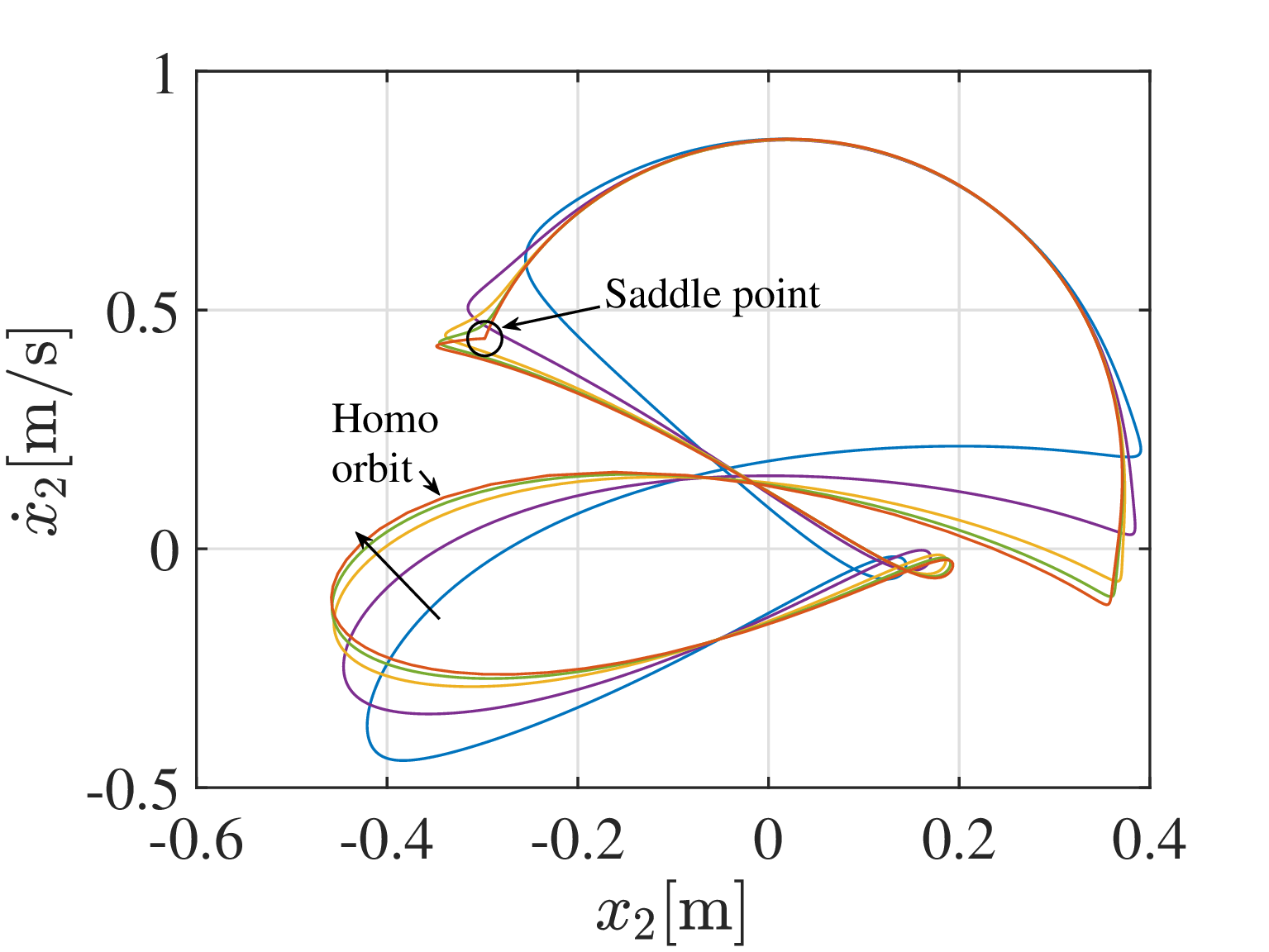}
    \includegraphics[width=0.45\textwidth]{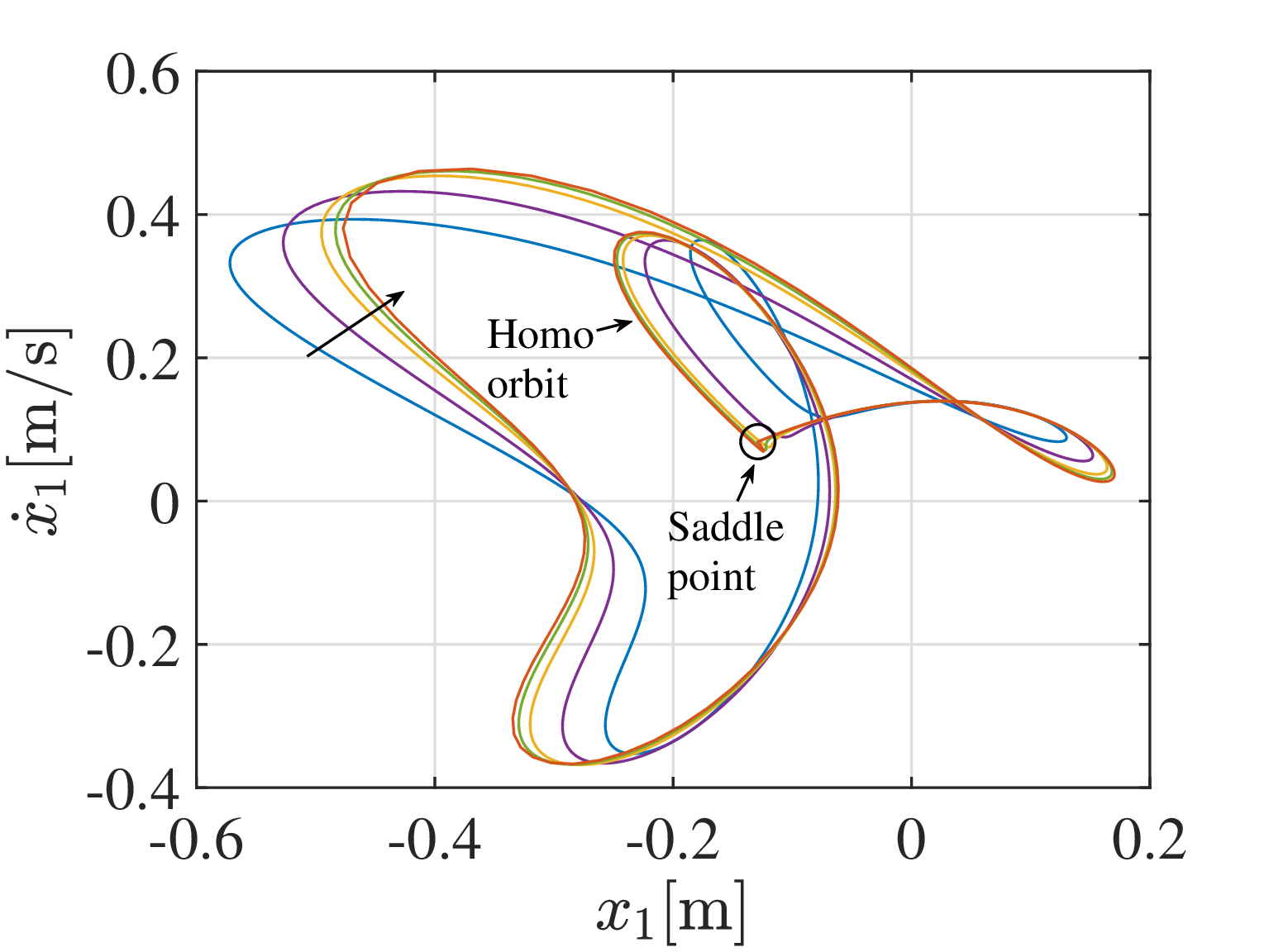}
    \caption{\small Intersections of the period-$2\pi/\Omega$ map of sampled tori of two coupled oscillators with $\epsilon=0.03$ and $\Omega\to\Omega_\mathrm{A}$ (cf. Fig.~\ref{fig:homo_x1}), obtained by limit cycles from the SSM-based analysis. The arrow denotes an increase of $\Omega$ from 0.9952 to $\Omega_{\rm{homo}}$ = 0.9963. The homo orbit in the figure denotes the homoclinic orbit. The upper and lower panels give the projection of the intersections onto $(x_2,\dot{x}_2)$ and $(x_1,\dot{x}_1)$ planes.}
    \label{fig:homo_saddle}
\end{figure}

To have a close look at the infinite-period bifurcation, we compute the intersection of the period-$2\pi/\Omega$ Poincar{\'e} map for some of these quasi-periodic orbits as $\Omega\to\Omega_\mathrm{A}$. We note that these quasi-periodic orbits stay on some two-dimensional invariant tori, and the intersection of these tori with the Poincar{\'e} section results in closed curves, which can be obtained by mapping the limit cycles of the SSM-based ROM~\eqref{eq:red-nonauto-th} back to physical coordinates~\cite{liNonlinearAnalysisForced2022c}. The obtained results are presented in Fig.~\ref{fig:homo_saddle}, from which we see that the limit cycle with an infinite period has a kink. This kink corresponds to a saddle fixed point. Therefore, we infer that the infinite-period bifurcation corresponds to a homoclinic bifurcation of quasi-periodic orbits.  

\begin{figure}[!ht]
    \centering
    \includegraphics[width=0.45\textwidth]{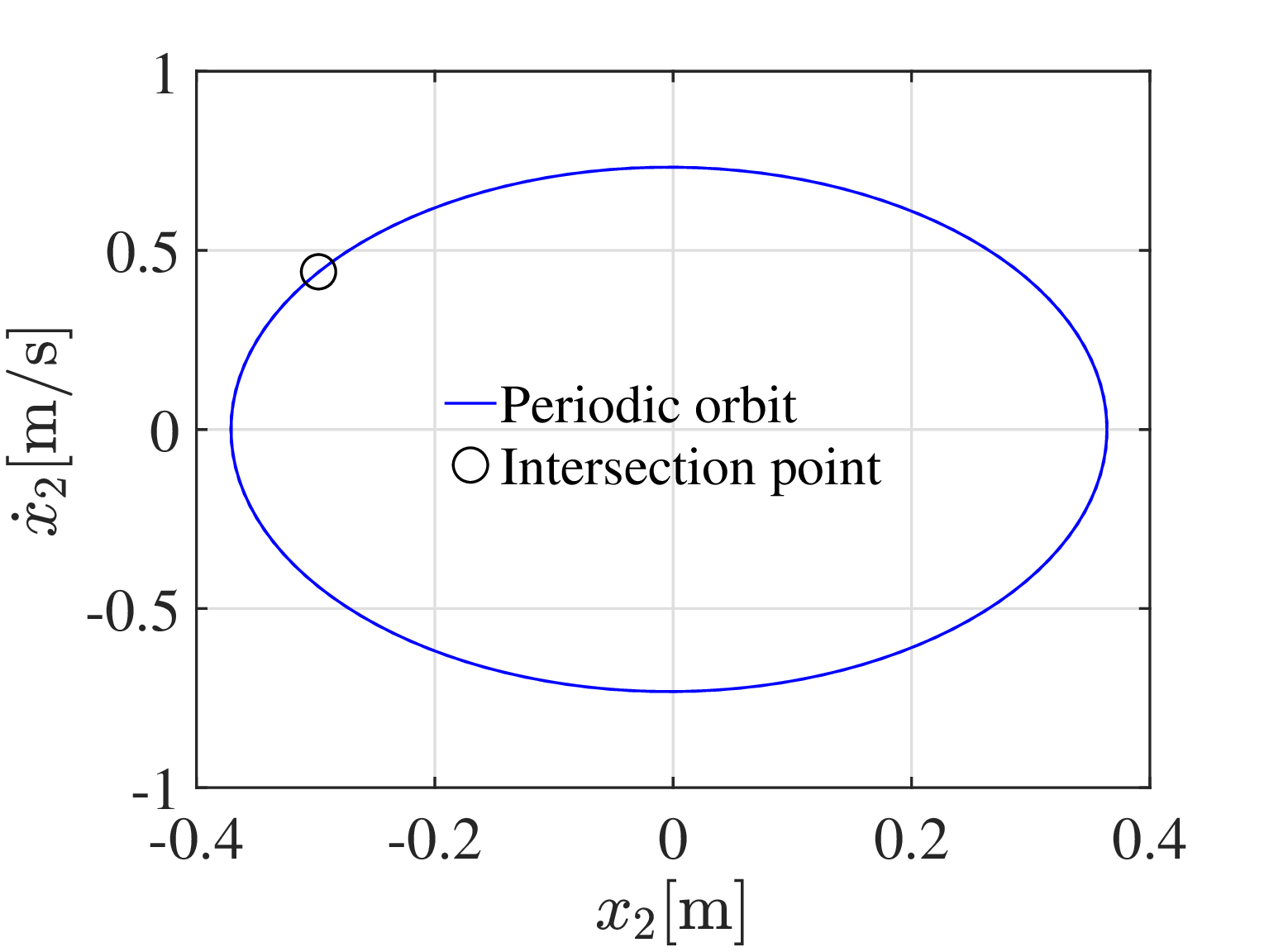}
    \includegraphics[width=0.45\textwidth]{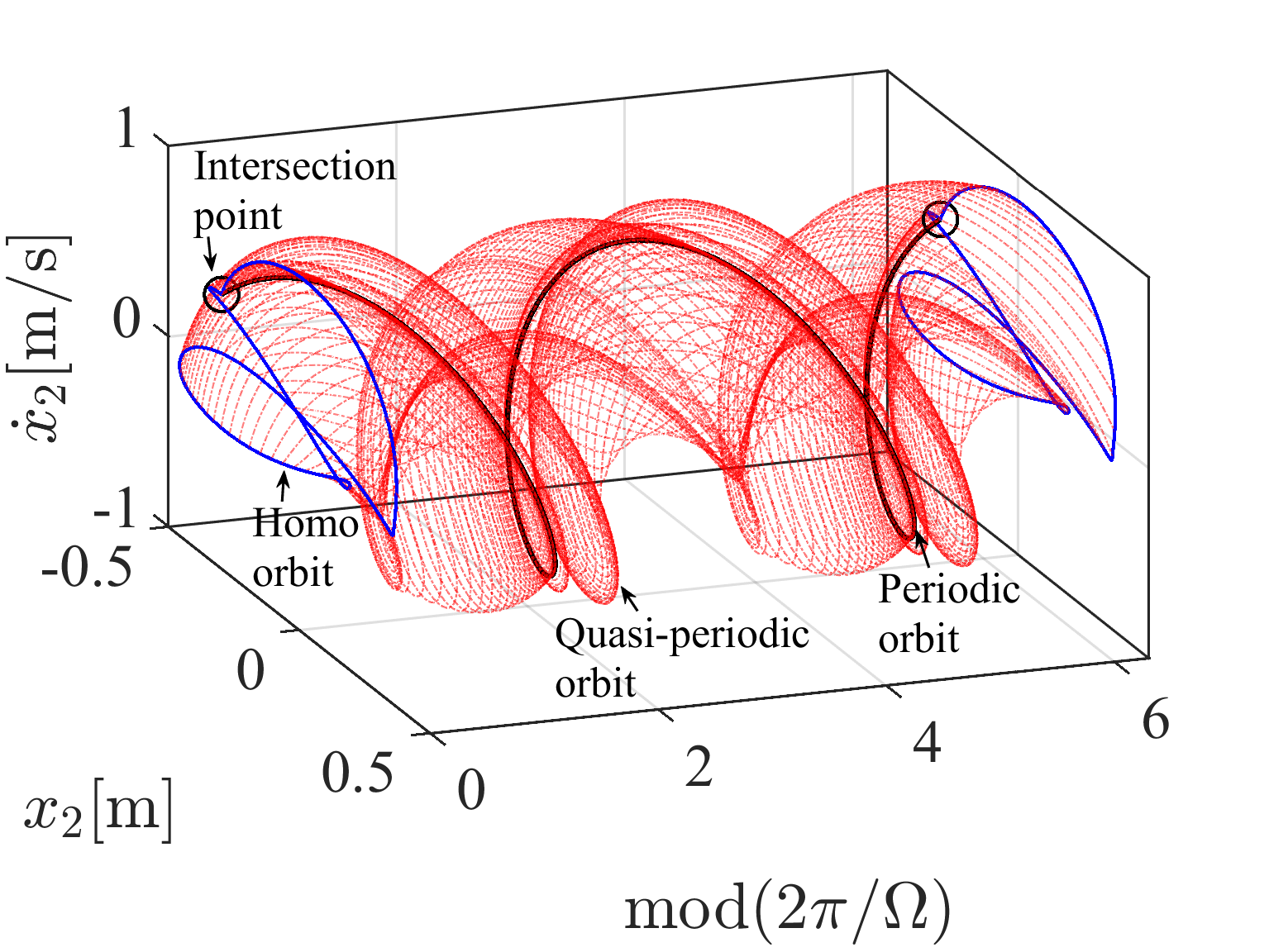}
    \caption{\small (Upper panel) A periodic orbit of the coupled oscillator \eqref{eq:system1} corresponding to the saddle fixed point on the Poincar{\'e} section shown in Fig.~\ref{fig:homo_saddle}. The black cycle denotes the intersection point of the periodic orbit with the Poincar{\'e} section. (Lower panel) A visualization of the torus on which the quasi-periodic orbit A stays and the periodic orbit B (cf. the upper panel of Fig.~\ref{fig:homo_x1}). Their intersections with the Poincar{\'e} section illustrate a homoclinic bifurcation.}
    \label{fig:saddle&period}
\end{figure}

A saddle fixed point on the Poincar{\'e} section corresponds to a periodic orbit of the full system with period-$2\pi/\Omega$. We plot the projection of this periodic orbit shown in the upper panel of Fig.~\ref{fig:saddle&period}. In fact, this periodic orbit is the one marked by B on the FRC-PO given by the upper panel of Fig.~\ref{fig:homo_x1}. The two markers A and B in the upper panel of Fig.~\ref{fig:homo_x1} have not coincided because the vertical axis denotes the amplitude of periodic or quasi-periodic orbits. Indeed, as seen in the upper panel of Fig.~\ref{fig:saddle&period}, the intersection point for the periodic orbit B at the Poincar{\'e} section does not coincide with the peak position of the orbit. We plot the torus A along with the periodic orbit B in the lower panel of Fig.~\ref{fig:saddle&period} for a direct visualization of the intersection, which illustrates clearly the global bifurcation, i.e., homoclinic bifurcation of quasi-periodic orbit.

\begin{figure}[!ht]
    \centering
    \includegraphics[width=0.45\textwidth]{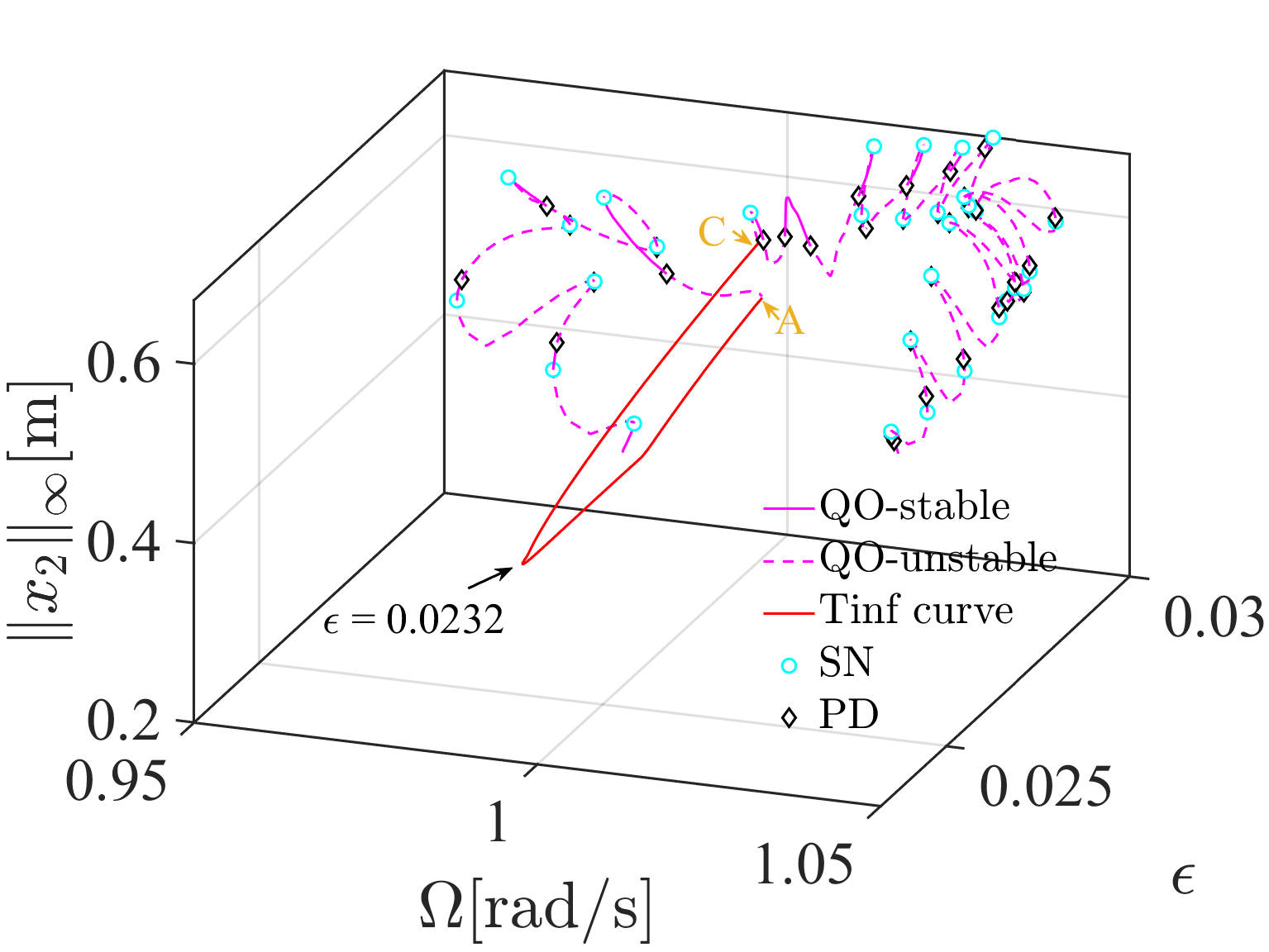}
    \caption{\small The continuation path of homoclinic bifurcated quasi-periodic orbits (red line) along with the forced response curve of quasi-periodic orbits at $\epsilon$ = 0.03.}
    \label{fig:poTinf_x2}
\end{figure}

We conclude this subsection by investigating how this global bifurcation evolves as $\epsilon$ changes. We expect that the homoclinic bifurcation will disappear when $\epsilon$ is less than a critical value. We perform parameter continuation of the homoclinic orbit under the variations in $(\Omega,\epsilon)$~\cite{liNonlinearAnalysisForced2022c,li2023nonlinear}. In particular, we fix the period of the associated limit cycle of the SSM-based ROM to be 5000, perform continuation of the limit cycle, and then map these limit cycles back to homoclinic bifurcated quasi-periodic orbits of the original system. The obtained continuation path of the homoclinic bifurcations is shown in Fig.~\ref{fig:poTinf_x2} (marked as Tinf curve). We observe that the continuation path starting at point A terminates at point C, suggesting that there are two homoclinic bifurcations when $\epsilon = 0.03$. In fact, this point C can also obtained from FRC-QO starting at HB2, as seen in Fig.~ \ref{fig:homo_x1}. We also find that there is no limit cycle with an infinite period when $\epsilon$ is less than 0.0232. Thus, the critical value of $\epsilon$ for the homoclinic bifurcation is 0.0232, below which there is no homoclinic bifurcations. 


\subsection{Periodic doubling bifurcations to chaos} \label{ssec:qusi2chaos}

\begin{figure}[!ht]
    \centering
    \includegraphics[width=0.48\textwidth]{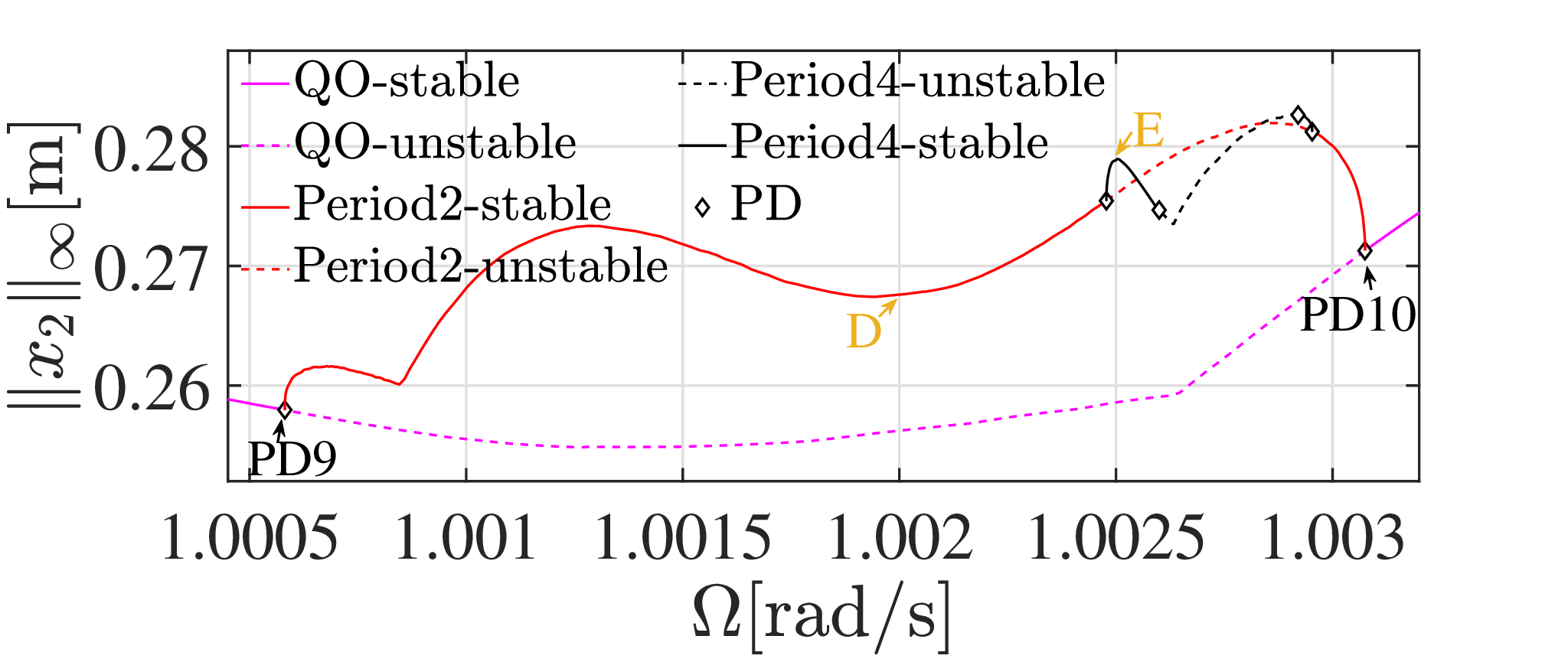}
    \caption{\small FRCs of quasi-periodic orbits of the second oscillator in \eqref{eq:system1} with $\Omega \in [\Omega_{\rm{PD9}},\Omega_{\rm{PD10}}]$. The magenta lines denote the FRC-QO obtained by switching at HB1 in FRC-PO; red lines denote the FRC of period2 quasi-periodic orbits obtained by switching at PD9 in the FRC-QO; black lines denote the FRC of period4 quasi-periodic orbits obtained by switching at PD bifurcations in period2 FRC-QO. The diamonds denote PD bifurcations. This figure illustrates clearly a cascade of periodic doubling bifurcations.}
    \label{fig:L3_x2}
\end{figure}

Now we revisit Fig.~\ref{fig:HB2po_x1} and discuss the periodic doubling bifurcation of quasi-periodic orbits. In particular, we extract the new families of quasi-periodic orbits born out of these PD bifurcations. For compactness, we focus on the region where $\Omega\in[\Omega_\mathrm{PD9},\Omega_\mathrm{PD10}]$. We switch to the continuation of limit cycles born out of PD9 of the ROM~\eqref{eq:red-nonauto-th}, and then map them back to quasi-periodic orbits of the original system. The FRC of these quasi-periodic orbits is plotted in red lines in Fig.~\ref{fig:L3_x2} and is denoted by 'Period2'. As expected, it also intersects with the main FRC-QO (magenta lines) at PD10. Along this FRC, two new PD bifurcation points are detected at $\Omega\approx1.0025$ and $\Omega\approx1.003$, marking the existence of a new family of periodic orbits of doubled period. We switch the parameter continuation to this new family of quasi-periodic orbits and obtain the FRC plotted in blacked lines in Fig.~\ref{fig:L3_x2}. The internal period-$T_s$ gets further doubled for these quasi-periodic orbits, and hence they are denoted as 'Period4' in the figure. Again, we detect two PD bifurcation points along the FRC of 'Period4' quasi-periodic orbits. In short, the system displays a cascade of period-doubling bifurcations for the quasi-periodic orbits between the frequency internal $\Omega\in[\Omega_\mathrm{PD9},\Omega_\mathrm{PD10}]$.

\begin{figure}[!ht]
    \centering
    \includegraphics[width=0.45\textwidth]{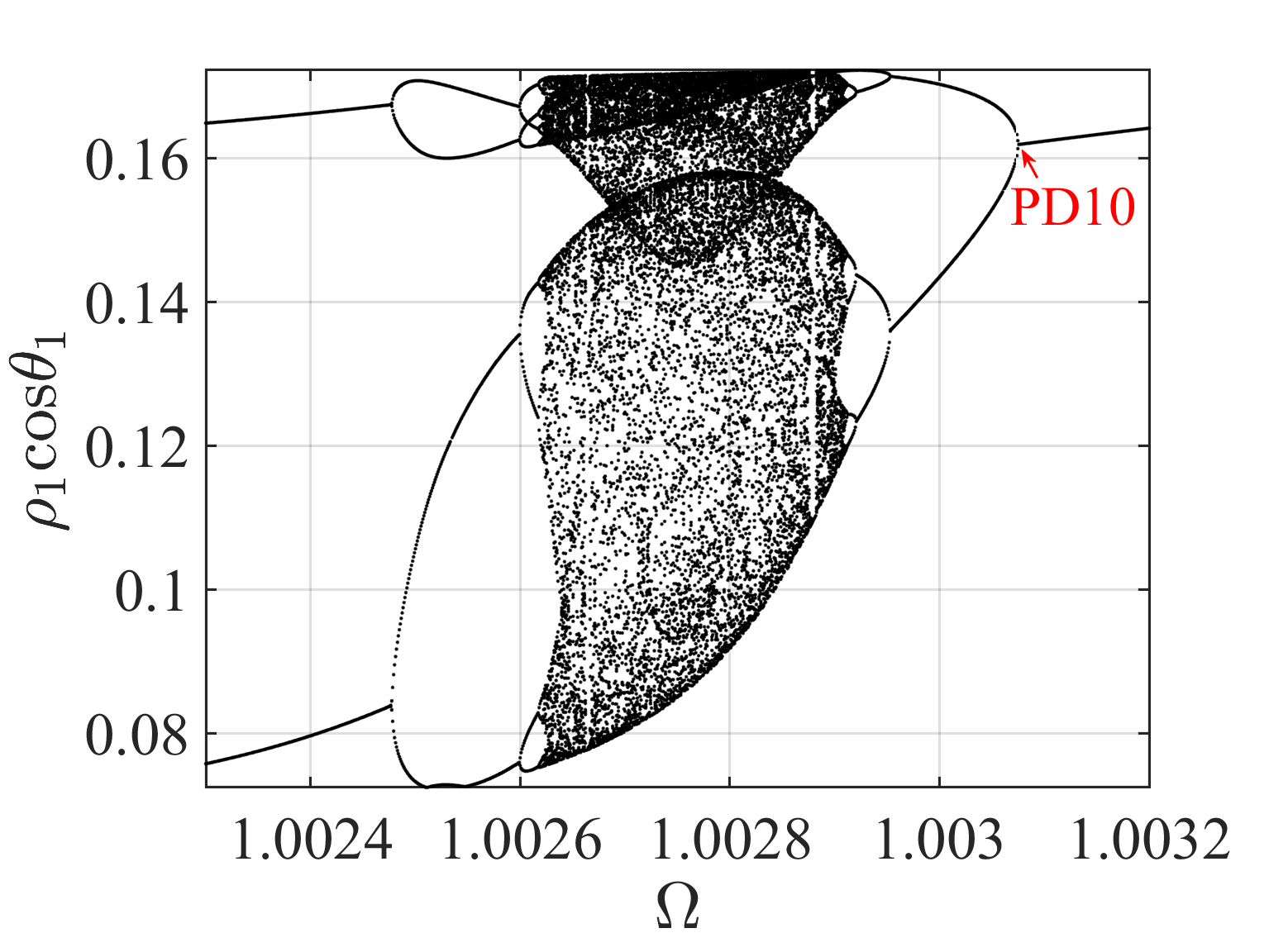}
    \caption{\small Bifurcation diagram of attractors for the SSM-based ROM \eqref{eq:red-nonauto-th} of the second oscillator in \eqref{eq:system1} with $\Omega \in [1.0023,1.0032]$. Here, the vertical axis gives the intersections of the attractors with the Poincar{\'e} section $\{(\boldsymbol{\rho},\boldsymbol{\theta}): q_1=\rho_1\cos\theta_1, \dot{q_1}\equiv0,\ddot{q}_1>0\}$ of the ROM, namely, the maximum of $\rho_1\cos\theta_1$ associated with the attractors. For a given $\Omega$, the attractor is a period-1/period-2/period-4/period-8 limit cycle of the ROM if the number of intersection points is 1/2/4/8. Likewise, the number of intersection points is infinite if the attractor is chaotic.}
    \label{fig:bif-chaos-ex1}
\end{figure}

A cascade of period-doubling bifurcations is a typical route to chaos. To have a close look at the transition from quasi-periodic to chaotic attractors, we present the bifurcation diagram for the attractors of the SSM-based ROM~\eqref{eq:red-nonauto-th} for $\Omega\in[1.0023,1.0032]$ (cf. Fig.~\ref{fig:L3_x2}) in Fig.~\ref{fig:bif-chaos-ex1}. This diagram is obtained via performing forward simulations of the ROM~\eqref{eq:red-nonauto-th} at given sampled frequencies $\Omega$, extracting corresponding attractors in steady state, and locating the intersections of these attractors with appropriately constructed Poincar{\'e} map. This bifurcation diagram displays the transition to chaos via the cascade of period-doubling bifurcations. This bifurcation diagram is consistent with that of Fig.~\ref{fig:L3_x2} but provides a more complete picture of the complex dynamics in this parameter region.

\begin{figure*}[!ht]
    \centering
   \includegraphics[width=0.31\textwidth]{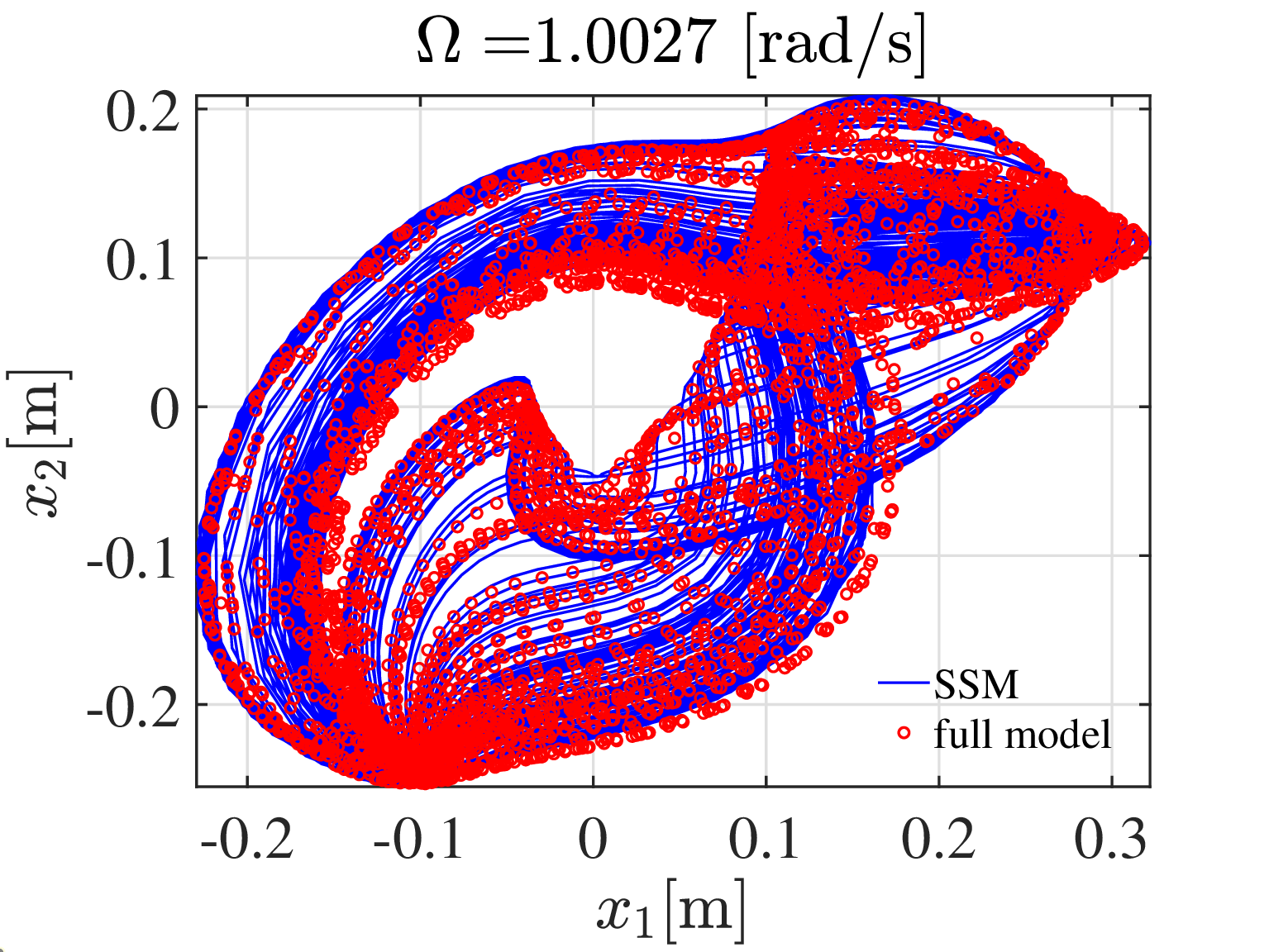} 
   \includegraphics[width=0.31\textwidth]{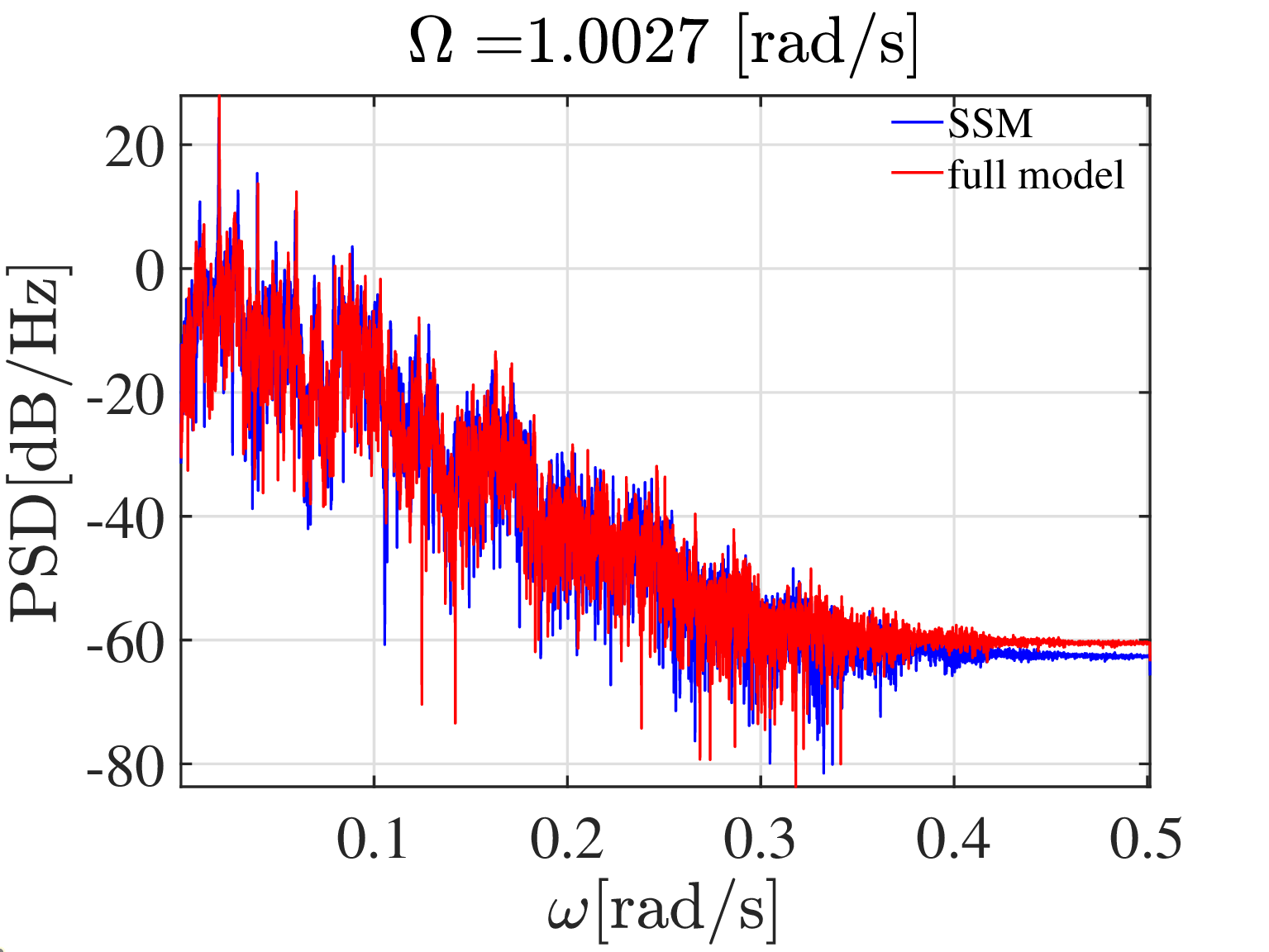}
   \includegraphics[width=0.31\textwidth]{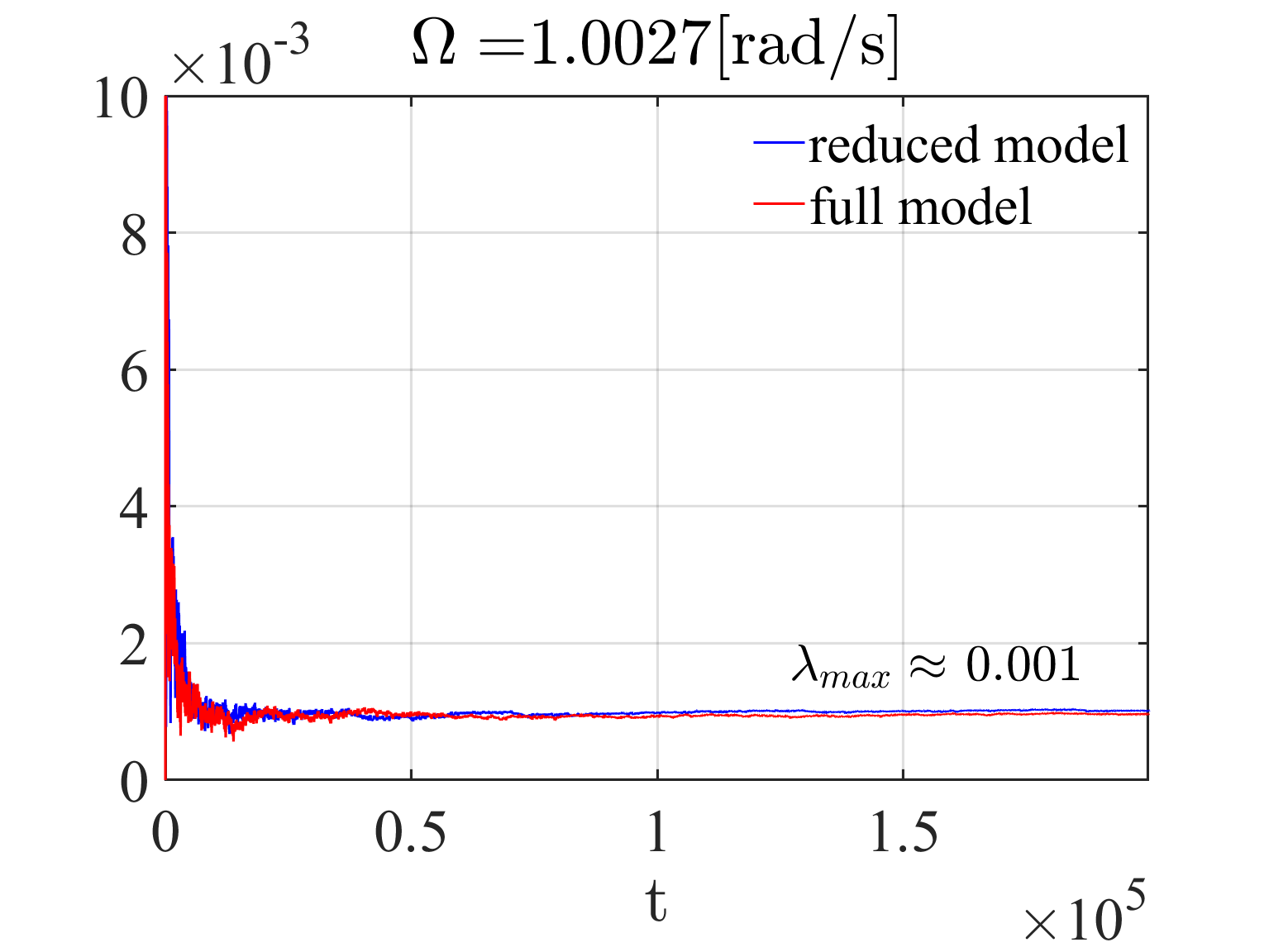}
    \caption{\small (Left panel) Intersections of the period-$2\pi/\Omega$ map of the chaotic attractor of the two coupled nonlinear oscillators~\eqref{eq:system1} with $\Omega$ = 1.0027. The blue lines denote the SSM-based prediction, and the red cycles denote reference results of the original (full) model obtained using numerical integration. (Middle panel) The power spectral density (PSD) of the intersected trajectories at the left panel. (Right panel) The time history of maximum Lyapunov exponent of the chaotic attractor.}
    \label{fig:chaos_x2}
\end{figure*}

To further check these strange attractors are indeed chaotic, we randomly select one representative at $\Omega$ = 1.0027 to conduct a detailed analysis. We map the attractor back to physical coordinates and then plot the intersection of this attractor with the Poincar{\'e} section induced by the period-$2\pi/\Omega$ map in the upper panel of Fig.~\ref{fig:chaos_x2}. As seen in the middle panel of the figure, the power spectral density of the intersected trajectory has significant values for a broad frequency bandwidth. We observe from the right panel of Fig.~\ref{fig:chaos_x2} that the maximum Lyapunov exponent of the attractor is 0.001. These three panels together indicate that the attractor is indeed chaotic. Here, the Lyapunov exponent is obtained via a discrete method detailed in~\cite{geist1990comparison}. In this method, variational equations are integrated, and QR decomposition is utilized to perform reorthonormalization.

To verify the effectiveness of the SSM-based prediction, we use numerical integration of the original system to generate reference results for the purpose of comparison. We take points D and E in Fig.~\ref{fig:L3_x2} as representatives of Period2 and Period4 quasi-periodic orbits and plot their Poincar{\'e} intersections in Fig.~\ref{fig:verify_L2_stable} of Appendix section. The associated power spectral density (PSD) plots for these two quasi-periodic orbits are presented in Fig.~\ref{fig:PSD-qusia}. We observe that the results from the numerical integration match well with that of SSM-based prediction. As for the chaotic attractor, we follow a similar validation procedure using numerical integration, and the results are presented in Fig.~\ref{fig:chaos_x2}. The original system indeed admits a chaotic attractor that coincides well with the SSM-based prediction.

\section{A shallow curved beam}  
\label{sec:beam}

\begin{figure}[!ht]
    \centering
    \includegraphics[width=0.45\textwidth]{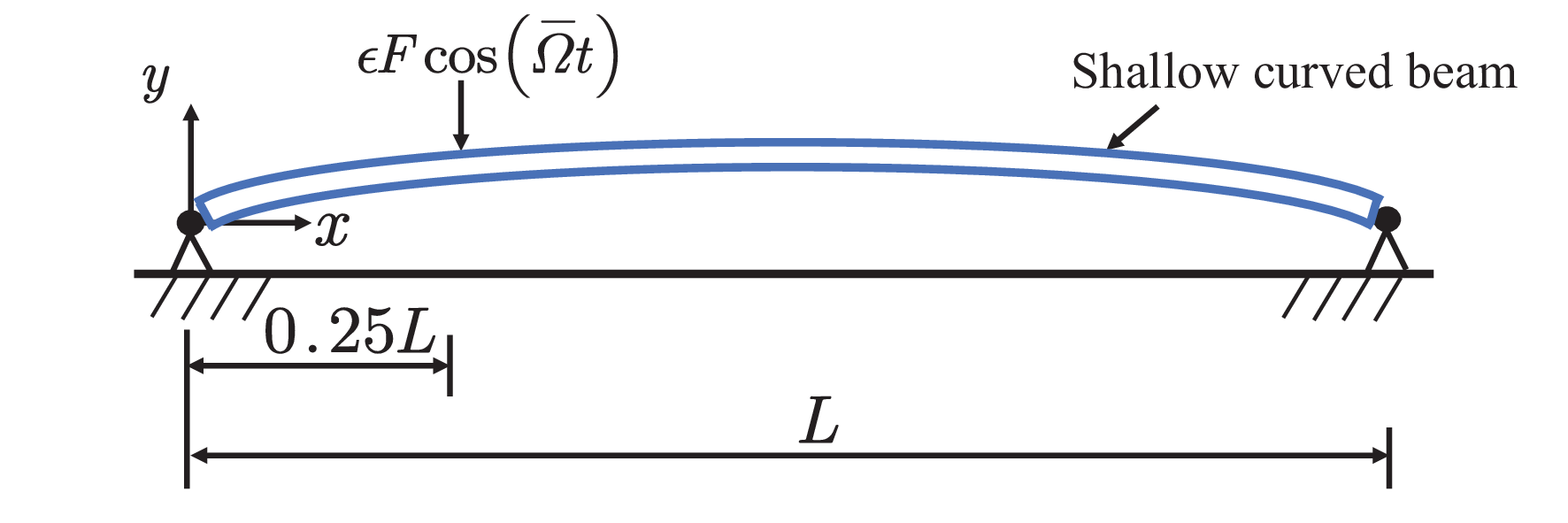}
    \caption{\small The schematic of a shallow curved beam.}
    \label{fig:beam}
\end{figure}

As our second mechanical system with 1:2 internal resonance, we consider a pinned-pinned shallow curved beam subject to a harmonic excitation shown in Fig.~\ref{fig:beam}. By tuning the curvature of the beam, the natural frequencies of the first two bending modes admit a 1:2 internal resonance. Here, we consider von K\'arm\'an type of geometric nonlinearity. This is a continuous system, and we will use the Galerkin approach to transform this infinite-dimensional system into a finite-dimensional nonlinear system. Then we apply the SSM reduction framework shown in Fig.~\ref{fig:FulltoROM} to study the bifurcation of quasi-periodic orbits of this system. As we will see, the FRC of quasi-periodic orbits for this curved beam consists of several isolated branches, which merge with the main branch via simple bifurcations. Again, we observe a cascade of period-doubling bifurcations of quasi-periodic orbits leading to chaotic motion. Such a chaotic attractor coexists with a stable quasi-periodic orbit on the isolas.

\subsection{Setup}

Let $Z_0(x)$ be the initial, imperfect configuration of the curved beam, $w$ be the  deflection relative to the initial configuration, the governing equation of the curved beam is given by~\cite{oz2006TwotooneInternalResonances}
\begin{equation}
\begin{aligned}
&EI\Partials{w}{x}{4}+\rho A \Partials{w}{t}{2} +c\frac{\partial w}{\partial t}= \frac{EA}{L}(\Partials{w}{x}{2}+\Partials{Z_0}{x}{2}) \\
&\int_{0}^{L} (\frac{1}{2}(\Partial{w}{x})^2+ \Partial{Z_0}{x} \Partial{w}{x})  dx + \epsilon F\delta(x-0.25L)\cos\bar{\Omega}t\label{eq:beam motion}
\end{aligned}
\end{equation}
along with boundary conditions
\begin{equation}
    w(0,t)=w(L,t)=0,\quad \Partials{w}{x}{2}(0,t)=\Partials{w}{x}{2}(L,t)=0.
\end{equation}
Here, $EI$ is flexural rigidity, $\rho$ is density, $A$ is the area of the cross-section, and $L$ is the length of the beam. We define the following dimensionless quantities~\cite{wang2012DynamicsSimplySupported}
\begin{align}
&\eta=\frac{w}{L},\ \xi =\frac{x}{L},\ \varphi _0=\frac{Z_0}{L},\ \tau =\frac{t}{L^2}\sqrt{\frac{EI}{\rho A}},\ \kappa =\frac{AL^2}{I}, \nonumber\\ &2\bar{\mu} = \frac{c L^2}{\sqrt{EI \rho A}},\,\ \bar{F}=\frac{FL^3}{EI},\ \Omega = {\bar{\Omega} L^2}{\sqrt{\frac{\rho A}{EI}}}\label{eq:overL}
\end{align}
and then \eqref{eq:beam motion} can be rewritten as
\begin{align}
   &\Partials{\eta}{\tau}{2}+2\bar{\mu}\Partial{\eta}{\tau}+\Partials{\eta}{\xi}{4}=\epsilon \bar{F}\delta(\xi-0.25)\cos\Omega \tau \nonumber\\
   &+\kappa (\Partials{\eta}{\xi}{2}+\Partials{\varphi_{0}}{\xi}{2}) \int_{0}^{1}(\frac{1}{2} (\Partial{\eta}{\xi})^{2}+\Partial{\varphi_{0}}{\xi} \Partial{\eta}{\xi}) d\xi ,
 \label{eq:dimensionless system} 
\end{align}
along with boundary conditions
\begin{equation}
    \eta(0,\tau)=\eta(1,\tau)=0,\quad \Partials{\eta}{\xi}{2}(0,\tau)=\Partials{\eta}{\xi}{2}(1,\tau)=0.
 \label{eq:dimensionless boundary} 
\end{equation}

We apply a Galerkin approach to discretize the partial-integral differential equation~\eqref{eq:dimensionless system}. Specifically, we substitute
\begin{equation}
\eta(\xi,\tau)=\sum_{j=1}^{n}\varPhi_j(\xi)q_j(\tau),\quad \varPhi _j ( \xi  ) =\sqrt{2}\sin j\pi\xi
 \label{eq:Galerkin}
\end{equation}
into \eqref{eq:dimensionless system} and then apply a Galerkin projection, yielding a system of second-order ordinary differential equations below
\begin{align}
\boldsymbol{M}&\ddot{\boldsymbol{q}}+\boldsymbol{C}\dot{\boldsymbol{q}}+ ( \boldsymbol{K}+\boldsymbol{H}  ) \boldsymbol{q}+\frac{1}{2}\boldsymbol{q}^T\boldsymbol{BqBq} +\frac{1}{2} \boldsymbol{q}^T\boldsymbol{Bqd}\nonumber\\
&+ \boldsymbol{d}^T\boldsymbol{qBq}=\epsilon\boldsymbol{f}\cos \Omega \tau, \ \  \ 0<\epsilon \ll 1,
 \label{eq:system 2}
\end{align}
where $\boldsymbol{q}\in\mathbb{R}^n$ is a generalized coordinate vector $\boldsymbol{M}$, $\boldsymbol{C}$, $\boldsymbol{K}$, $\boldsymbol{H}$, and $\boldsymbol{B}\in\mathbb{R}^{n\times n}$ are constant matrices, $\boldsymbol{d},\boldsymbol{f}\in\mathbb{R}^n$ and are constant vectors. Detailed expressions for these constant matrices and vectors are presented in~\eqref{eq:matrix} in the Appendix.

We use a parabolic function, namely
$Z_0 (x ) =4a_0x  \\ (L-x)$, to characterize the initial imperfect configuration. The associated $\varphi_0(\xi)=4a_0\xi(1-\xi)$. We tune $a_0$ such that the ratio of the first two undamped natural frequencies is 1:2. We take a five-mode truncation, i.e., $n=5$, because it is sufficient to generate converged solutions in our setting. Thus, the state space of the full system is 10, which is reduced to 4 via the SSM-based ROM~\eqref{eq:red-nonauto-th}. Numerical experiments show that the natural frequencies of the first two modes are given as $\omega_1=39.4784$ and $\omega_2=78.9580\approx2\omega_1$ when $a_0$ = 14.2365. In the following computations, other system parameters are chosen as $\bar{\mu}=0.01$, $\kappa=1$, $\epsilon=0.01$, and $\bar{F}=5$, unless otherwise stated.

\subsection{Periodic and quasi-periodic orbits}

\begin{figure}[!ht]
    \centering
    \includegraphics[width=0.45\textwidth]{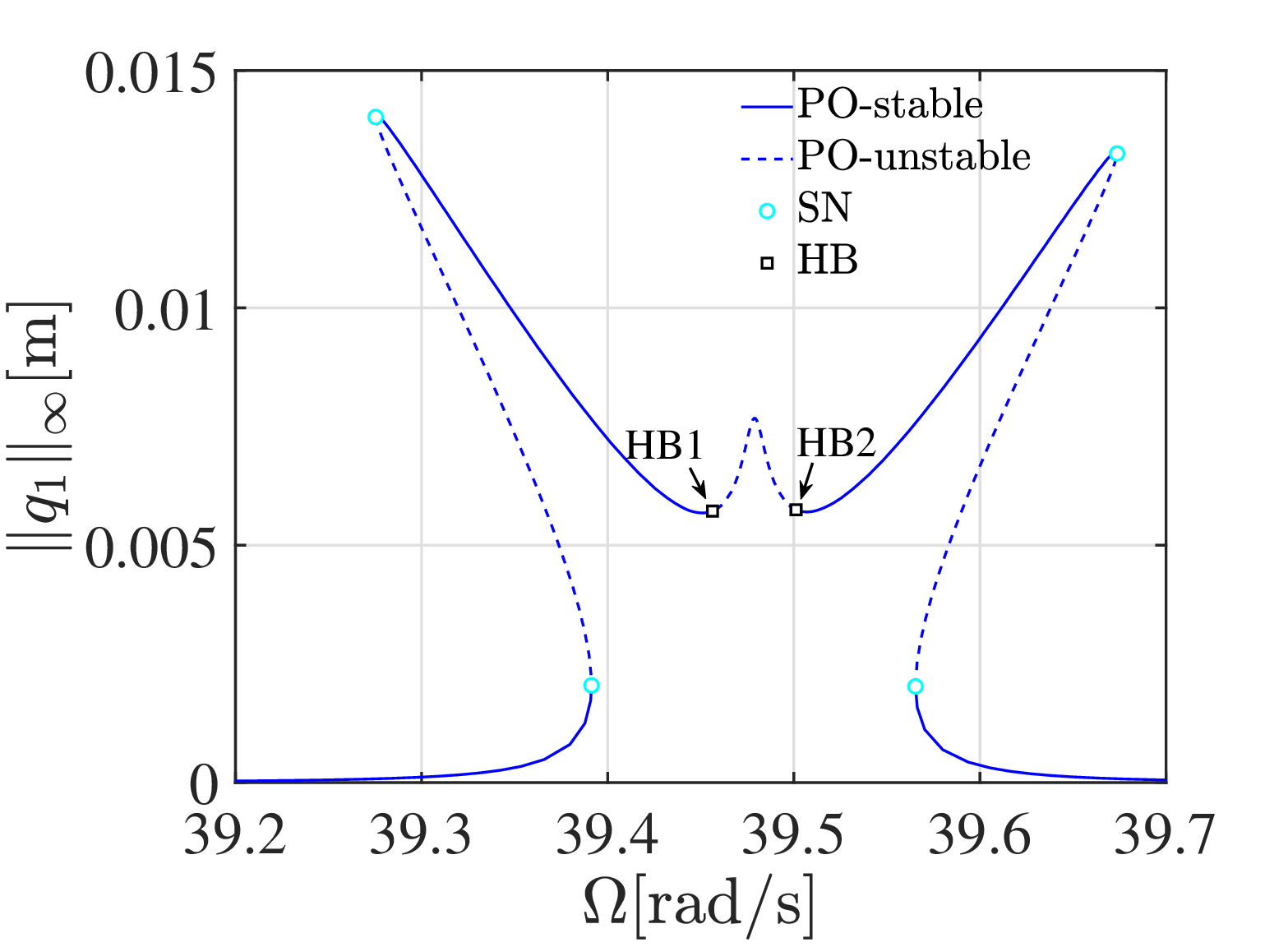}
    \includegraphics[width=0.45\textwidth]{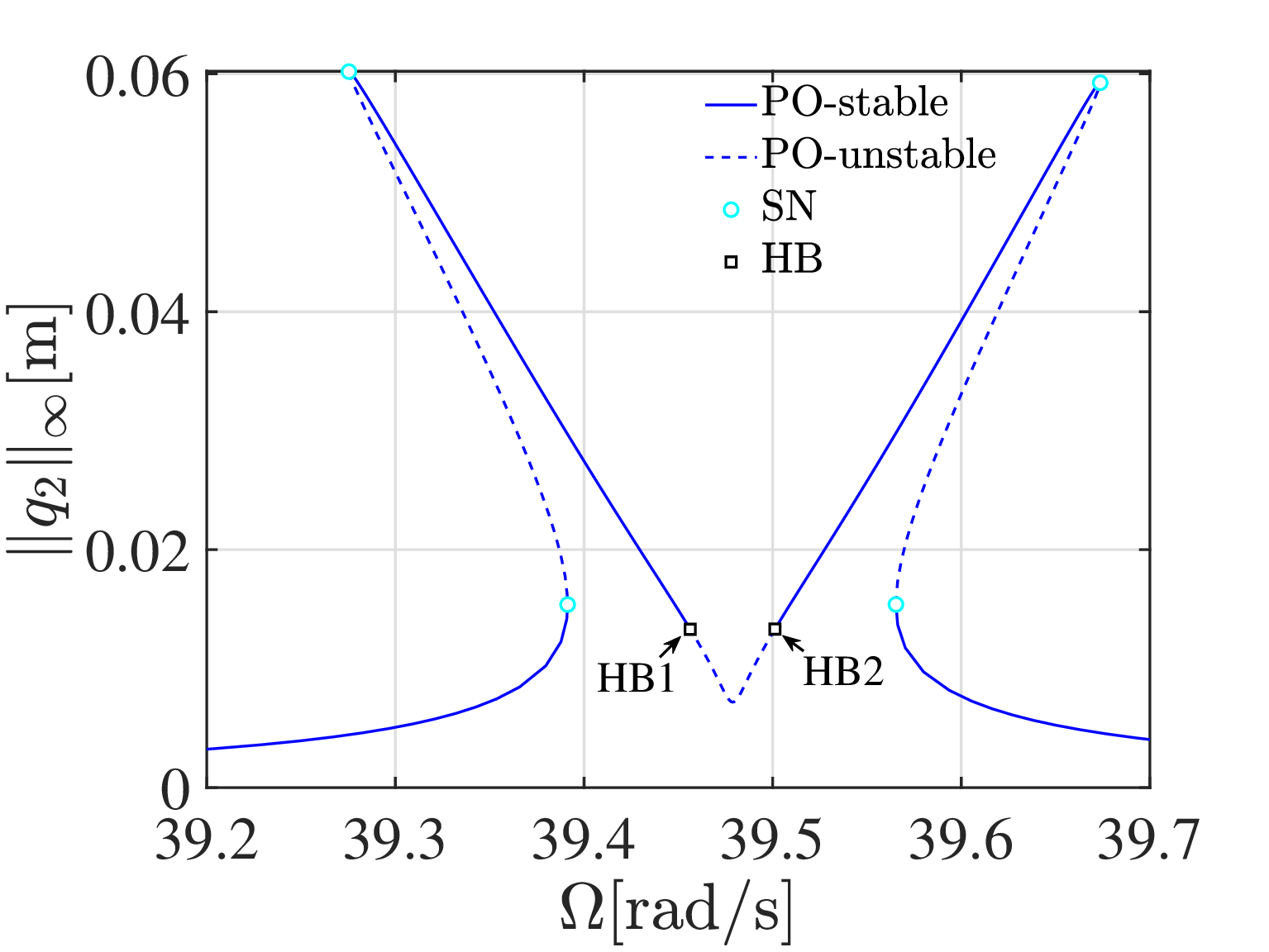}
    \caption{\small FRCs of periodic orbits of the shallow curved beam system \eqref{eq:system 2} with $\Omega \in$ [39.2,39.7]. The blue solid/dashed lines denote the amplitudes of stable/unstable periodic solutions. The circles and squares denote saddle-node (SN) and Hopf bifurcation (HB) points, respectively.}
    \label{fig:ep_q1_q2}
\end{figure}

We first compute the forced response curve of periodic orbits for the system with $\Omega \in [39.2, 39.7]$. Similar to the previous example, this FRC-PO is obtained by parameter continuation of the fixed point of the SSM-based ROM~\eqref{eq:red-nonauto-th}. In this example, an $\mathcal{O}(3)$ expansion is used because it is sufficient to yield converged solutions for the selected parameters. The obtained FRC-PO is shown in Fig.~\ref{fig:ep_q1_q2}. Here $\Vert q_1 \Vert_\infty$ and $\Vert q_2 \Vert_\infty$ represent the infinite norm of  the modal coordinates $q_1(\tau),q_2(\tau)$, giving the amplitude of the periodic response. Similar to the system of two coupled oscillators, there are two HB points (HB1 and HB2) and four SN bifurcation points on the FRC-PO. Moreover, the periodic motion for $\Omega\in [\Omega_\mathrm{HB1},\Omega_\mathrm{HB2}]$ is unstable, and we will focus on quasi-periodic motions born out of these two HB points of the system.

\begin{figure}[!ht]
    \centering
    \includegraphics[width=0.45\textwidth]{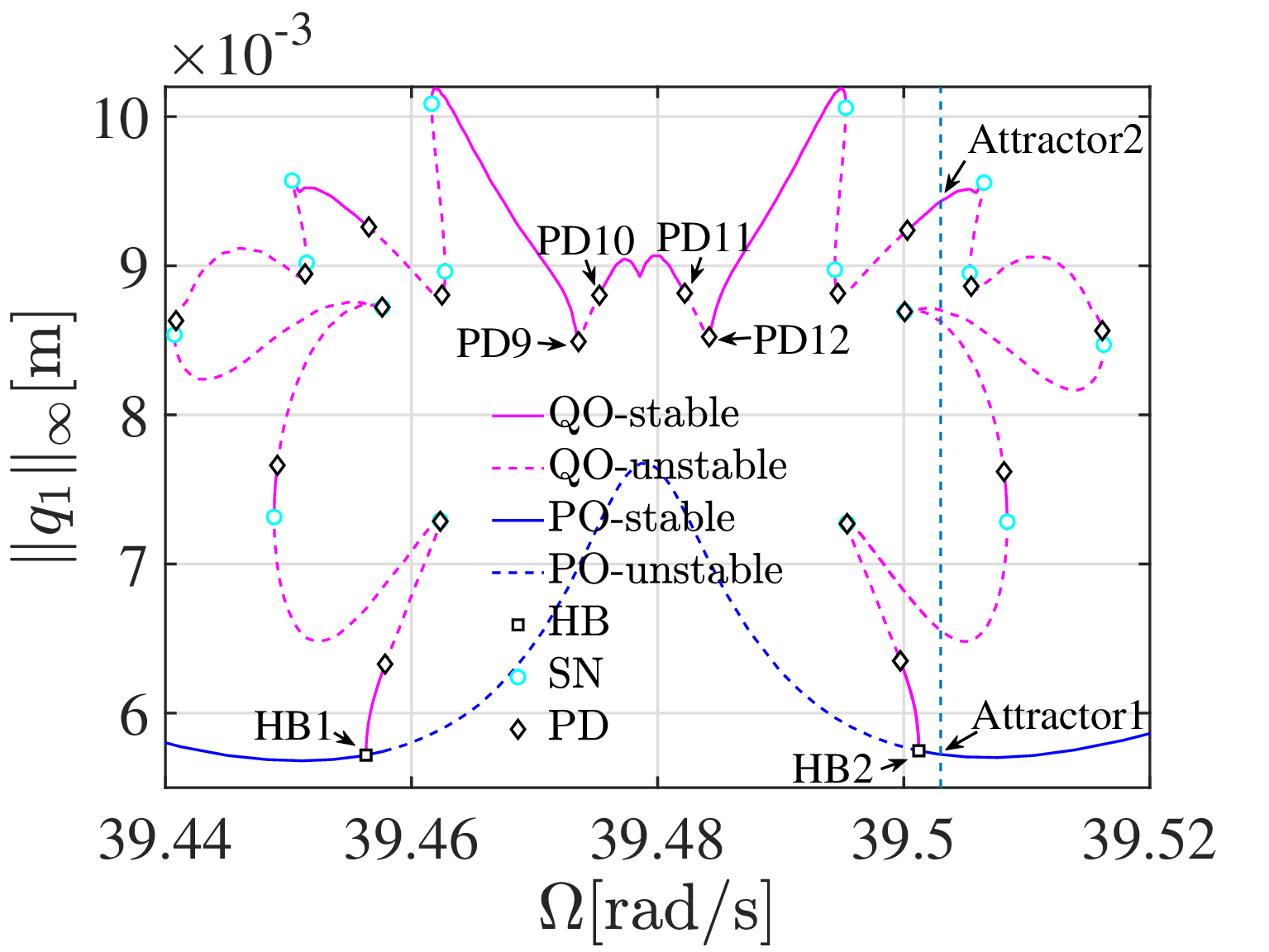}
    \includegraphics[width=0.45\textwidth]{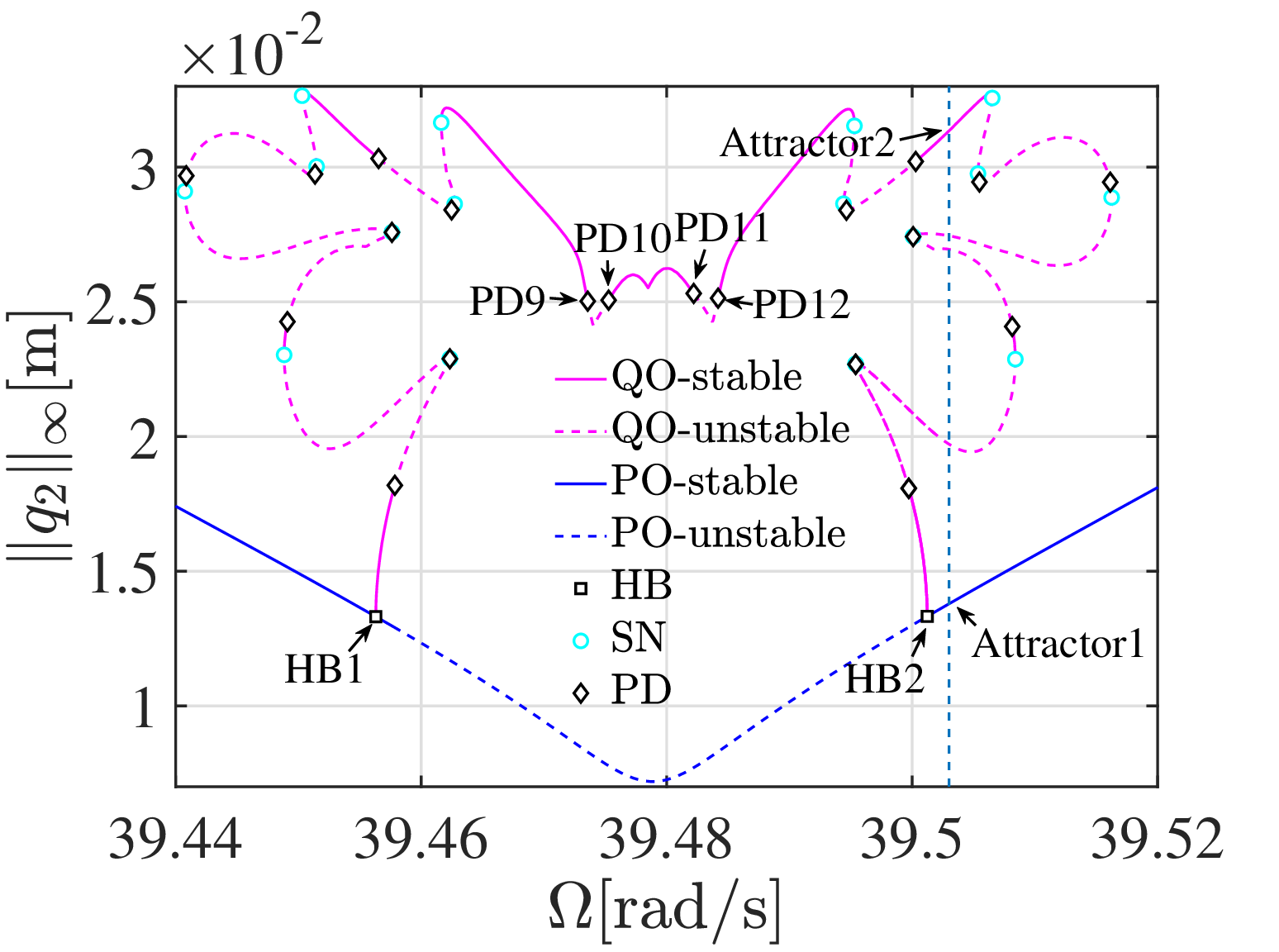}
    \caption{\small FRCs of quasi-periodic and periodic orbits of the shallow curved beam system \eqref{eq:system 2} with $\Omega\in[39.44,39.52]$. The magenta solid/dashed lines denote stable/unstable quasi-periodic solutions. The blue solid/dashed lines denote stable/unstable periodic solutions. The circles, diamonds, and squares denote saddle-node (SN), period-double (PD), and Hopf bifurcation (HB) points, respectively.}
    \label{fig:HB2po_q1_q2}
\end{figure}

We switch to the parameter continuation of limit cycles of the SSM-based ROM~\eqref{eq:red-nonauto-th} at HB1 and then extract the forced-response curve of quasi-periodic orbits (FRC-QO) under variations in $\Omega$. This continuation run proceeds until it reaches HB2. The obtained FRC-QO along with FRC-PO are shown in Fig.~\ref{fig:HB2po_q1_q2}. 20 PD bifurcation points and 20 SN bifurcation points are detected on the FRC-QO, leading to the coexistence of stable and unstable quasi-periodic motions. In addition, we also observe the coexistence of periodic and quasi-periodic attractors in Fig.~\ref{fig:HB2po_q1_q2}. For example, coexisting periodic attractor (Attractor1) and quasi-periodic attractor (Attractor2) can be found at $\Omega$ = 39.503. The validation of the SSM-based predictions for the FRC-PO and FRC-QO is provided in Appendix \ref{sec:app-beam}, as seen in Fig.~\ref{fig:verify PO_q1q2} and Figs.~\ref{fig:verify QO_q1q2}-\ref{fig:verify stability_q1q2}.

\subsection{Coexistence of quasi-periodic and chaotic attractors}

\begin{figure}[!ht]
    \centering
    \includegraphics[width=0.45\textwidth]{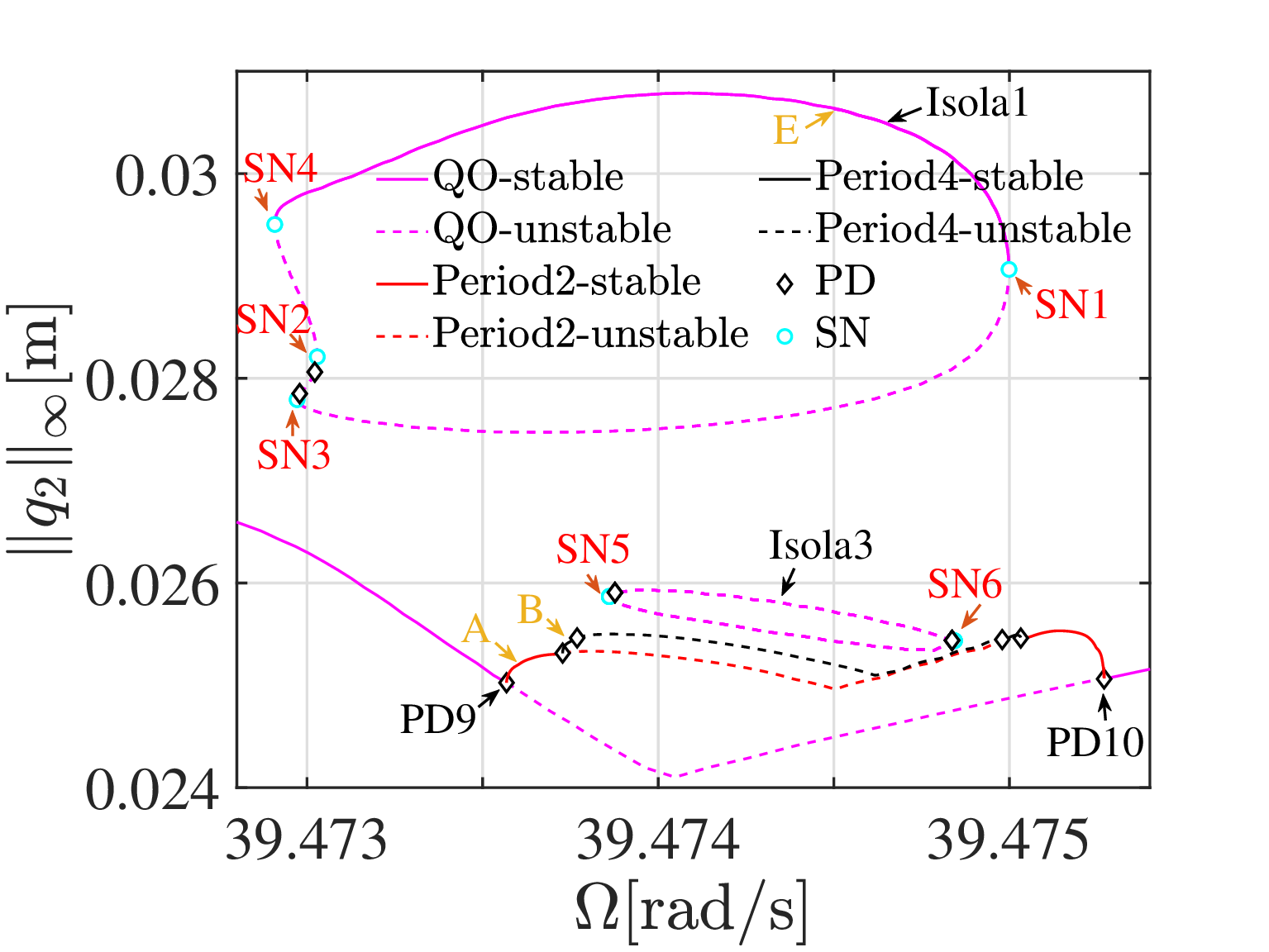}
    \includegraphics[width=0.45\textwidth]{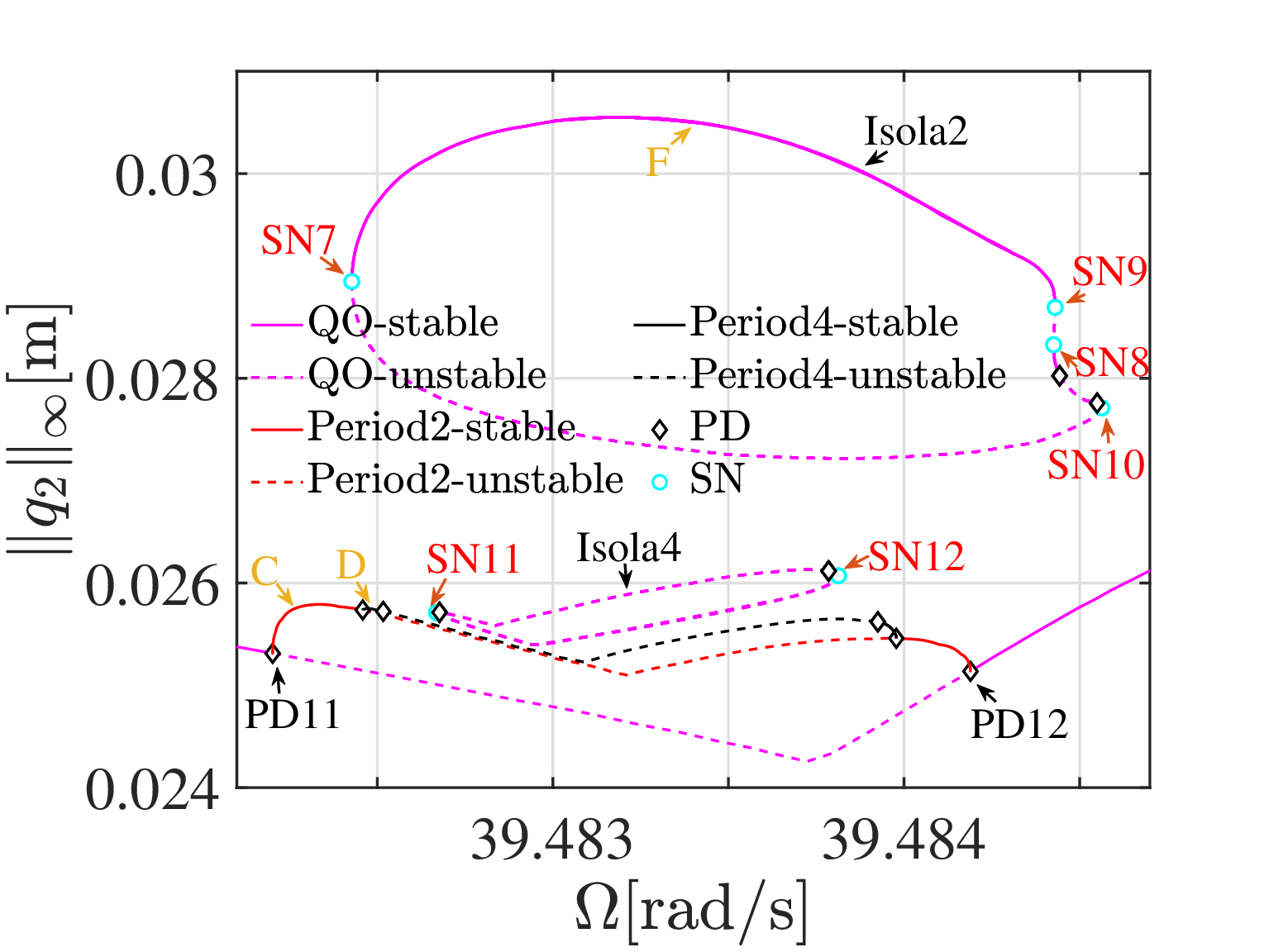}
    \caption{\small Quasi-periodic orbits of shallow curved beam in Eq. (\ref{eq:system 2}) with $\Omega \in [\Omega_{\rm{PD9}},\Omega_{\rm{PD10}}]$ and $[\Omega_{\rm{PD11}},\Omega_{\rm{PD12}}]$. The magenta solid/dashed lines denote the isolated branches of quasi-periodic orbits and FRC-QO obtained by continuing HB1 in FRCs; red solid/dashed lines denote period2 quasi-periodic orbits obtained by continuing PD9 in FRC-QO; black solid/dashed lines denote period4 quasi-periodic orbits obtained by continuing period-double (PD) bifurcations in period2 quasi-periodic orbits. The circles and diamonds denote SN and PD bifurcations, respectively.}
    \label{fig:isola_q2&L3}
\end{figure}

Since PD bifurcation points were detected, we expect that this curved beam system also has a cascade of PD bifurcations for the quasi-periodic orbits, leading to chaotic motion. This is indeed the case here. We focus on the region $\Omega \in [\Omega_{\rm{PD9}},\Omega_{\rm{PD10}}]$ and $[\Omega_{\rm{PD11}},\Omega_{\rm{PD12}}]$ to illustrate the cascade of PD bifurcations. We switch to the parameter continuation of limit cycles of the ROM~\eqref{eq:red-nonauto-th} with doubled period (Period2) at PD9 and obtain the continuation path in red lines in the upper panel of Fig.~\ref{fig:isola_q2&L3}. Along this Period2 FRC-QO, two new PD bifurcation points are detected, indicating the appearance of quasi-periodic motions with a quadrupled period (Period4). We switch to the continuation of these Period4 tori and obtain the Period4 FRC-QO plotted as black lines in the figure. We once again detect two new PD points on this black line, which illustrates the cascade of PD bifurcations. This cascade of PD bifurcations is also observed for $\Omega \in [\Omega_{\rm{PD11}},\Omega_{\rm{PD12}}]$, as can be seen in the lower panel of Fig.~\ref{fig:isola_q2&L3}, where the red and black lines again denote the FRCs of Period2 and Period4 quasi-periodic orbits.

\begin{figure*}[!ht]
    \centering
   \includegraphics[width=0.45\textwidth]{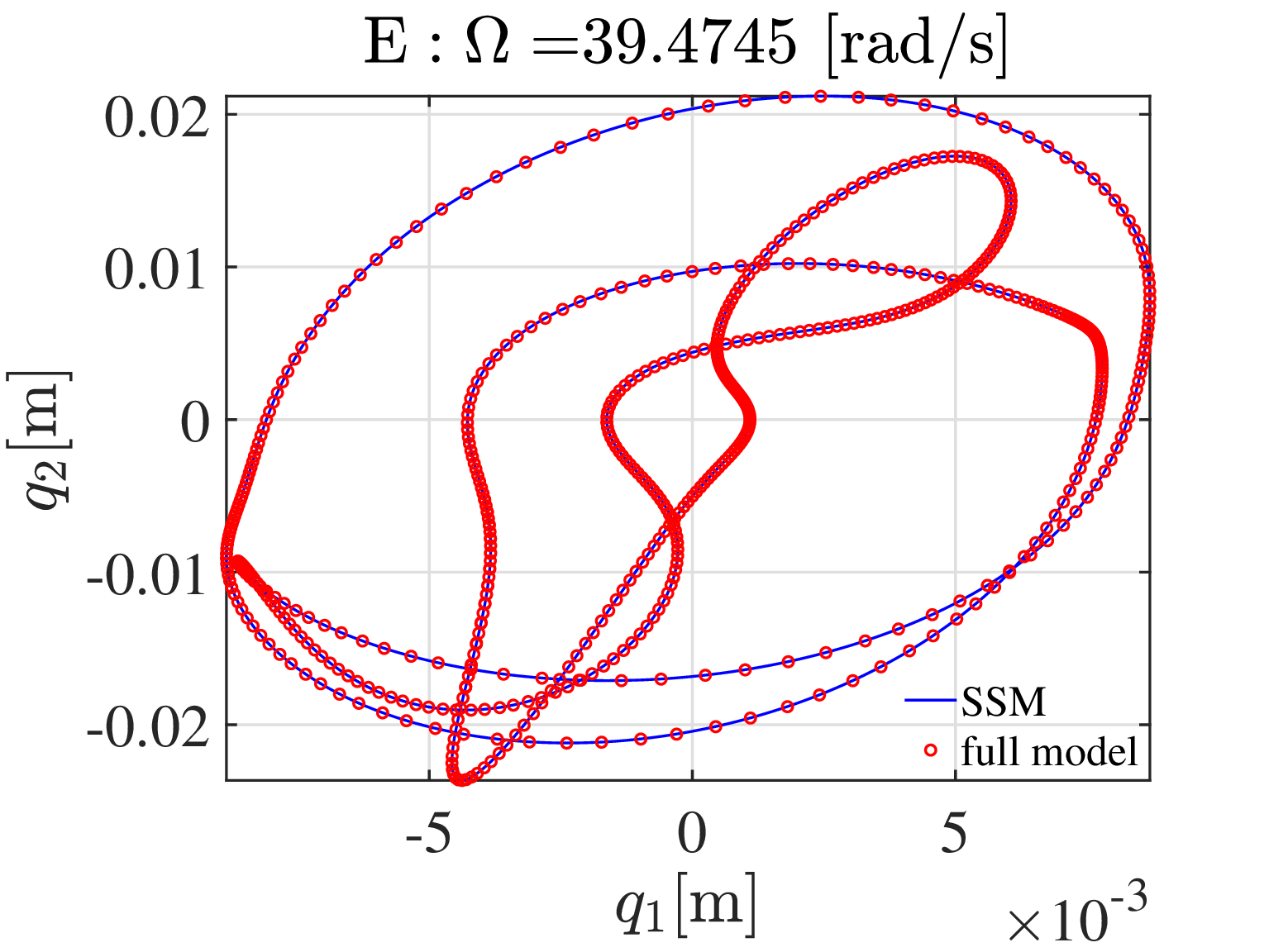}
   \includegraphics[width=0.45\textwidth]{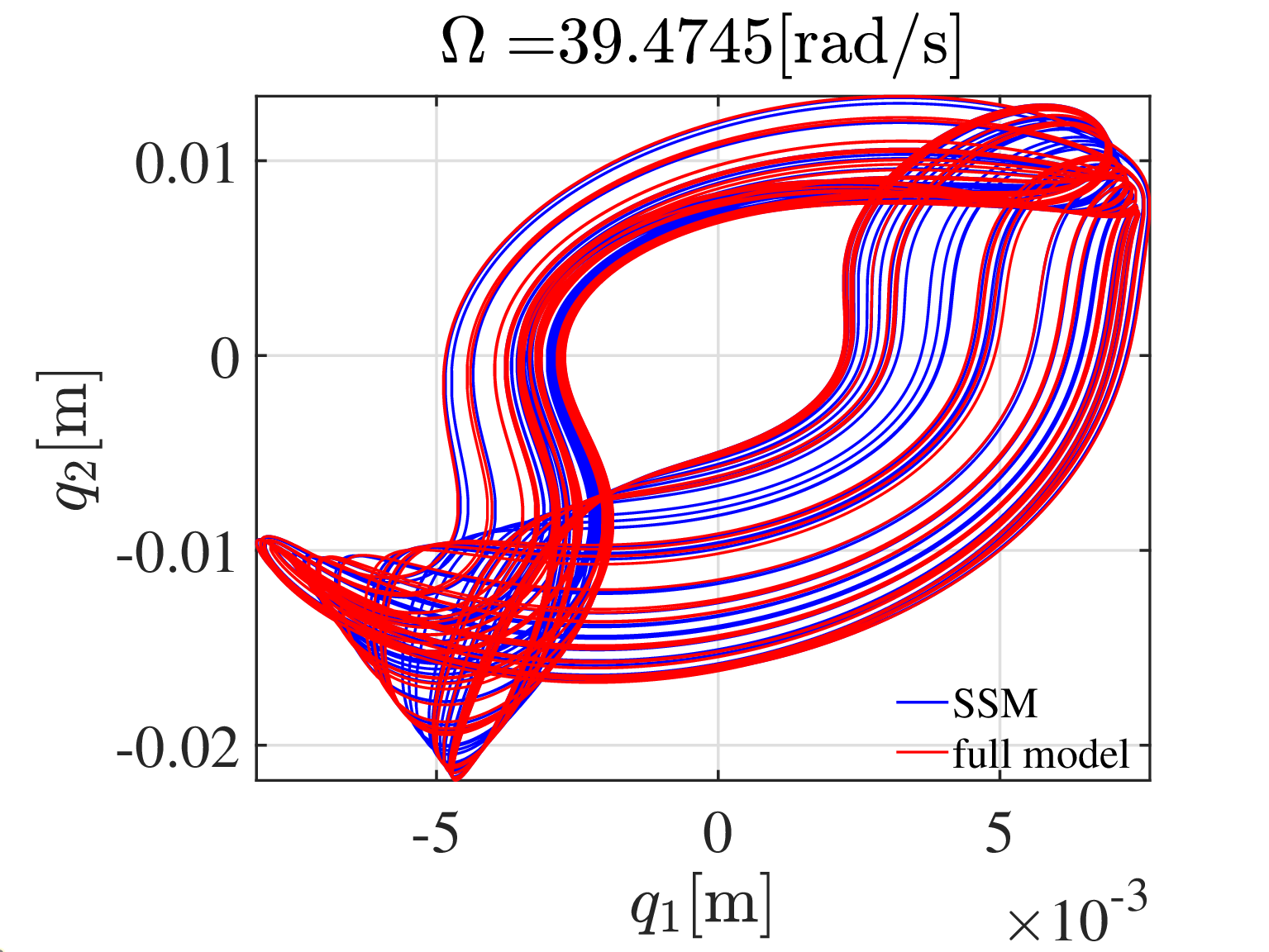} \\
   \includegraphics[width=0.45\textwidth]{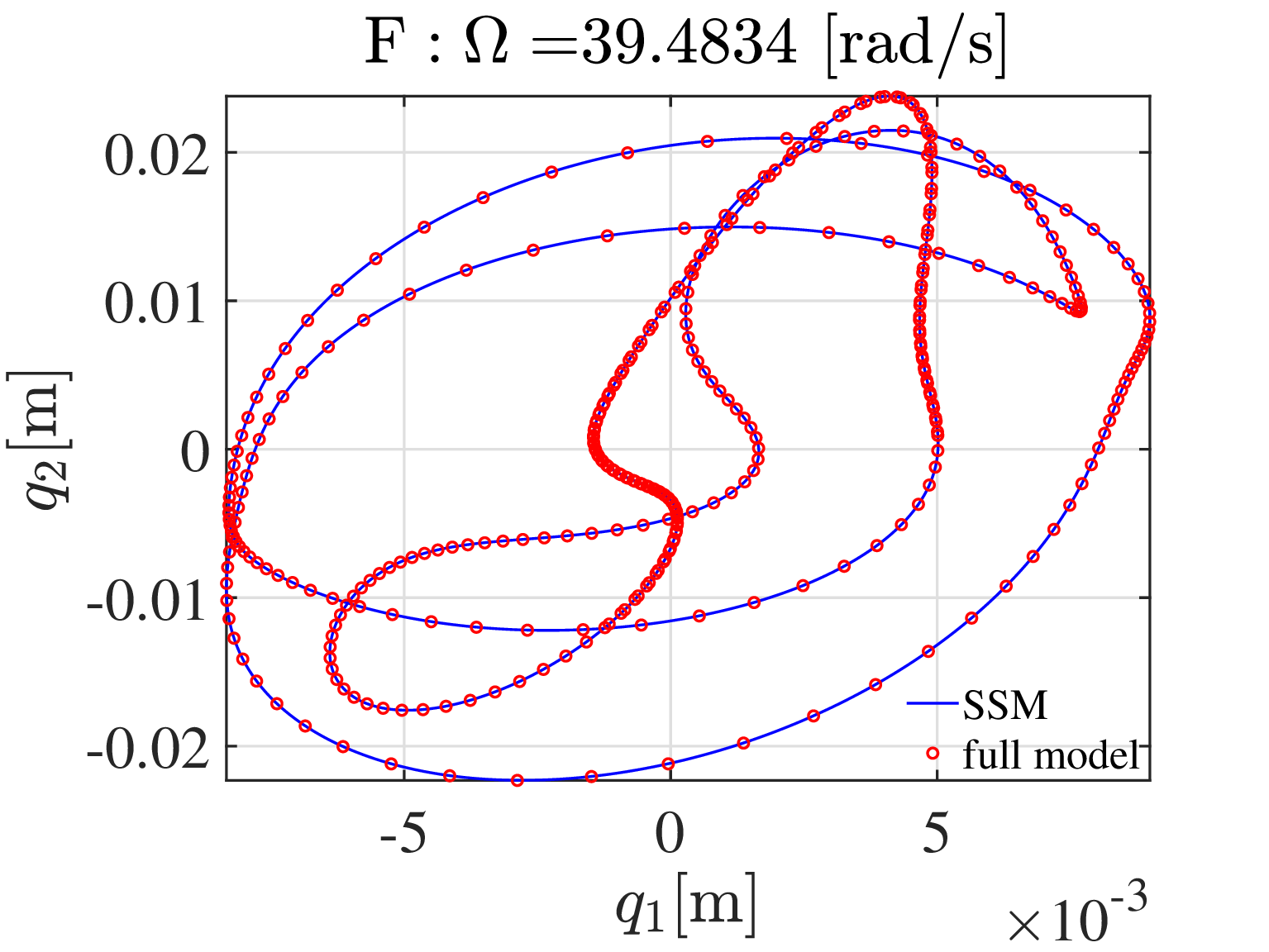}
   \includegraphics[width=0.45\textwidth]{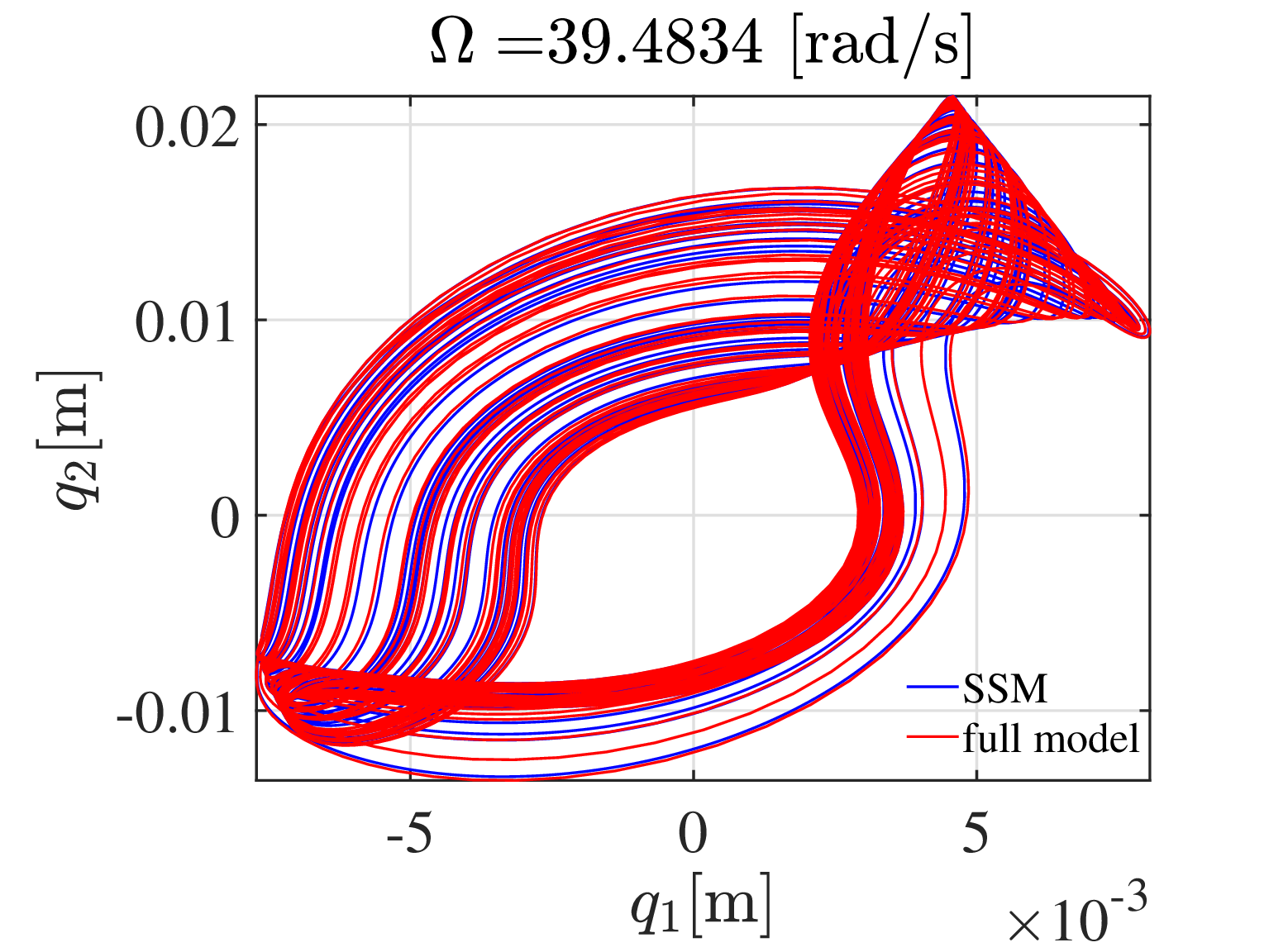} 
    \caption{\small Intersections of the period-$2\pi/\Omega$ map of coexisting attractors of the curved beam with $\Omega = 39.4745$ (upper panels) and 39.4834 (lower panels). Here, the left panels denote quasi-periodic attractors, while the right panels illustrate chaotic attractors. The blue lines denote the results obtained via the SSM-based reduced-order model. The red lines/cycles denote reference results of the full model obtained using forward simulations. Here, the two quasi-periodic attractors correspond to points E and F on isolas of quasi-periodic orbits in Fig.~\ref{fig:isola_q2&L3}.}
    \label{fig:chaos&isola}
\end{figure*}

The above-observed cascades of PD bifurcations are expected to lead to chaotic motions. To confirm it, we check whether the system~\eqref{eq:system 2} exists chaotic attractors at two sampled excitation frequencies $\Omega=39.4745$ and $\Omega=39.4834$. Specifically, we perform forward simulations to the SSM-based ROM~\eqref{eq:red-nonauto-th} and then map the generated trajectories back to the full system. Interestingly, we find that the steady state of the forward simulations can be either chaotic or quasi-periodic, depending on the choice of initial conditions. This differs from the previous example's case (cf. Sect.~\ref{ssec:qusi2chaos}). We present the Poincar{\'e} intersections of these coexisting quasi-periodic and chaotic attractors with period-$2\pi/\Omega$ map in Fig.~\ref{fig:chaos&isola}, where the upper two panels show the two coexisting attractors at $\Omega=39.4745$, while the lower two panels display the two coexisting attractors at $\Omega=39.4834$. The two strange attractors in the right panels of Fig.~\ref{fig:chaos&isola} are indeed chaotic because their associated maximum Lyapunov exponents are 0.0038 and 0.0041, as detailed in Fig.~\ref{fig:lyapuv-ex2-chaos}.

We use numerical integration of the full system to validate the predicted cascade of PD bifurcations and the coexisting quasi-periodic and chaotic attractors. We take points A and C in Fig.~\ref{fig:isola_q2&L3} as representative of Period2 quasi-periodic orbits, and points B and D in Fig.~\ref{fig:isola_q2&L3} as representative of Period4 quasi-periodic orbits. We plot the Poincar{\'e} section of these quasi-periodic orbits in Fig.~\ref{fig:verify L2&3 of system2} in Appendix, from which we see that SSM-based predictions match perfectly with the reference solutions obtained via the numerical integration. In addition, as seen in Fig.~\ref{fig:chaos&isola}, the predicted coexisting quasi-periodic and chaotic attractors are also successfully validated using the numerical integration of the full system. Further, we observe from Fig.~\ref{fig:lyapuv-ex2-chaos} in Appendix that the predicated maximum Lyapunov exponents match well with the reference solutions as well. These demonstrate the effectiveness of the SSM-based ROM~\eqref{eq:red-nonauto-th}.

\begin{figure}[!ht]
    \centering
    \includegraphics[width=0.45\textwidth]{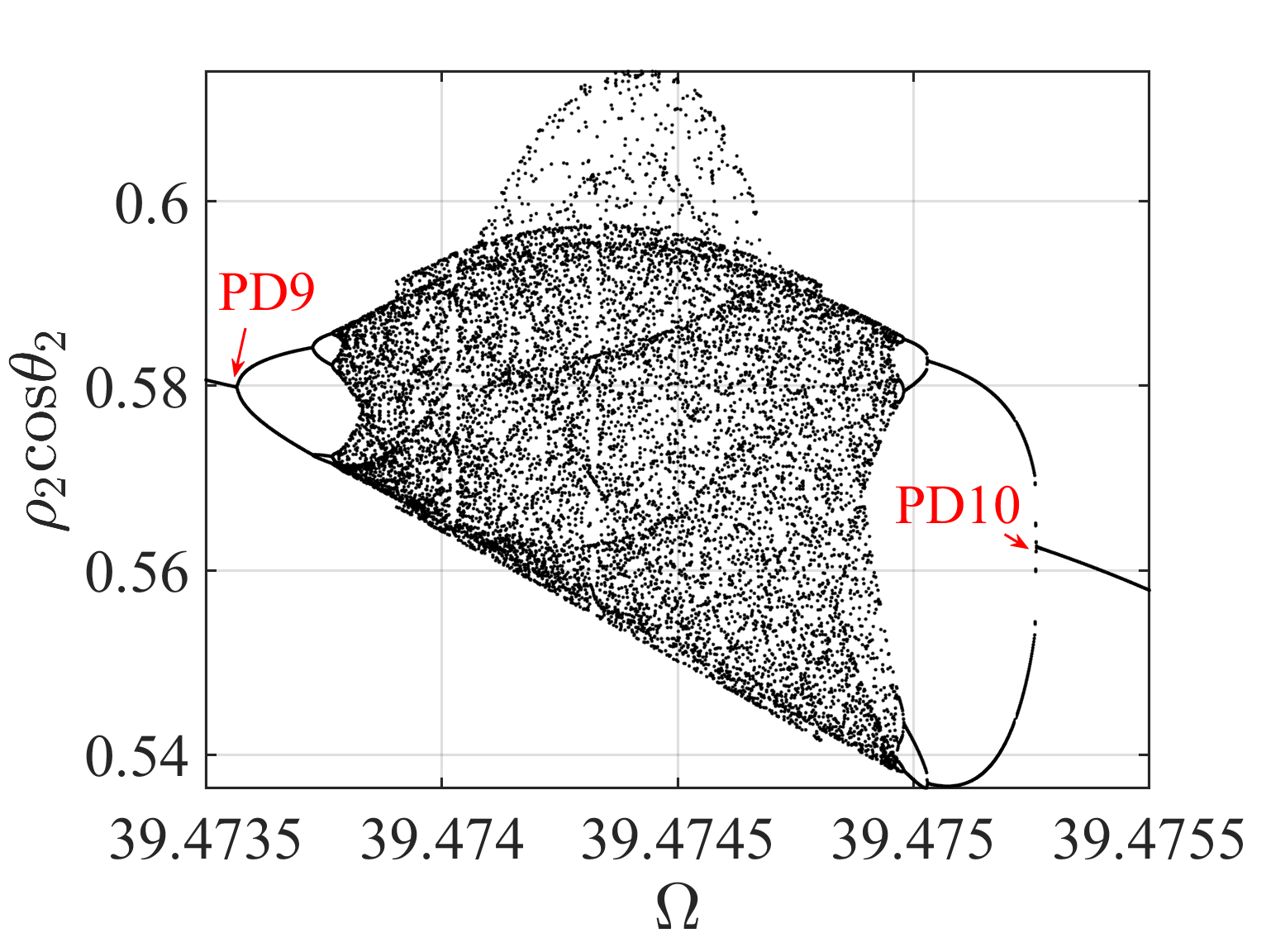}
    \includegraphics[width=0.45\textwidth]{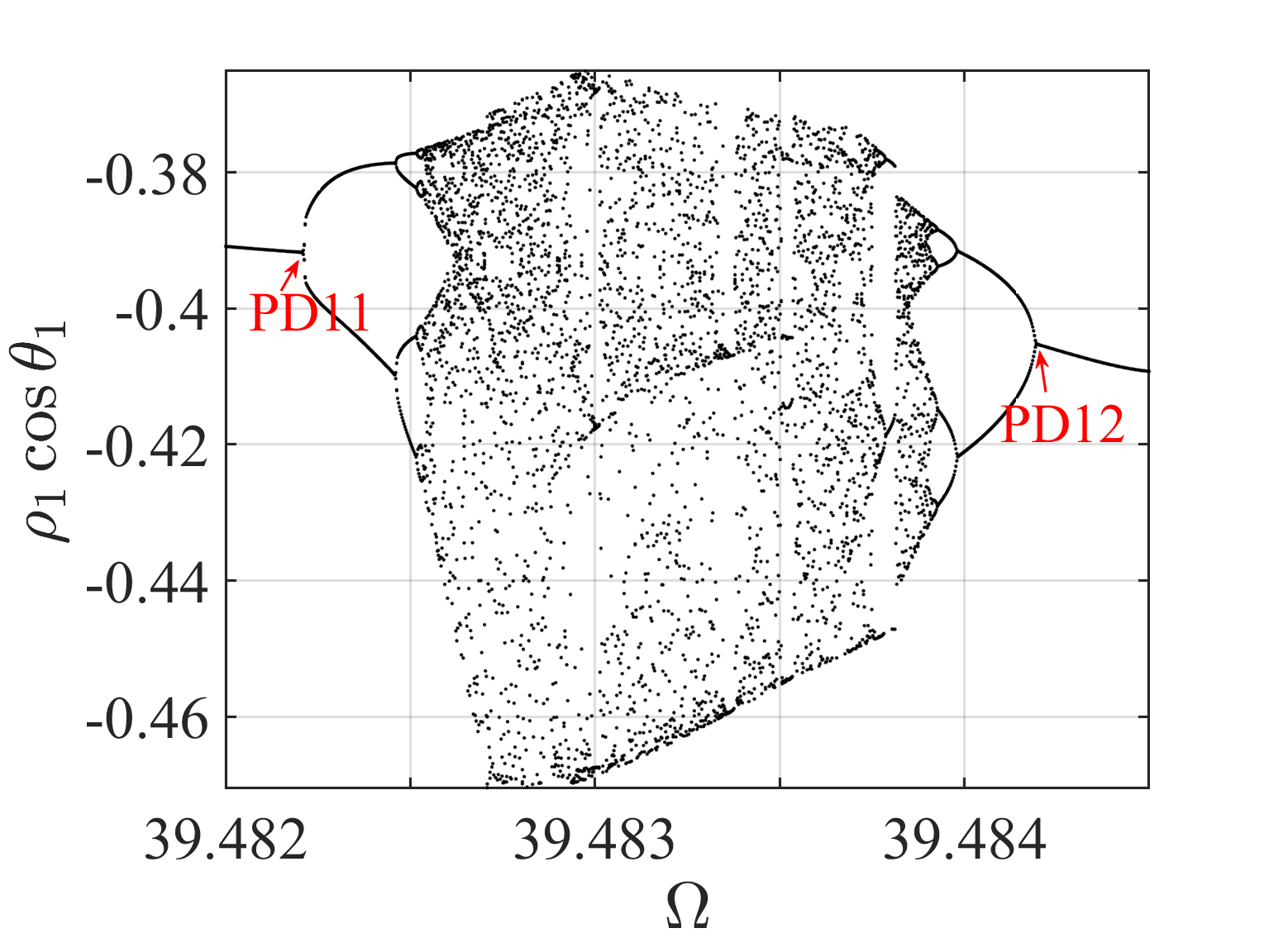}
    \caption{\small Bifurcation diagram of attractors for the SSM-based ROM \eqref{eq:red-nonauto-th} of the shallow curved beam \eqref{eq:system 2} with $\Omega \in [39.4735,39.4755]$ (upper panel) and $\Omega \in [39.482,39.4845]$ (lower panel). Here, the vertical axis in the upper panel gives the intersections of the attractors with the Poincar{\'e} section $\{(\boldsymbol{\rho},\boldsymbol{\theta}): q_2=\rho_2\cos\theta_2, \dot{q_2}\equiv0,\ddot{q}_2>0\}$ of the ROM, namely, the maximum of $\rho_2\cos\theta_2$ associated with the attractors. Likewise, the vertical axis in the lower panel gives the intersections of the attractors with the Poincar{\'e} section $\{(\boldsymbol{\rho},\boldsymbol{\theta}): q_1=\rho_1\cos\theta_1, \dot{q_1}\equiv0,\ddot{q}_1>0\}$ of the ROM.}
    \label{fig:bif-chaos-ex2}
\end{figure}

We conclude this subsection with bifurcation diagrams of attractors on the region $\Omega \in [\Omega_{\rm{PD9}},\Omega_{\rm{PD10}}]$ and $[\Omega_{\rm{PD11}},\Omega_{\rm{PD12}}]$. In particular, we follow the same method to produce Fig.~\ref{fig:bif-chaos-ex1} to extract the bifurcation diagram of attractors of the SSM-based ROM~\eqref{eq:red-nonauto-th}. The obtained bifurcation diagrams are shown in Fig.~\ref{fig:bif-chaos-ex2}, from which we see the complete picture of the transition to chaos via the cascade of period-doubling bifurcations. These bifurcation diagrams are also consistent with that of Fig.~\ref{fig:isola_q2&L3}.


\subsection{Isolated branches of quasi-periodic orbits}

The coexisting quasi-periodic and chaotic attractors at $\Omega=39.4745$ and $\Omega=39.4834$ indicate that the curved beam admits stable quasi-periodic orbits at these two sampled excitation frequencies. As seen in Fig.~\ref{fig:isola_q2&L3}, the quasi-periodic orbits for  $\Omega=39.4745$ and $\Omega=39.4834$ on the main branch of FRC-QO and its bifurcated branches with Period2 and Period4 quasi-periodic orbits (red and black lines) are all unstable. This implies that isolated branches of quasi-periodic orbits exist. To locate the isolas, we recall that each quasi-periodic orbit corresponds to a limit cycle of the SSM-based ROM~\eqref{eq:red-nonauto-th} (cf.~Fig.~\ref{fig:FulltoROM}). So we perform two continuation runs of limit cycles of the ROM~\eqref{eq:red-nonauto-th} under the variation in $\Omega$. These continuation runs are initialized with the limit cycles at $\Omega=39.4745$ and $\Omega=39.4834$ respectively. Indeed, we locate an isola in each of these continuation runs and then map the isolas of the two continuation runs back to the full system, yielding the isolas of quasi-periodic orbits that are marked as Isoa1 and Isola2 in Fig.~\ref{fig:isola_q2&L3}. We observe that each of these two isolas has 2 PD bifurcation points and 4 SN bifurcation points. Moreover, the two quasi-periodic attractors in Fig.~\ref{fig:chaos&isola} correspond to point E on Isola1 and point F on Isola2, as seen in Fig.~\ref{fig:isola_q2&L3}.

\begin{figure}[!ht]
    \centering
    \includegraphics[width=0.45\textwidth]{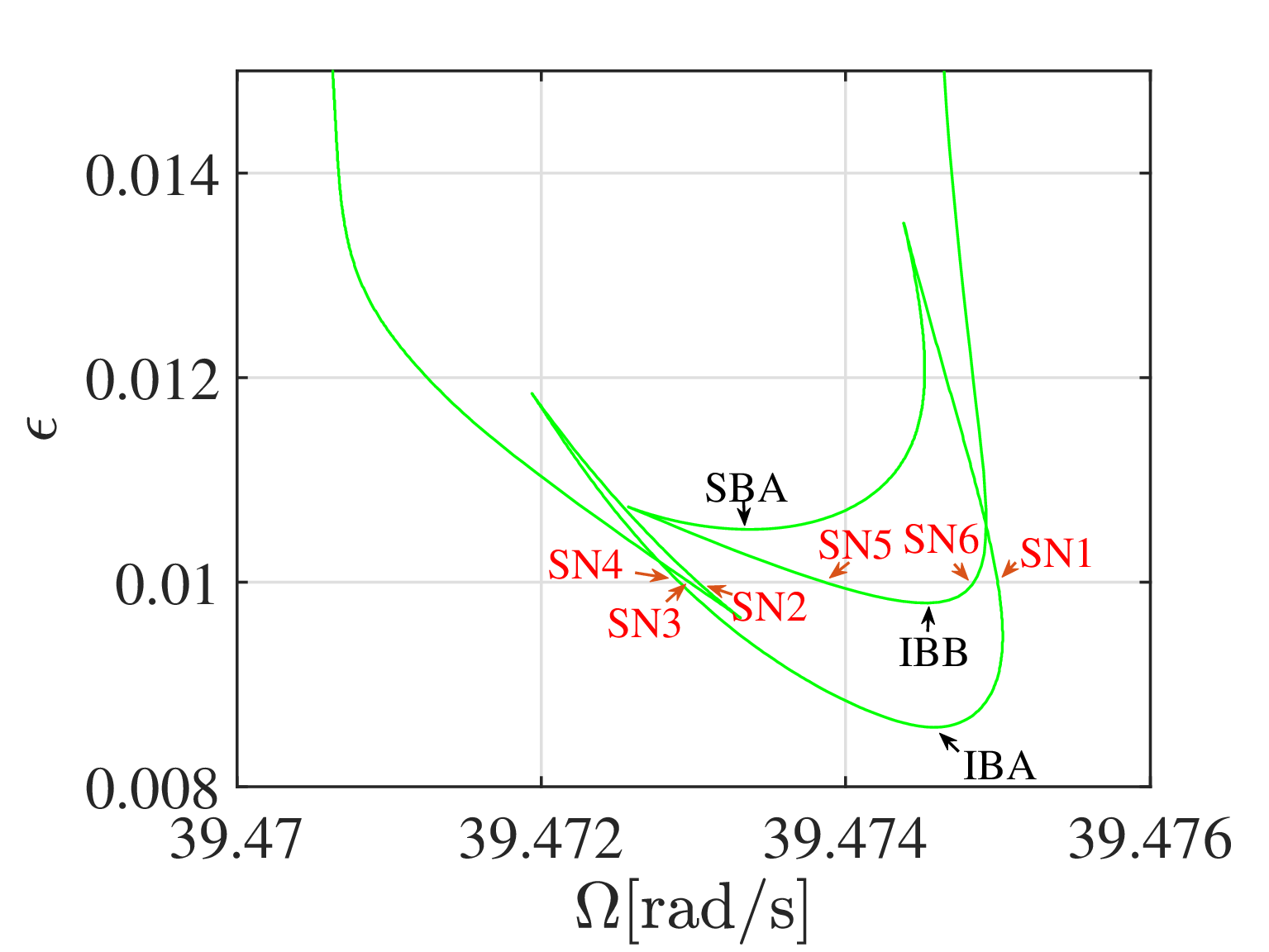}
    \includegraphics[width=0.45\textwidth]{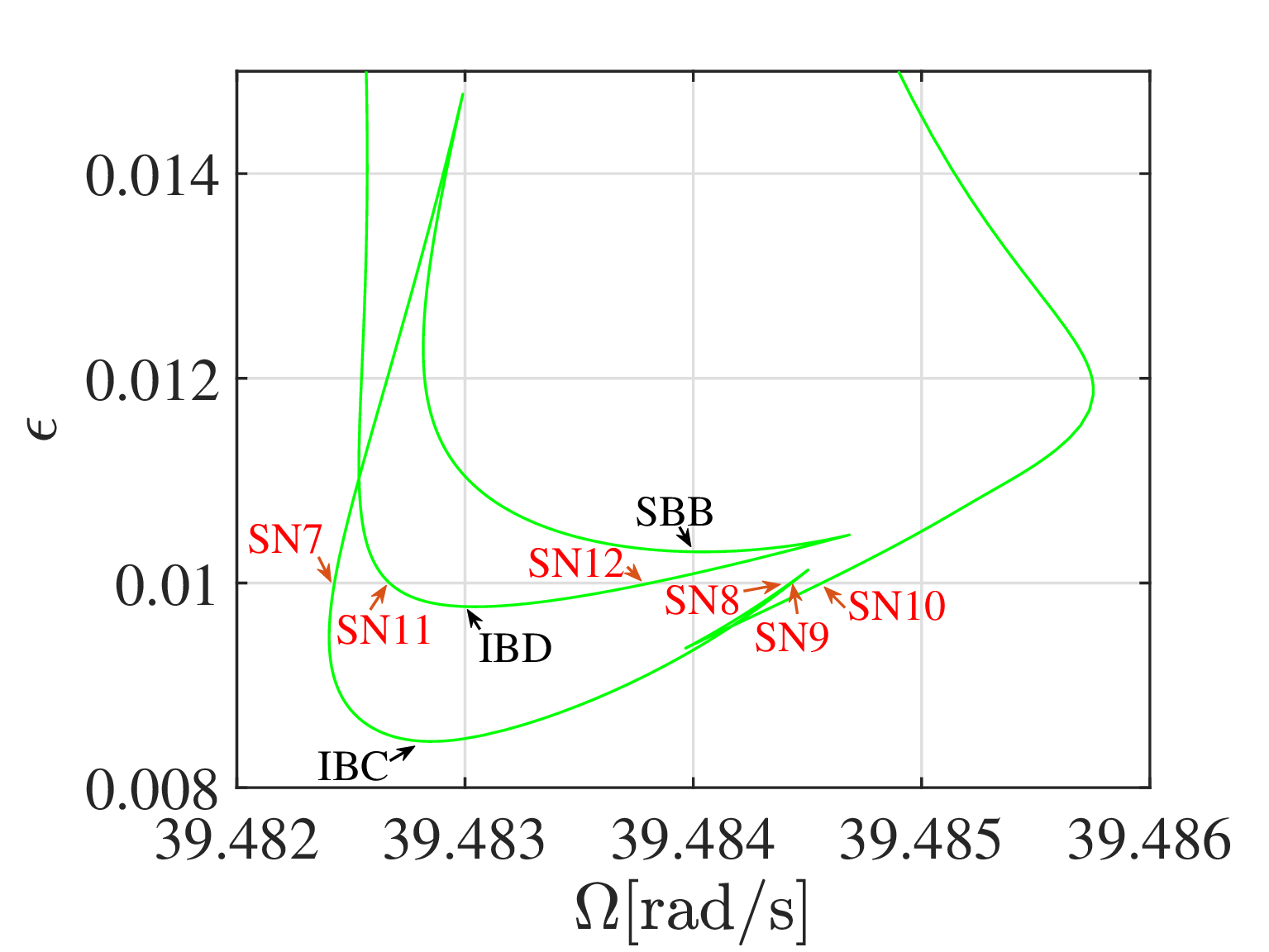}
    \caption{\small Projections of saddle-node (SN) bifurcation curves of quasi-periodic orbits of the shallow curved beam system with $\Omega \in [39.47,39.476]$ (upper panel) and $\Omega \in [39.482,39.486]$ (lower panel). The SN1-SN6 in the upper panel correspond to the SN points on Isola1 and Isola3 in Fig.~\ref{fig:isola_q2&L3}. The SN7-SN12 in the lower panel correspond to the SN points on Isola2 and Isola4 in Fig.~\ref{fig:isola_q2&L3}.}
    \label{fig:isolapo2SN}
\end{figure}

Next, we study how these isolas evolve under the variations in the forcing amplitude $\epsilon$. We perform the continuation of SN bifurcated quasi-periodic orbits with SN1 or SN7 as the initial solution. We restrict to the computational domain $\epsilon\in[0,0.015]$. The projection of the SN bifurcation curves of these two continuation runs are shown in Fig.~\ref{fig:isolapo2SN}. Here, the upper and lower panels show the SN bifurcation curves with SN1 and SN7 respectively. Interestingly, we observe from the upper panel that there exist two more SN points, i.e., SN5 and SN6, other than the four SN points on Isola1 (cf. the upper panel of Fig.~\ref{fig:isola_q2&L3}), at $\epsilon=0.01$. This indicates the existence of other isolas at $\epsilon=0.01$. Indeed, we take SN5 as an initial point and perform continuation of quasi-periodic orbits with fixed $\epsilon$ and varying $\Omega$, resulting in the Isola3 in the upper panel of Fig.~\ref{fig:isola_q2&L3}. Similarly, we observe from the lower panel that there are two more SN points, i.e., SN11 and SN12, other than the four SN points on Isola2 (cf. the lower panel of Fig.~\ref{fig:isola_q2&L3}.), at $\epsilon=0.01$. We follow the same procedure and locate the Isola4 in the lower panel of Fig.~\ref{fig:isola_q2&L3}.

\begin{figure}[!ht]
    \centering
    \includegraphics[width=0.45\textwidth]{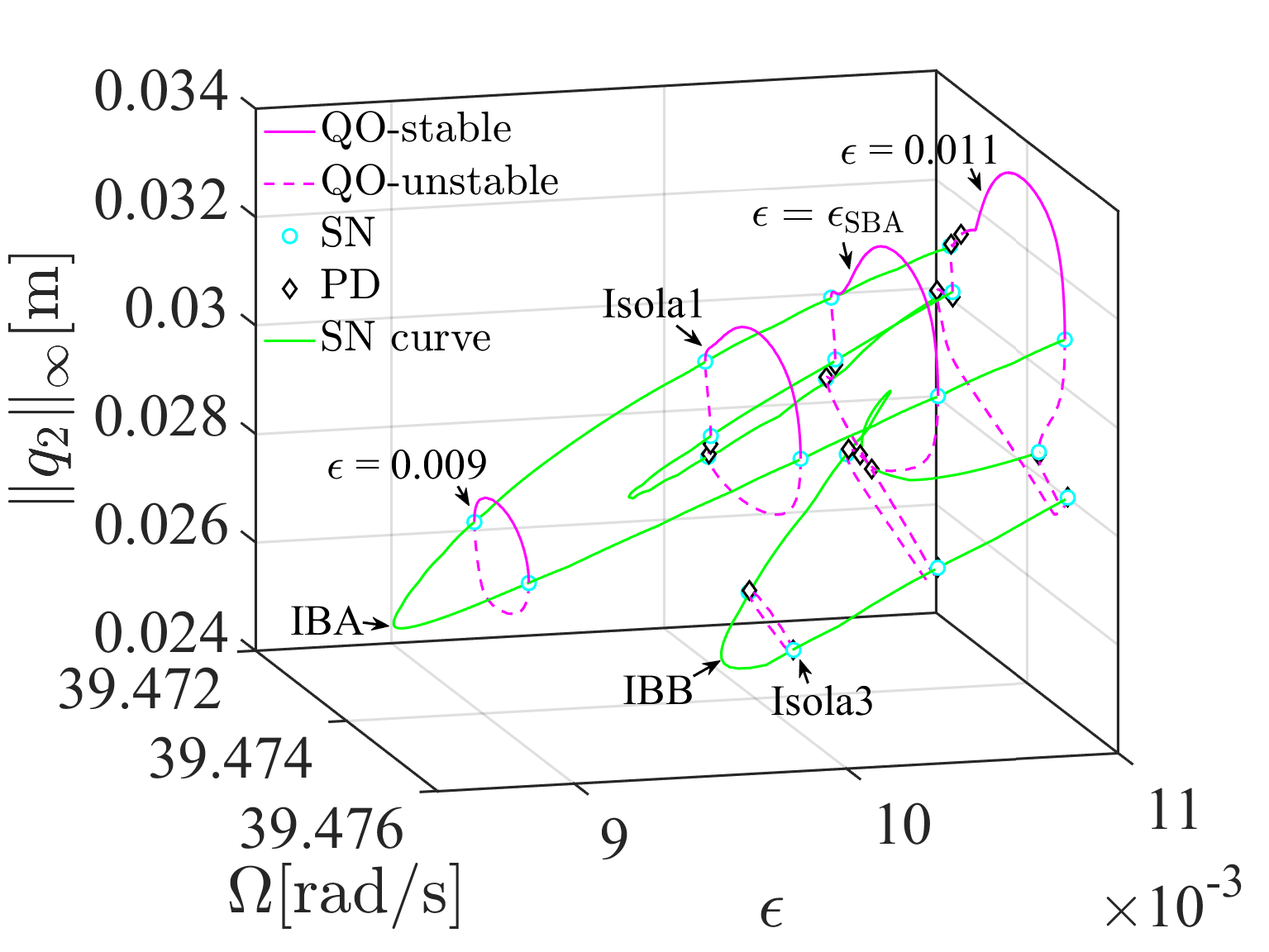}
    \includegraphics[width=0.45\textwidth]{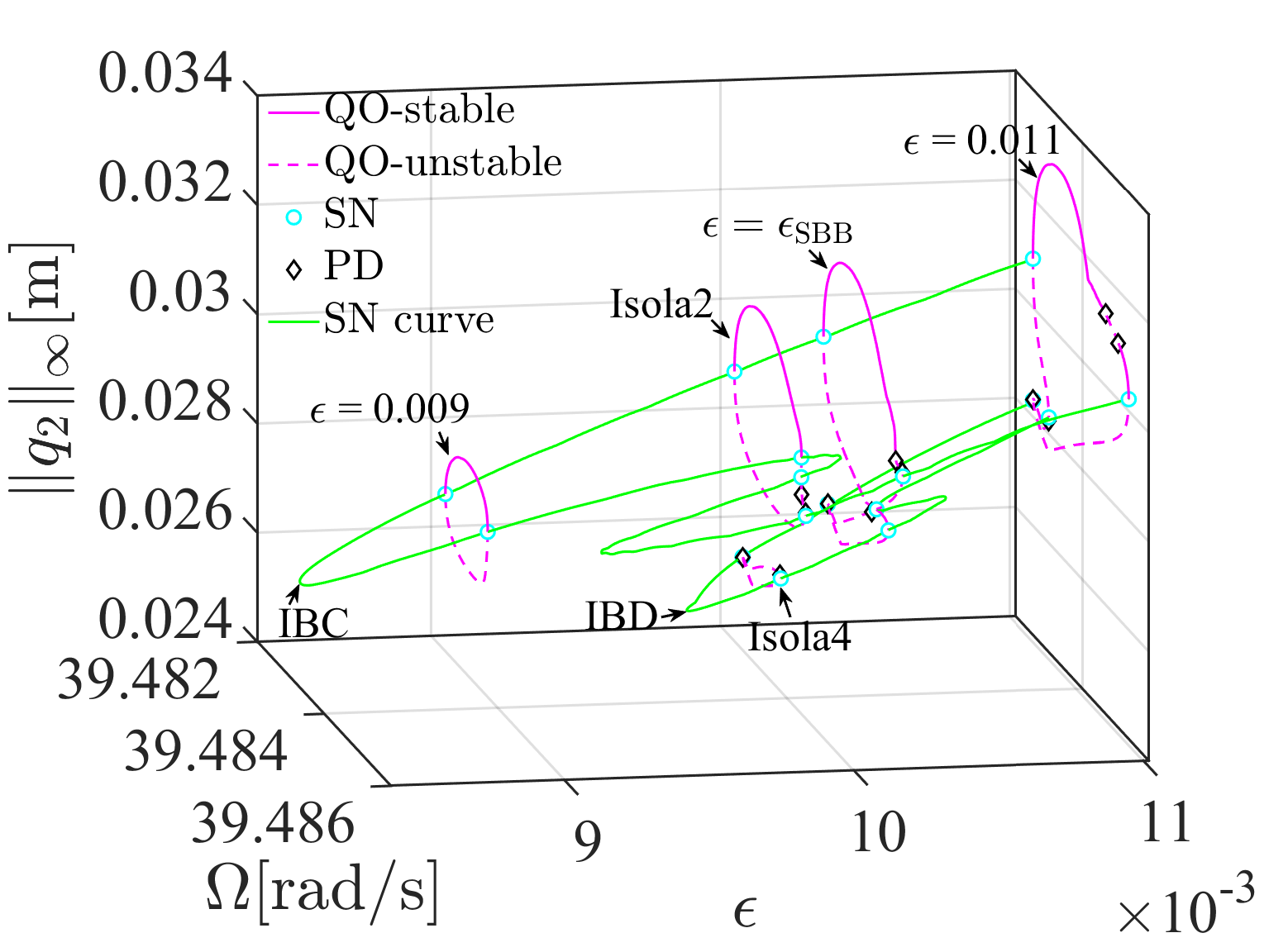}
    \caption{\small Projection of saddle-node (SN) bifurcation curves of quasi-periodic orbits of the shallow curved beam into  $(\Omega,\epsilon,||q_2||_\infty)$ along with four sampled forced response curves of quasi-periodic orbits at various forcing amplitude, namely, $\epsilon=0.009$, $\epsilon=0.01$, $\epsilon=\epsilon_\mathrm{SBA}$ (upper panel) or $\epsilon=\epsilon_\mathrm{SBB}$ (lower panel), and $\epsilon=0.011$. Here, the green lines represent the curve of SN bifurcated quasi-periodic orbits, while the magenta solid/dashed lines denote stable/unstable quasi-periodic solutions on FRC-QO. The circles and diamonds denote SN and PD bifurcations, respectively. The projection of the green lines onto $(\Omega,\epsilon)$ plane is shown in Fig.~\ref{fig:isolapo2SN}.}
    \label{fig:isolapo2SN_3D}
\end{figure}

We can further infer cusp, simple, and isola bifurcations from the continuation path shown in Fig.~\ref{fig:isolapo2SN}. Indeed, the kink points in Fig.~\ref{fig:isolapo2SN} mark cusp bifurcations where two SN points merge and then disappear. Meanwhile, we see the emergence of two SN points at which $\partial\epsilon/\partial\Omega=0$ along the continuation path in Fig.~\ref{fig:isolapo2SN}. These critical points where $\partial\epsilon/\partial\Omega=0$ holds are either isola bifurcation points on which isolas are born or simple bifurcation points where two isolas merge. To illustrate them more clearly, we plot these continuation paths in $(\Omega,\epsilon,||q_2||_\infty)$ along with some sampled FRC-QO in Fig.~\ref{fig:isolapo2SN_3D}. We infer from these plot that Isola1 (Isola3) is born out of the isola bifurcation point IBA (IBB). In addition, Isola1 and Isola3 merge as a single isola via the simple bifurcation point SBA, and then the single isola persists as $\epsilon$ increases, as seen in the upper panel of Fig.~\ref{fig:isolapo2SN_3D}. We observe similar bifurcation patterns for the lower panels of Fig.~\ref{fig:isolapo2SN} and Fig.~\ref{fig:isolapo2SN_3D}. In particular, Isola2 and Isola4 are born out of IBC and IBD and then merge via SBB.

\section{A shallow shell}
\label{sec:shell}

\begin{figure}[!ht]
    \centering
    \includegraphics[width=0.4\textwidth]{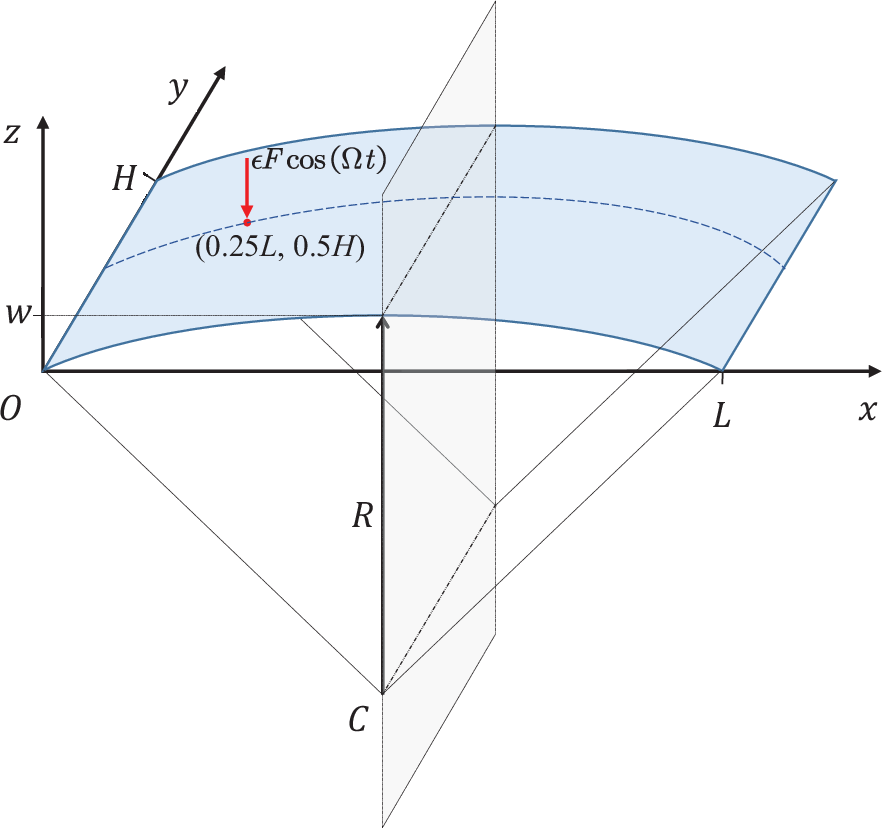}
    \caption{\small The schematic of a shallow shell structure \cite{jain2022HowComputeInvariant}.}
    \label{fig:system3}
\end{figure}

As our third mechanical system with 1:2 internal resonance, we consider the finite element (FE) model of a geometrically nonlinear shallow shell structure subject to a harmonic excitation shown in Fig.~\ref{fig:system3}. The shorter ends of the shell are fixed. Similarly to the system of shallow curved beam, we tune a curvature parameter $w$ representing the height of the midpoint relative to the support ends of the shell to trigger a 1:2 internal resonance between the first two modes of the FE model with more than 1000 degrees of freedom (DOF). Then we present a bifurcation analysis of quasi-periodic orbits of the FE model under the variations in both excitation frequency and amplitude. In particular, we apply our SSM-based computational framework shown in Fig.~\ref{fig:FulltoROM} to reduce this high-dimensional mechanical system to a four-dimensional ROM. We then use this SSM-based ROM to perform the bifurcation analysis. We further validate the SSM-based predictions using reference solutions of the high-dimensional full system.

\subsection{Setup}

Following \cite{li2022NonlinearAnalysisForceda}, the geometric parameters of the shell in Fig.~\ref{fig:system3} are set to be: the length $L$ = 2 m, width $H$ = 1 m, thickness $t$ = 0.01 m, and the curvature parameter $w$ = 0.041 m.  Material properties are specified with the density $\rho$ = 2700 kg/m$^3$, Young's modulus $E$ = 70 $\times$ 10$^9$ Pa, and Poisson's ratio $\nu$ = 0.33. The forcing amplitude is given as $F$ = 100 N.

The shallow shell model is discretized using flat, triangular shell elements. The FE model here contains 400 elements, and each node in the elements has six DOFs, resulting in $n$ = 1320 DOF~\cite{li2022NonlinearAnalysisForceda}. Thus, the dimension of state space is reduced from 2640 for the full system to 4 of SSM-based ROM~\eqref{eq:red-nonauto-th}. The ODEs of the system obtained by the finite element method can be expressed as
\begin{equation}
\boldsymbol{M}\ddot{\boldsymbol{z}}+\boldsymbol{C}\dot{\boldsymbol{z}}+\boldsymbol{K}\boldsymbol{z}+\boldsymbol{f}(\boldsymbol{z})=\epsilon \boldsymbol{f}^{\mathrm{ext}}(\Omega t),\quad 0<\epsilon\ll1
 \label{eq:system 3}
\end{equation}
where $\boldsymbol{z} \in \mathbb{R}^{1320}$ is a displacement vector, $\boldsymbol{M}, \boldsymbol{C}, \boldsymbol{K}$ are the mass, damping, and stiffness matrices, $\boldsymbol{f}(\boldsymbol{z})$ is an analytic vector-valued function characterizing nonlinear internal force and $\boldsymbol{f}^{\mathrm{ext}}$ is the external load vector. Here we adopt a Rayleigh damping model such that the damping matrix can be expressed as $\boldsymbol{C} = \alpha \boldsymbol{M} + \beta \boldsymbol{K}$. The natural frequencies of the first two modes are given as $\omega_{1} = 149.22$ and $\omega_{2} = 298.78 \approx 2 \omega_1$. We choose $\alpha$ and $\beta$ such that the damping ratios of the first two modes are equal to 0.001. In the following computations, we set $\epsilon=0.03$ unless otherwise stated. We take the transverse displacements at two nodes with coordinates ($x$, $y$) = (0.25$L$, 0.5$H$) and (0.5$L$, 0.5$H$) to characterize the nonlinear vibration of the shell. They are denoted as $z_{\rm{out1}}$ and $z_{\rm{out2}}$ in this example.

\subsection{Periodic and quasi-periodic orbits}

\begin{figure}[!ht]
    \centering
    \includegraphics[width=0.45\textwidth]{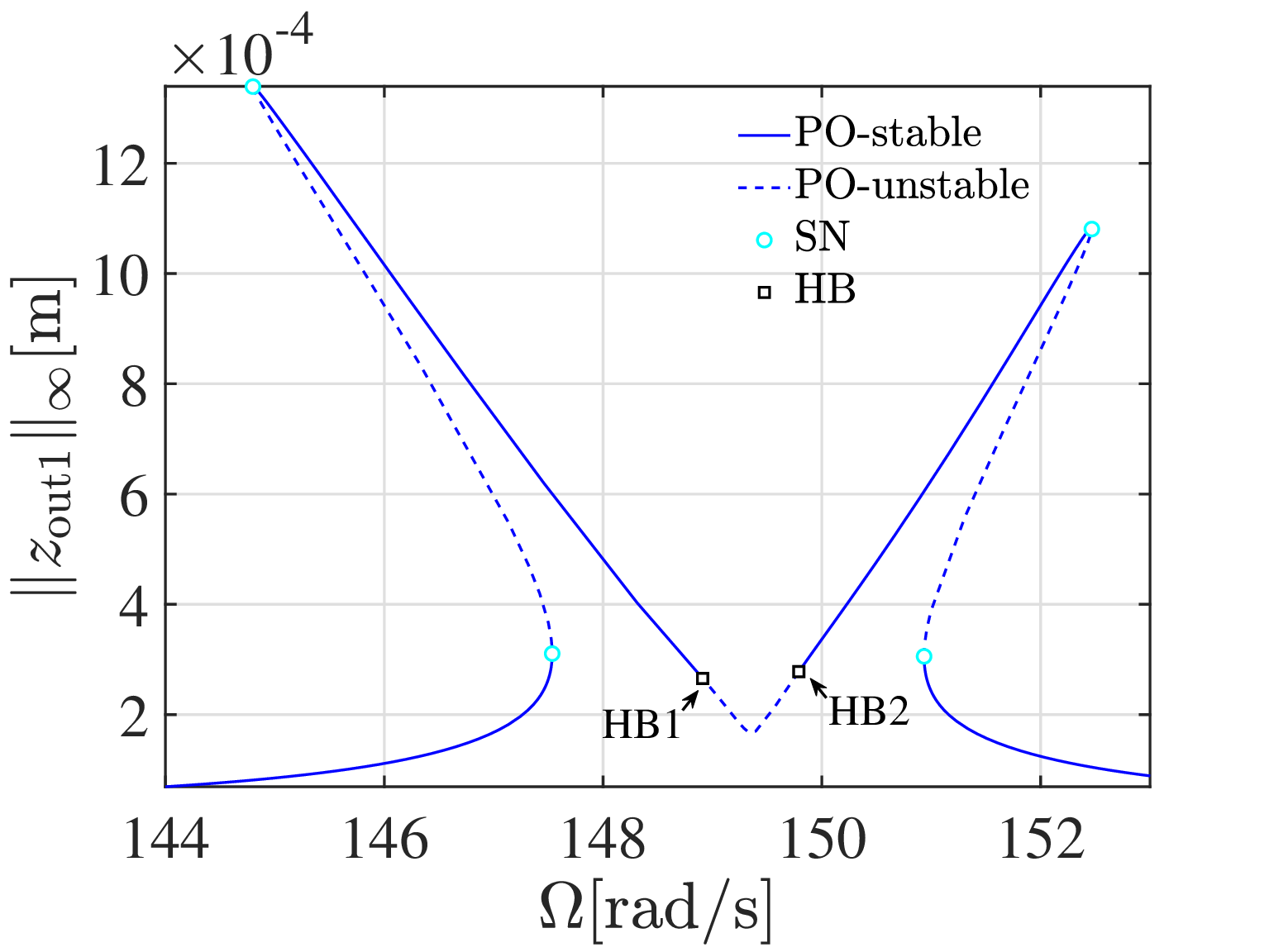}
    \includegraphics[width=0.45\textwidth]{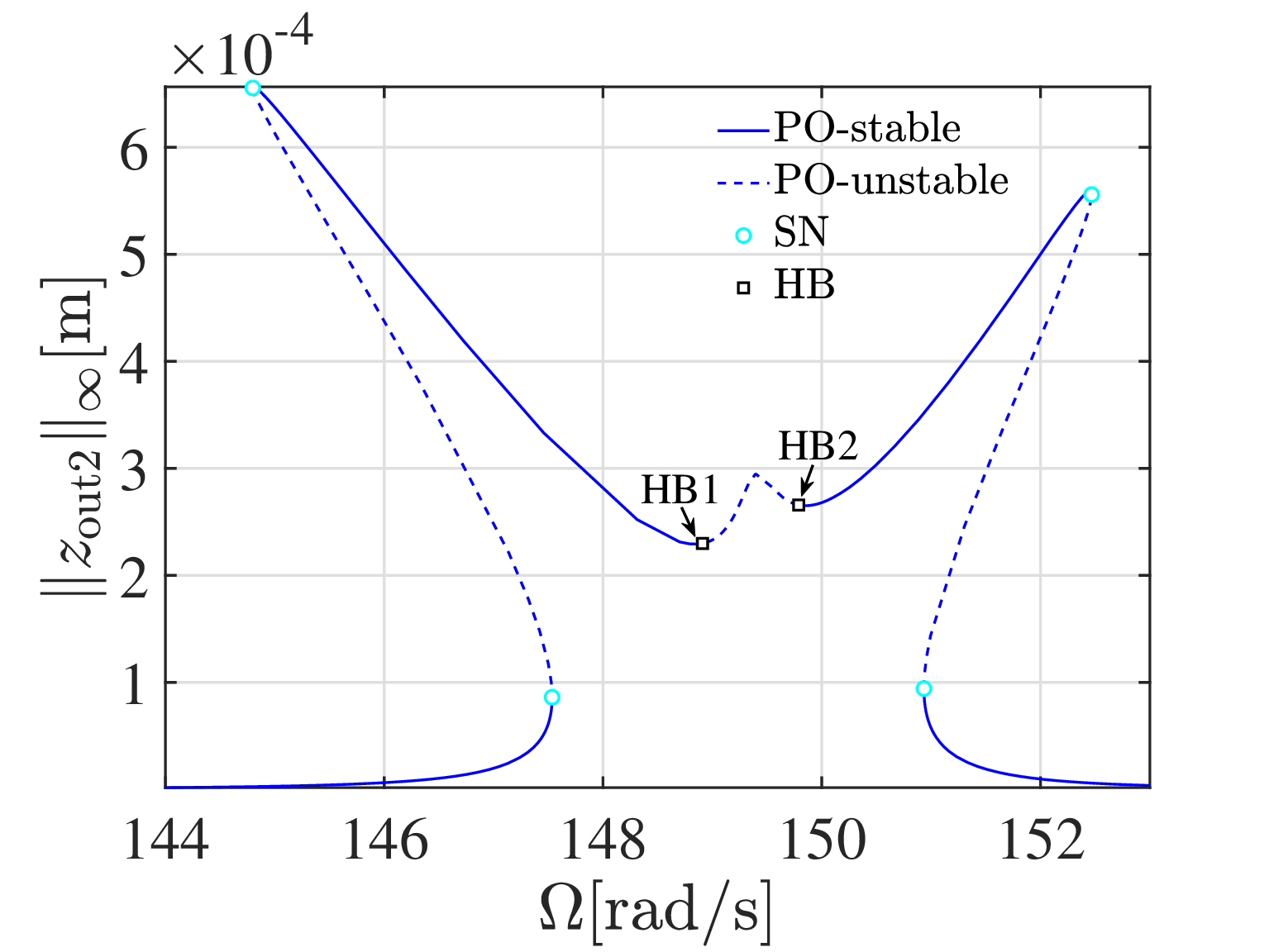}
    \caption{\small FRCs of periodic orbits of the shallow shell at the nodes with coordinates ($x$, $y$) = (0.25$L$, 0.5$H$) (upper panel) and (0.5$L$, 0.5$H$) (lower panel) with $\Omega \in$ [144, 153]. The solid lines and the dashed lines denote stable and unstable solutions, respectively. The circles and squares denote saddle-node (SN) and Hopf bifurcation (HB) points, respectively.}
    \label{fig:ep_z1_z2}
\end{figure}

We first compute the FRC-PO with $\Omega \in [144, 153]$. This FRC-PO is obtained by parameter continuation of the fixed point of the SSM-based ROM~\eqref{eq:red-nonauto-th}. In this example, an $\mathcal{O}(5)$ expansion is used because it is sufficient to yield converged solutions for the selected parameters. Here, $\Vert z_{\rm{out1}} \Vert_\infty$ and $\Vert z_{\rm{out2}} \Vert_\infty$ represent the infinite norm of $z_{\rm{out1}}$, $z_{\rm{out2}}$, giving the amplitude of periodic or quasi-periodic response of the system. The FRC-PO is shown in Fig.~\ref{fig:ep_z1_z2}. Similar to the previous two examples, there are two Hopf bifurcation (HB) points and four saddle-node (SN) bifurcation points on the FRC-PO. Since the periodic motion for $\Omega \in [\Omega_{\rm{HB1}},\Omega_{\rm{HB2}}]$ is unstable, we next analyze the quasi-periodic motion of the system between these two HB points.

\begin{figure*}[!ht]
    \centering
    \includegraphics[width=0.45\textwidth]{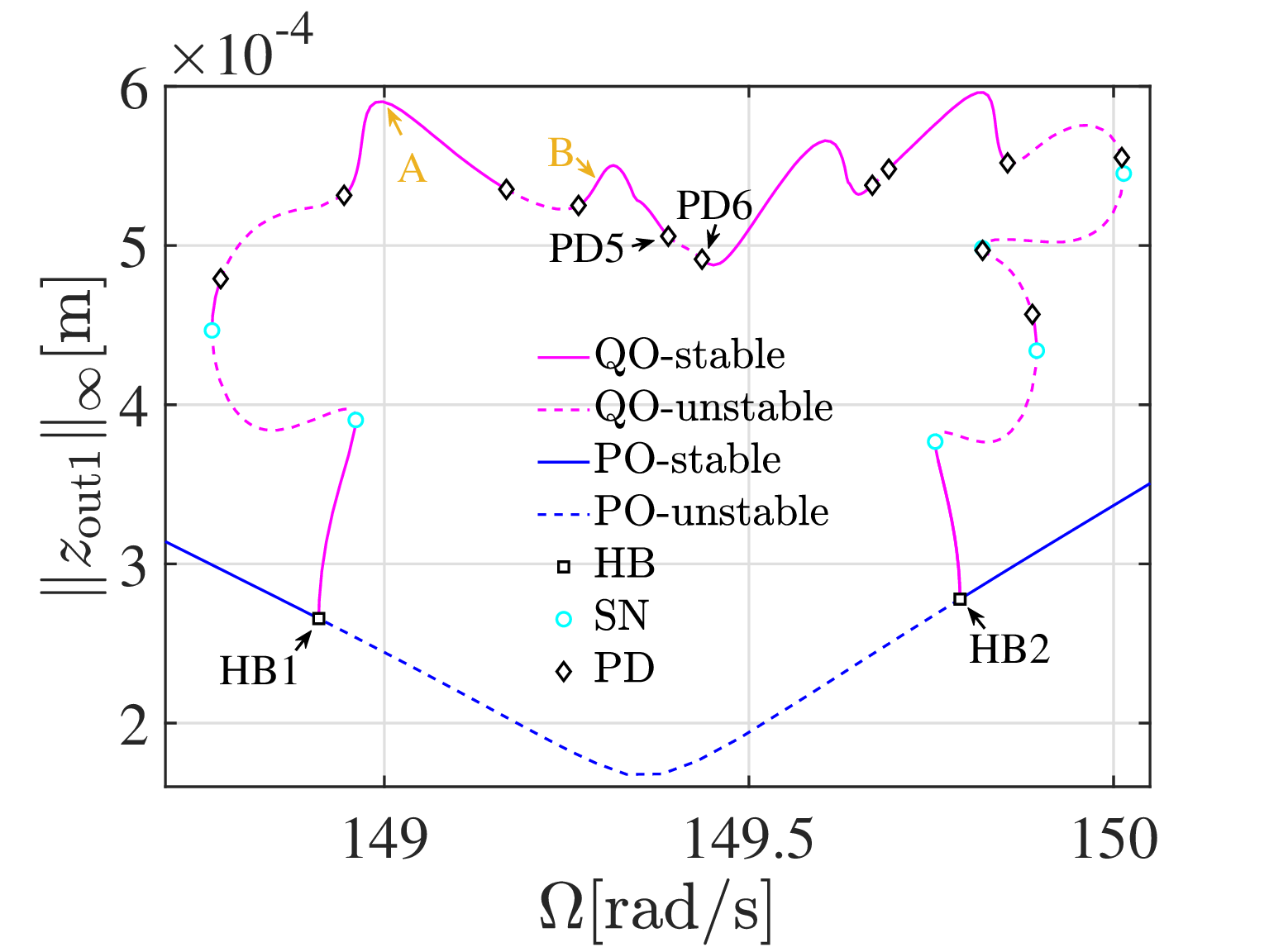}
    \includegraphics[width=0.45\textwidth]{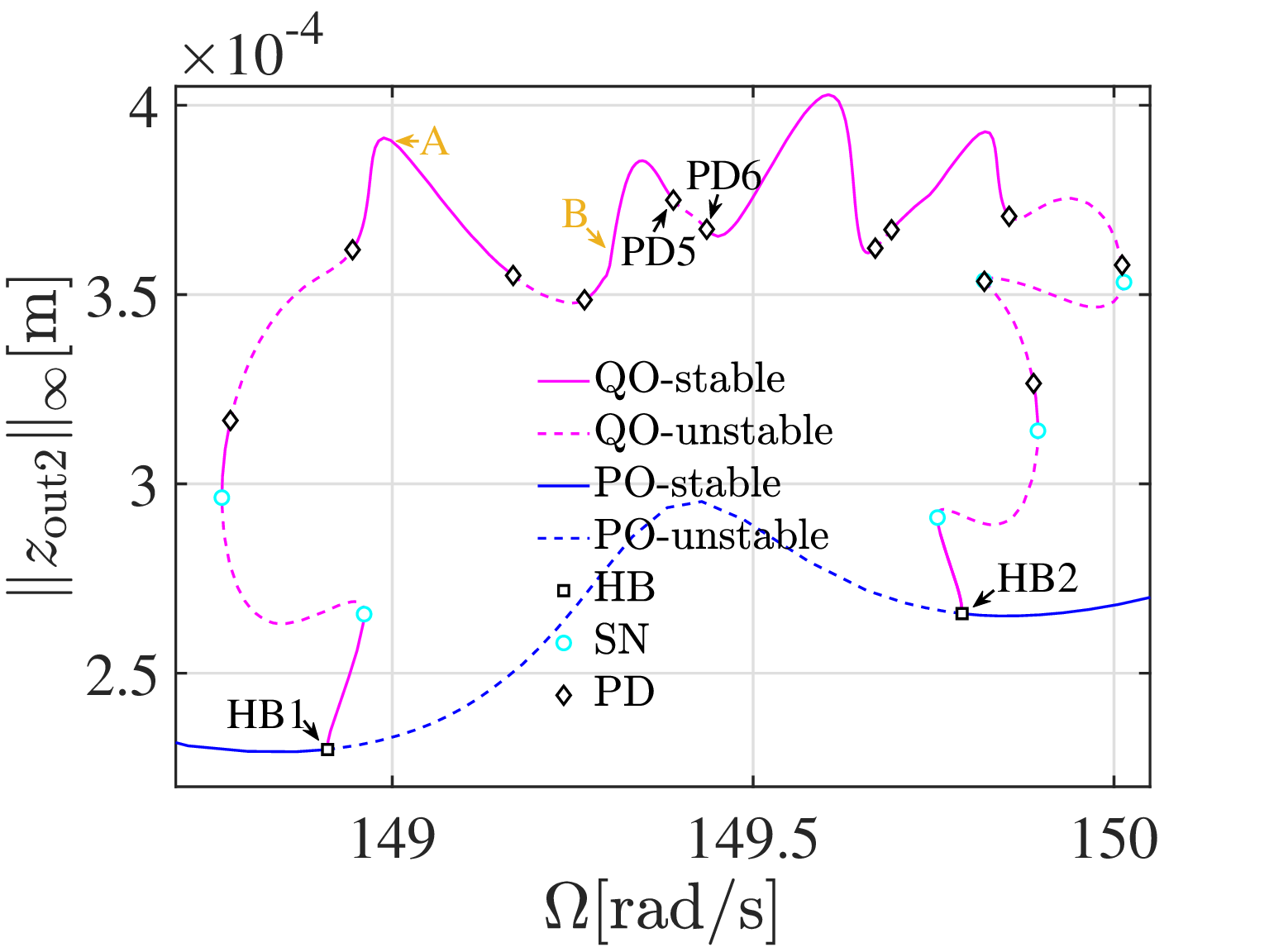}
    \caption{\small FRCs of quasi-periodic and periodic orbits of the shallow shell system at the nodes with coordinates ($x$, $y$) = (0.25$L$, 0.5$H$) (left panel) and (0.5$L$, 0.5$H$) (right panel) with $\Omega \approx \omega_1$. The magenta solid/dashed lines denote stable/unstable quasi-periodic solutions. The blue solid/dashed lines denote stable/unstable periodic solutions. The circles, diamonds, and squares denote saddle-node (SN), period-double (PD), and Hopf bifurcation (HB) points, respectively.}
    \label{fig:HB2po_z1_z2}
\end{figure*}

The FRC-QO under varying $\Omega$ is shown in Fig.~\ref{fig:HB2po_z1_z2}. Along this curve of quasi-periodic orbits, 12 periodic doubling (PD) bifurcation points and 6 saddle-node (SN) bifurcation points are detected. Similar to the previous two examples, the coexistence of stable and unstable quasi-periodic motions can be found. Moreover, we observe the coexistence of periodic and quasi-periodic attractors for $\Omega<\Omega_\mathrm{HB1}$ and $\Omega>\Omega_\mathrm{HB2}$ in Fig.~\ref{fig:HB2po_z1_z2}.

We provide a validation of the SSM-based predictions for the FRC-PO is provided in Appendix \ref{sec:app-shell}. In particular, we use a \textsc{COCO}-based shooting toolbox to extract the FRC-PO. As seen in Fig.~\ref{fig:verify PO_z1z2}, the FRC of periodic orbits obtained from SSM-based reduction matches well with that of the full system obtained via the shooting method. As detailed in Appendix \ref{sec:app-shell}, a significant speed-up gain is obtained using the SSM-based reduction. Specifically, the computational times for extracting the FRC-PO of the SSM-based prediction and the shooting method are about 1 minute and 2 days, respectively.

We note that it is challenging to extract quasi-periodic orbits of this high-dimensional full system. As detailed in Appendix~\ref{sec:app-shell}, we take two representative stable tori, i.e., points A and B in Fig.~\ref{fig:HB2po_z1_z2}, for the sake of verification. Specifically, we take a point on the invariant torus A or B obtained from SSM-based prediction as an initial condition, perform a forward simulation of the full system with the initial condition, and extract the associated quasi-periodic attractor in steady state. As seen in the upper and middle panels of Fig.~\ref{fig:verify QO-z1z2}, the SSM-based predictions match well with that of the reference solutions of the full system. Here, we again achieve a significant speed-up gain. Indeed, a forward simulation of the full system with 4500 cycles to ensure that the steady state arrived took more than 4 days of computational time. In contrast, we can obtain the whole FRC-QO shown in Fig.~\ref{fig:HB2po_z1_z2} in less than 4 minutes.

\subsection{PD and SN bifurcation curves}

\begin{figure*}[!ht]
    \centering
    \includegraphics[width=0.45\textwidth]{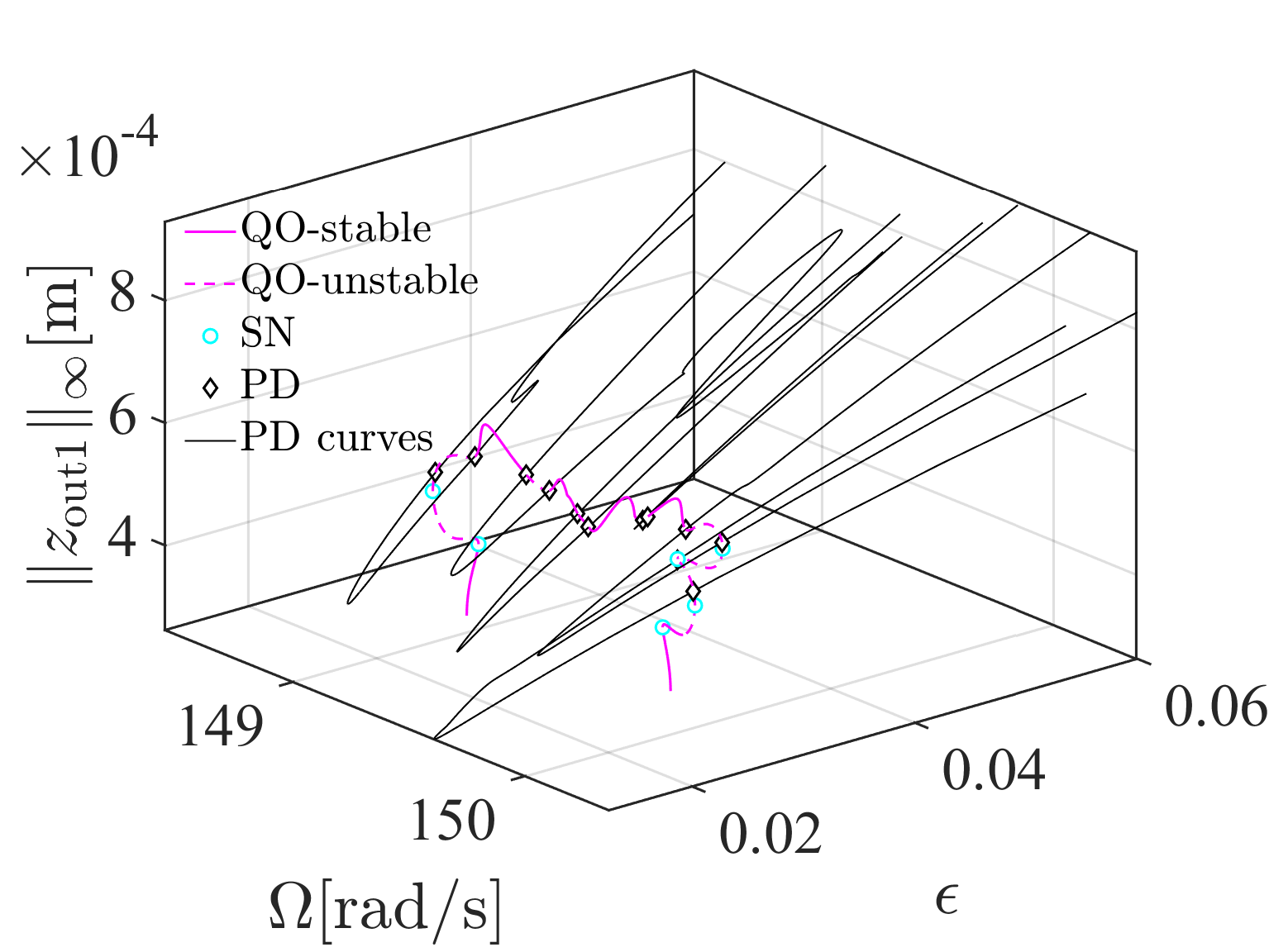} 
    \includegraphics[width=0.45\textwidth]{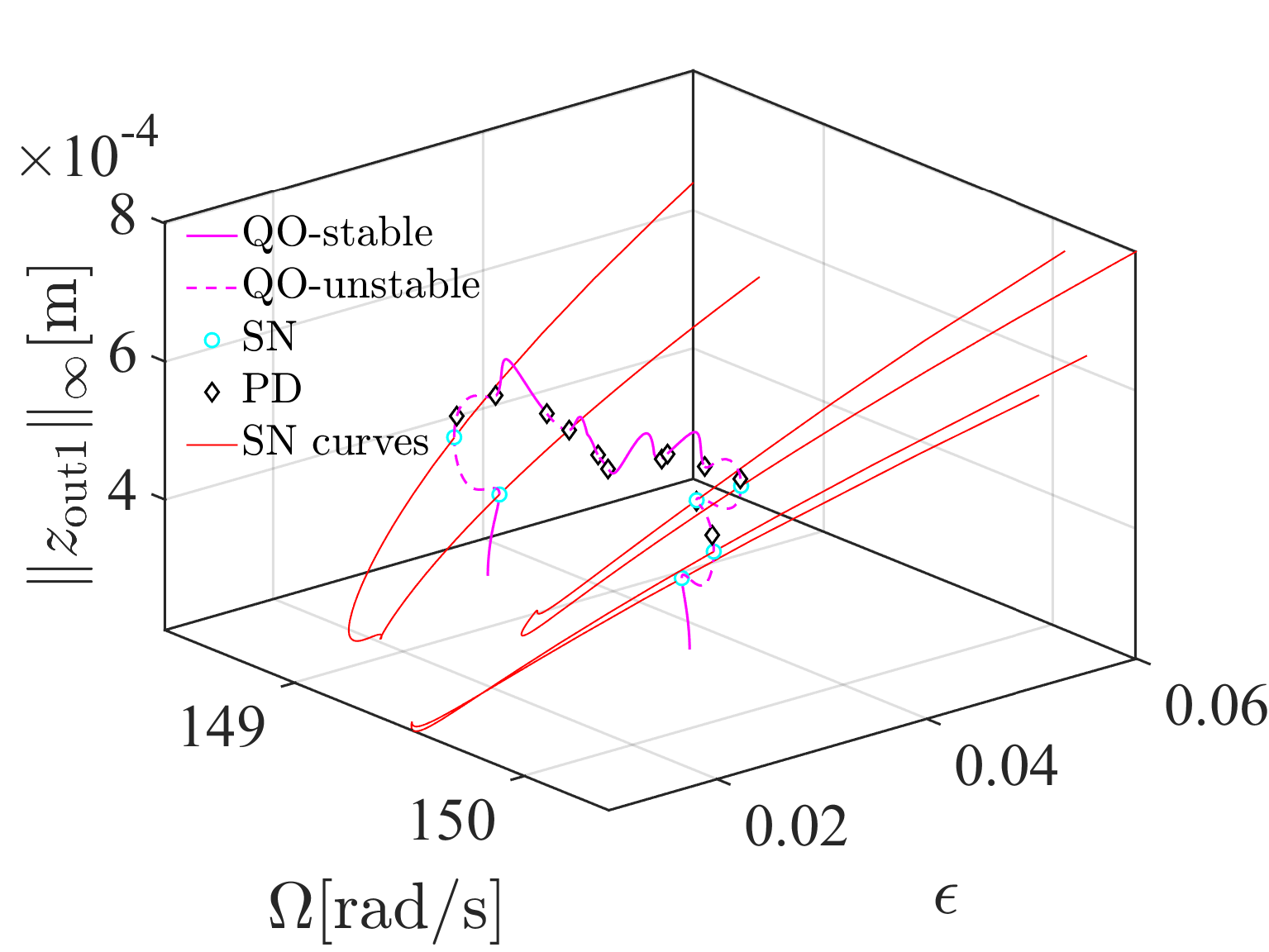}\\
    \includegraphics[width=0.45\textwidth]{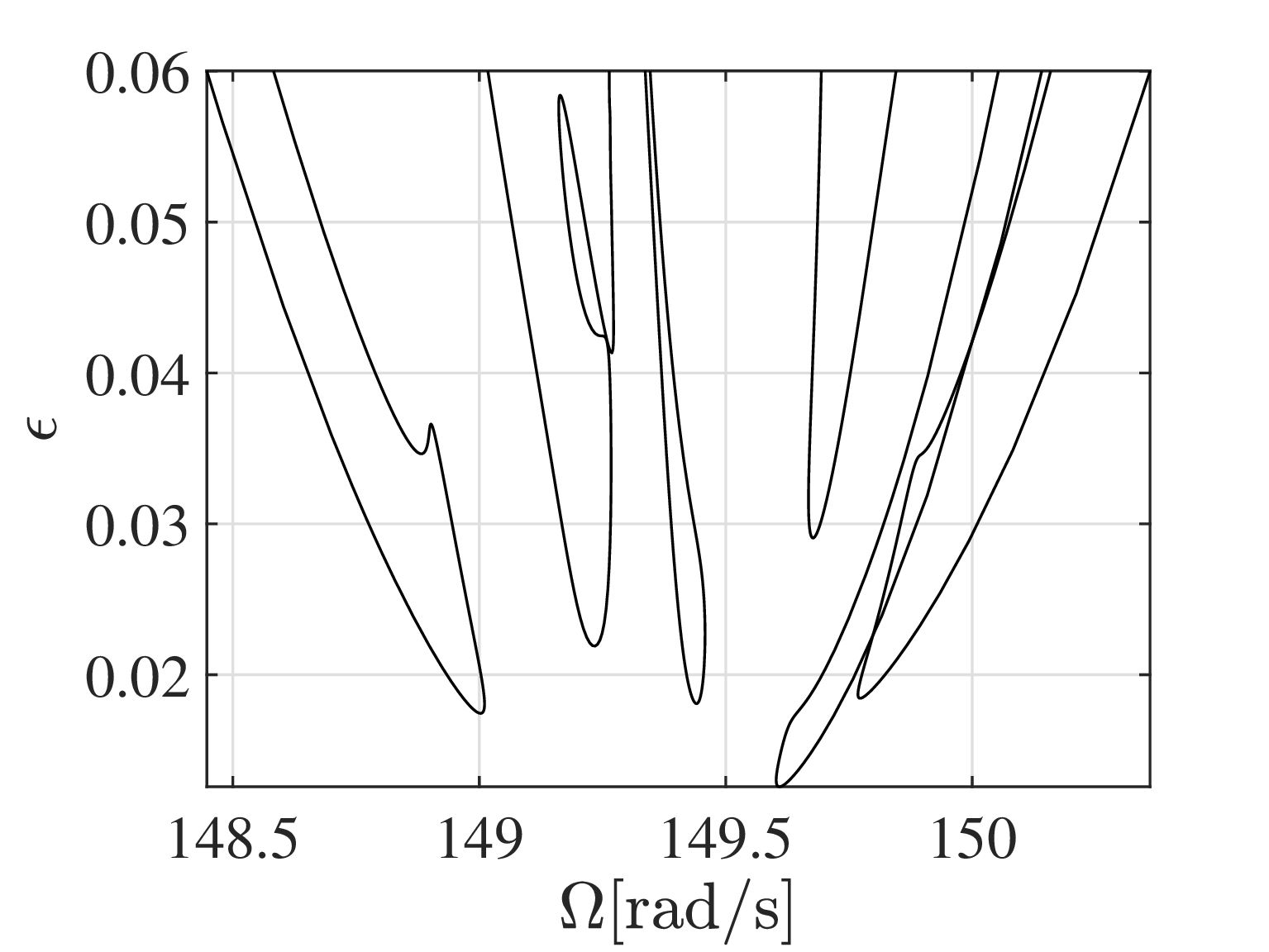}
    \includegraphics[width=0.45\textwidth]{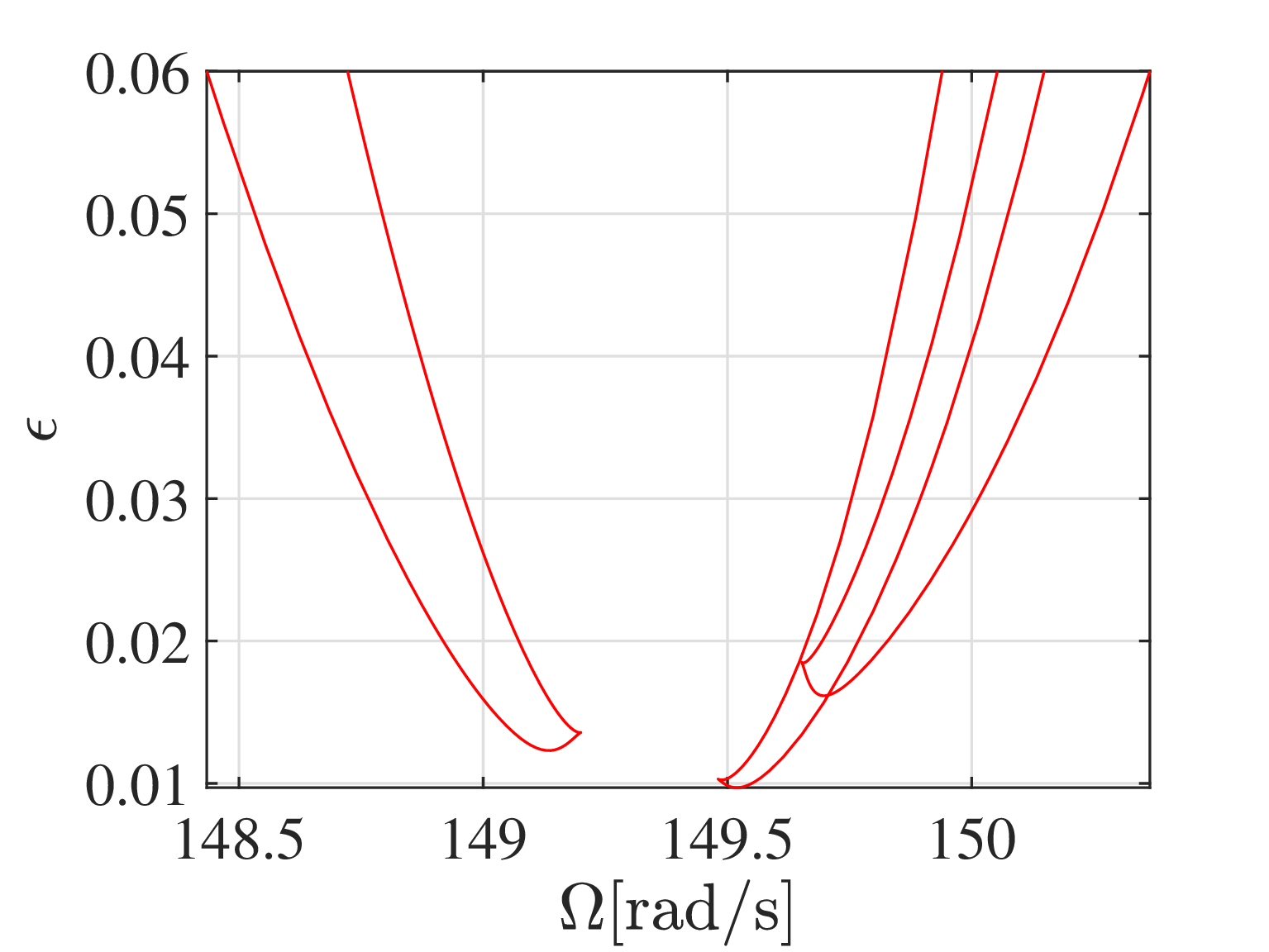}
    \caption{\small Bifurcation curves of quasi-periodic orbits of the shallow shell system. The upper-left and upper-right panels present the projection of the continuation paths of period-doubling (PD) and saddle-node (SN) bifurcated quasi-periodic orbit onto $(\Omega,\epsilon,||z_{\rm{out1}}||_{\infty})$. The lower two panels give the top view of the corresponding upper panels. In the upper two panels, the magenta solid/dashed lines denote the FRC-QO of stable/unstable quasi-periodic solutions with $\epsilon$ = 0.03 and varying $\Omega$. The circles and diamonds along the FRC-QO denote SN and PD bifurcations, respectively. The black/red solid lines denote the continuation of PD/SN bifurcations with $\epsilon \in$ [0.0001,0.06].}
    \label{fig:po2SN&PD_z1}
\end{figure*}

Similar to the case of two coupled oscillators, we perform parameter continuation of PD and SN bifurcated quasi-periodic orbits on the FRC-QO shown in Fig.~\ref{fig:HB2po_z1_z2} to study how these PD and SN bifurcated quasi-periodic orbits evolve with varying $\epsilon$. In particular, we take the PD and SN points on the FRC-QO in Fig.~\ref{fig:HB2po_z1_z2} as the initial solutions of the continuation runs. The projection of the continuation paths for the PD and SN bifurcation points with $(\Omega, \epsilon) \in [144, 153] \times [0.0001, 0.06]$ are shown in the left and right panels of Fig.~\ref{fig:po2SN&PD_z1}, respectively. Here, the lower panels are the projection of the corresponding upper panels onto the parameter plane $(\Omega,\epsilon)$. The amplitudes of SN and PD bifurcated quasi-periodic orbits is reduced as $\epsilon$ decreases, as seen in the upper two panels. We observe that there are 6 branches for the PD continuation path and 3 branches for the SN continuation path. Further, the number of PD or SN bifurcation points changes dynamically as $\epsilon$ varies. In particular, these bifurcations disappear for sufficiently small forcing amplitude $\epsilon$. These features are similar to those in the system of two coupled oscillators.

\subsection{Periodic doubling bifurcations to chaos}

\begin{figure}[!ht]
    \centering
    \includegraphics[width=0.48\textwidth]{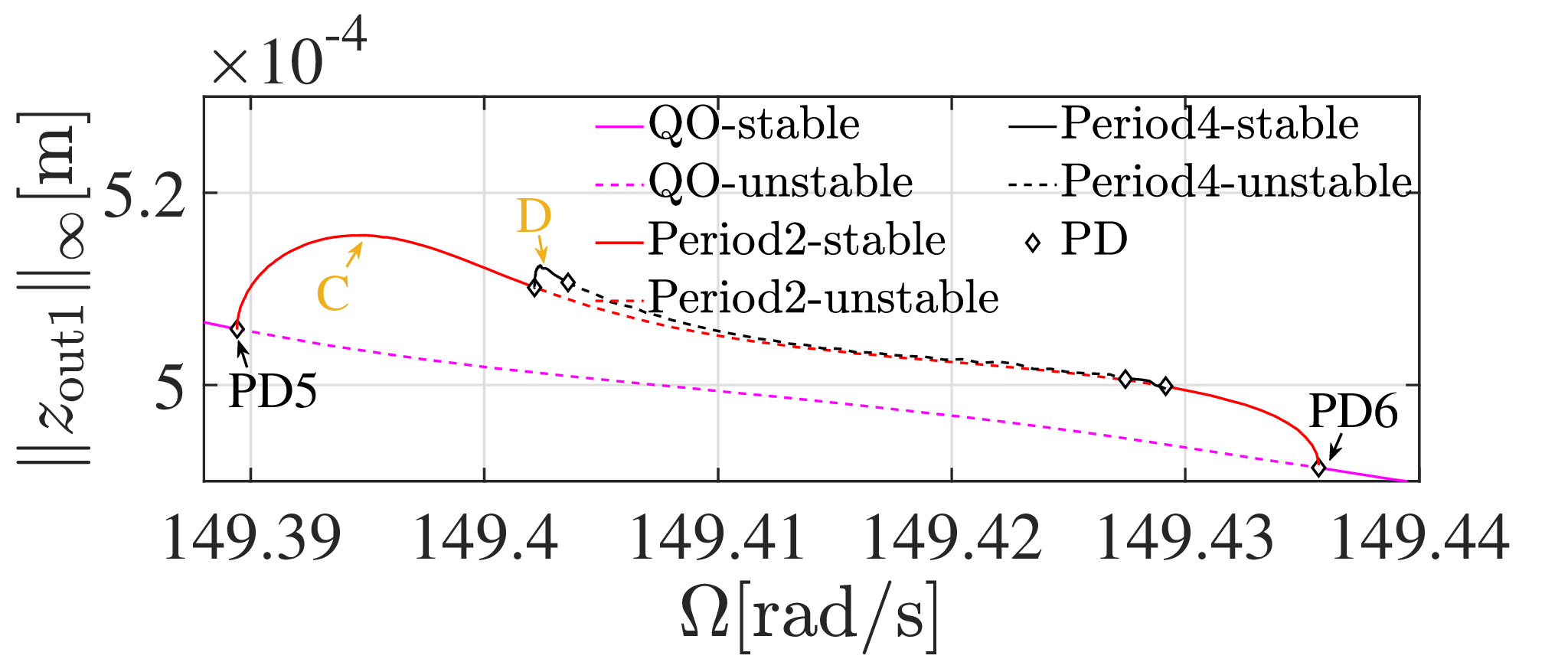}
    \caption{\small FRCs of quasi-periodic orbits of the shallow shell system at the mesh node ($x$, $y$) = (0.25$L$, 0.5$H$) with $\Omega \in [\Omega_{\rm{PD5}},\Omega_{\rm{PD6}}]$. The magenta solid/dashed lines denote the FRC-QO born out of HB1 in FRC shown in the upper panel of Fig.~\ref{fig:HB2po_z1_z2}; red solid/dashed lines denote Period2 quasi-periodic orbits born out of PD5 in the FRC-QO; black solid/dashed lines denote Period4 quasi-periodic orbits born out of a PD point on the FRC of Period2 quasi-periodic orbits. The diamonds denote PD bifurcations.}
    \label{fig:PD2po_z1}
\end{figure}

Similarly to the previous two mechanical systems, we expect that there is a cascade of PD bifurcations for quasi-periodic orbits because PD bifurcation points were detected in Fig.~\ref{fig:HB2po_z1_z2}. Indeed, the cascade of PD bifurcations is found on the region $\Omega \in [\Omega_{\rm{PD5}},\Omega_{\rm{PD6}}]$. By switching to parameter continuation of quasi-periodic orbits with doubled internal period (Period2) at PD5, we obtain the continuation path shown as the red line in Fig.~\ref{fig:PD2po_z1}. Along this Period2 FRC-QO, two new PD bifurcation points are detected. By switching to parameter continuation of quasi-periodic orbits with a quadrupled internal period (Period4) at one of the PD points on the Period2 FRC-QO, we obtain the continuation path of quasi-periodic orbits shown as a black line in Fig.~\ref{fig:PD2po_z1}. We once again detect two new PD points on this black curve of FRC-QO, which illustrates the cascade of PD bifurcations.

\begin{figure}[!ht]
    \centering
    \includegraphics[width=0.45\textwidth]{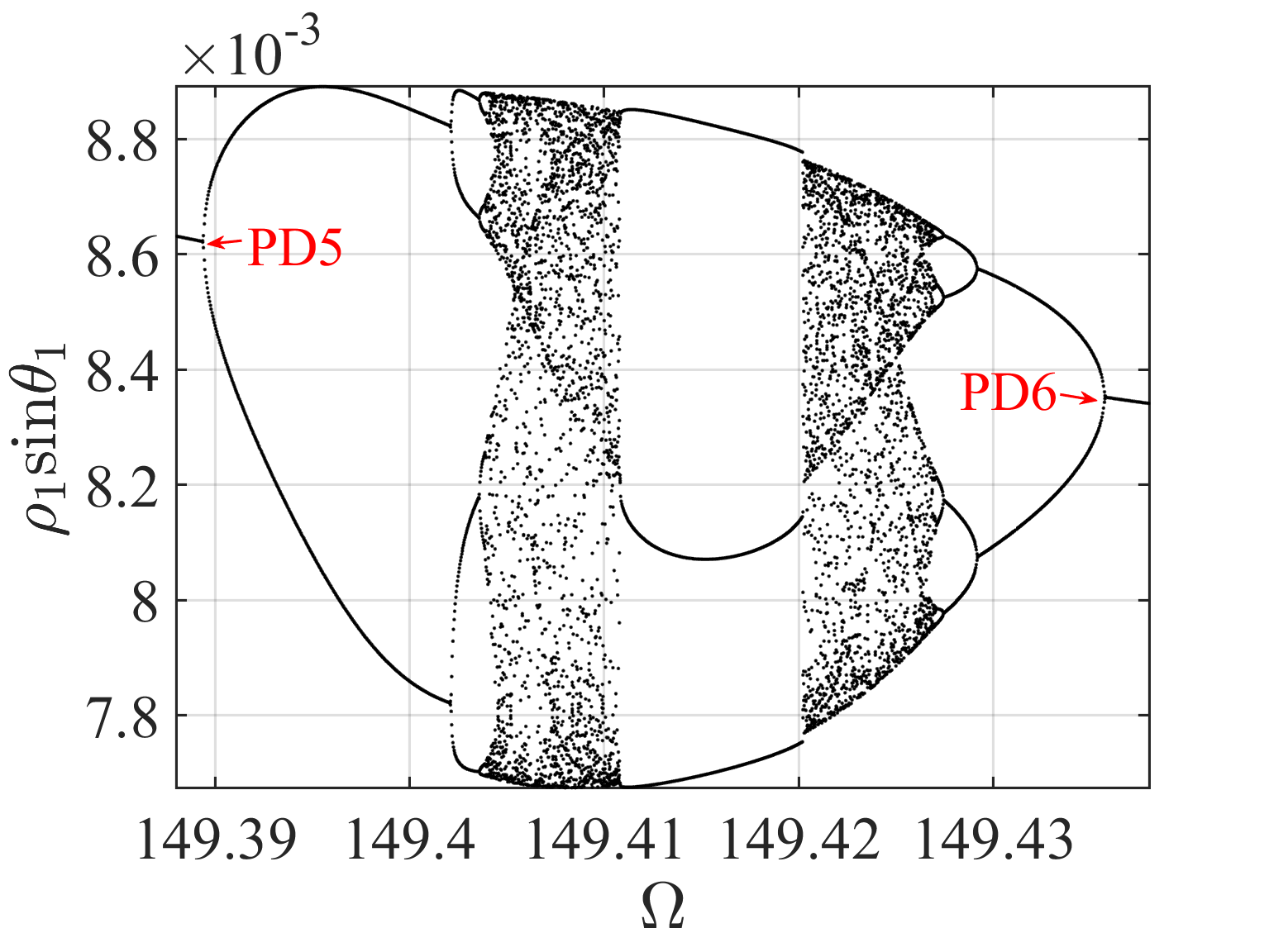}
    \caption{\small {Bifurcation diagram of attractors for the SSM-based ROM \eqref{eq:red-nonauto-th} of the shallow curved shell in \eqref{eq:system 3} with $\Omega \in [149.388,149.438]$. Here, the vertical axis gives the intersections of the attractors with the Poincar{\'e} section $\{(\boldsymbol{\rho},\boldsymbol{\theta}): q_3=\rho_1\sin\theta_1, \dot{q}_3\equiv0,\ddot{q}_3>0\}$ of the ROM, namely, the maximum of $\rho_1\sin\theta_1$ associated with the attractors.}}
    \label{fig:bif-chaos-ex3}
\end{figure}

In the previous two mechanical systems, a cascade of PD bifurcations finally leads to chaotic motions. To test whether this is a common feature for mechanical systems with 1:2 internal resonance, we perform forward simulations of the SSM-based ROM~\eqref{eq:red-nonauto-th} to extract attractors for $\Omega\in[149.388,149.438]$. We present the obtained bifurcation diagram for these attractors of the SSM-based ROM in Fig.~\ref{fig:bif-chaos-ex3} (cf. Fig.~\ref{fig:PD2po_z1}). This diagram is obtained following a similar procedure as that of Fig.~\ref{fig:bif-chaos-ex1}. This bifurcation diagram displays the transition to chaos via the cascade of period-doubling bifurcations. This bifurcation diagram is consistent with that of Fig.~\ref{fig:PD2po_z1} but provides a more complete picture of the complex dynamics in this parameter region. Interestingly, we observe a large window of period-3 limit cycles of the SSM-based ROM~\eqref{eq:red-nonauto-th}. These period-3 limit cycles correspond to period-3 tori of the full system.

\begin{figure}[!ht]
    \centering
    \includegraphics[width=0.45\textwidth]{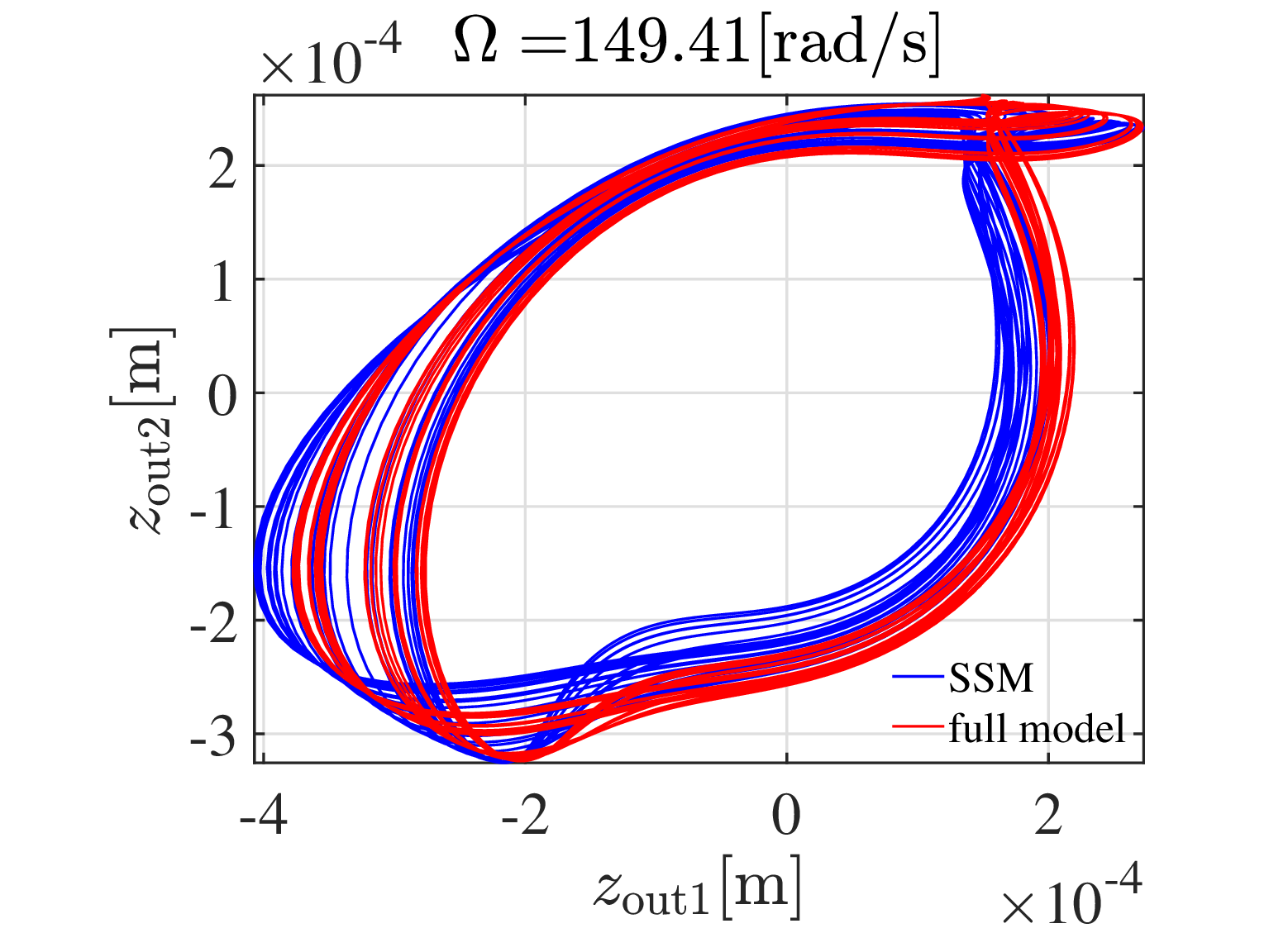}
    \includegraphics[width=0.45\textwidth]{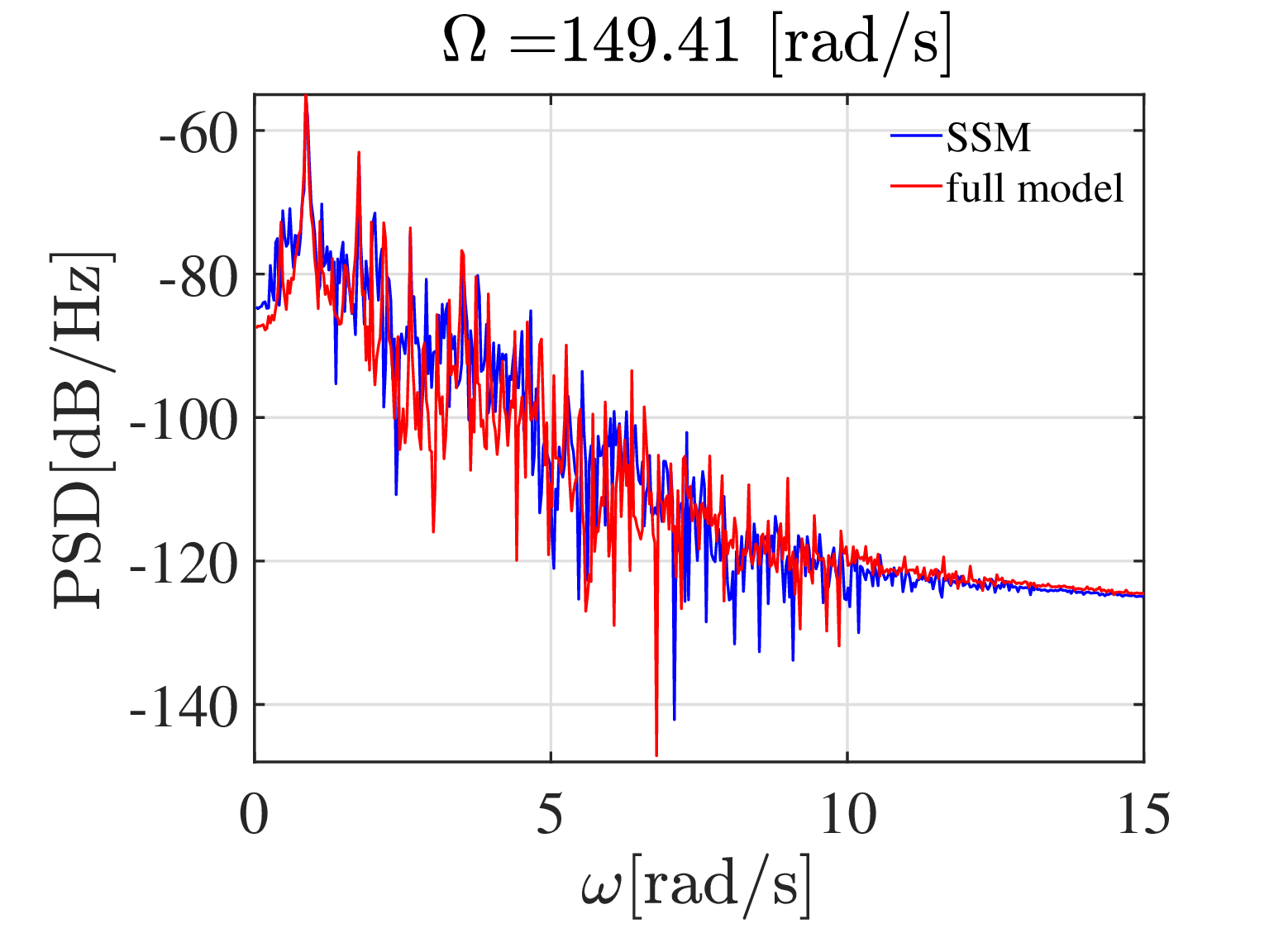}
    \caption{\small (Upper panel) Intersections of the period-$2\pi/\Omega$ map of a chaotic attractor of the shallow shell model with $\Omega$ = 149.41. The blue lines denote the results obtained via the SSM-based predictions. The red lines represent reference results of the full model obtained using numerical integration. (Lower panel) The power spectral density (PSD) plot of the intersected trajectories at the upper panel.~{The SSM-based prediction gives that the maximum Lyapunov exponent of this chaotic attractor is 0.0534.}}
    \label{fig:chaos_z1}
\end{figure}

{We again have a close look at a representative chaotic attractor at $\Omega = 149.41$. We map the simulated attractor} back to physical coordinates. The intersections of the simulation results with the Poincar{\'e} section induced by the period-$2 \pi/\Omega$ map are shown as a blue line on the upper panel of Fig.~\ref{fig:chaos_z1}. The associated plot of PSD is shown in the lower panel of Fig.~\ref{fig:chaos_z1}.~{We follow the same computational method as in the right panel of Fig.~\ref{fig:chaos_x2} to compute the maximum Lyapunov exponent of this strange attractor. The obtained maximum Lyapunov exponent is 0.0534. Therefore}, a chaotic attractor is indeed observed in steady state.

To verify the effectiveness of the SSM-based predictions for the cascade of PD bifurcations, we take representative Period2 and Period4 quasi-periodic orbits, namely, points C and D on the FRC-QO shown in Fig.~\ref{fig:PD2po_z1}, and also the chaotic attractor in Fig.~\ref{fig:chaos_z1} to perform the verification. Specifically, we apply forward simulations of the full system to obtain reference solutions. More details on the forward simulations can be found in Appendix~\ref{sec:app-shell}. As seen in the lower two panels of Fig.~\ref{fig:verify QO-z1z2}, SSM-based predictions for the Period2 quasi-periodic orbit C and the Period4 quasi-periodic orbit D match well with that of the reference solutions. Likewise, as seen in Fig.~\ref{fig:chaos_z1}, the predicted chaotic attractor is confirmed by the numerical integration of the full system.~{Here, computing the maximum Lyapunov exponent of the chaotic attractor via a long-time forward integration of the variational equations for the 2046-dimensional full system~\eqref{eq:system 3} is impractical, and thus, we do not report the reference result for this maximum Lyapunov exponent.}


\section{Conclusion}  
\label{sec:conclusion}

In this paper, we perform bifurcation analysis of quasi-periodic orbits in harmonically excited mechanical systems with 1:2 internal resonance. Reductions on spectral submanifolds (SSMs) are used to transform quasi-periodic orbits of high-dimensional nonlinear mechanical systems into limit cycles of four-dimensional reduced-order models (ROMs). This transformation enables accurate and efficient predictions of local and global bifurcations of quasi-periodic orbits. Specifically, we can uncover homoclinic bifurcations of quasi-periodic orbits and isola and simple bifurcations of the forced response curve of quasi-periodic orbits. The SSM-based ROMs also enable effective predictions on the cascade of period-doubling bifurcations of quasi-periodic orbits and the coexistence of quasi-periodic and chaotic attractors.

We have considered three representative 1:2 internally resonant mechanical systems to illustrate the effectiveness of SSM reductions in the bifurcation analysis of quasi-periodic orbits:
\begin{itemize}
    \item Two coupled nonlinear oscillators. This system serves as a benchmark study. We have extracted the bifurcation curve of periodic doubling and saddle-node quasi-periodic orbits for the system under the variations in excitation frequency and amplitude. We have also uncovered the homoclinic bifurcations of quasi-periodic orbits.
    \item A shallow curved beam. The forced response curve of quasi-periodic orbits (FRC-QO) of this system can have isolated branches, and we have identified isola and simple bifurcations where the isolas were born and merged. We have also uncovered the coexistence of quasi-periodic and chaotic attractors in this system.
    \item A shallow shell structure. This structure is discretized via a finite element model with more than 1300 degrees of freedom. We use this example to illustrate the effectiveness of SSM-based ROMs for the bifurcation analysis of quasi-periodic orbits in high-dimensional nonlinear systems. Indeed, we can extract the FRC-QO of this high-dimensional system in a few minutes.
\end{itemize}  
In all these three systems, we have consistently uncovered the cascade of PD bifurcations for quasi-periodic orbits that finally lead to chaotic motions. We have also validated the SSM-based predictions in these systems using reference solutions of the full systems.

In summary, we have demonstrated that the SSM-based reductions enable an efficient and accurate computational framework for the bifurcation analysis of quasi-periodic orbits for 1:2 internally resonant mechanical systems. We believe that this computational framework can be effectively applied to other mechanical systems other than the three systems considered in this study. We have also uncovered the complex local and global bifurcation mechanisms of quasi-periodic orbits of mechanical systems with 1:2 internal resonance. Our findings shed light on the utilization and design of these quasi-periodic motions.


\begin{acknowledgements}
HL and ML acknowledge the financial support of the National Natural Science Foundation of China (No. 12302014), Guangdong Natural Science Foundation (No. 2024A1515011709), and Shenzhen Science and Technology Innovation Commission (No. 20231115172355001). We are grateful to Junqing Wu for the helpful discussions on the computation of Lyapunov exponents.
\end{acknowledgements}

%

\section*{Data availability}
The data used to generate the numerical results included in this paper are available from the corresponding author on request.

\section*{Conflict of interest}
The authors declare that they have no conflict of interest.

\appendix

\section{Supplementary materials for the two coupled oscillators}
\label{sec:appendix-A}

\begin{figure}[!ht]
    \centering
    \includegraphics[width=0.45\textwidth]{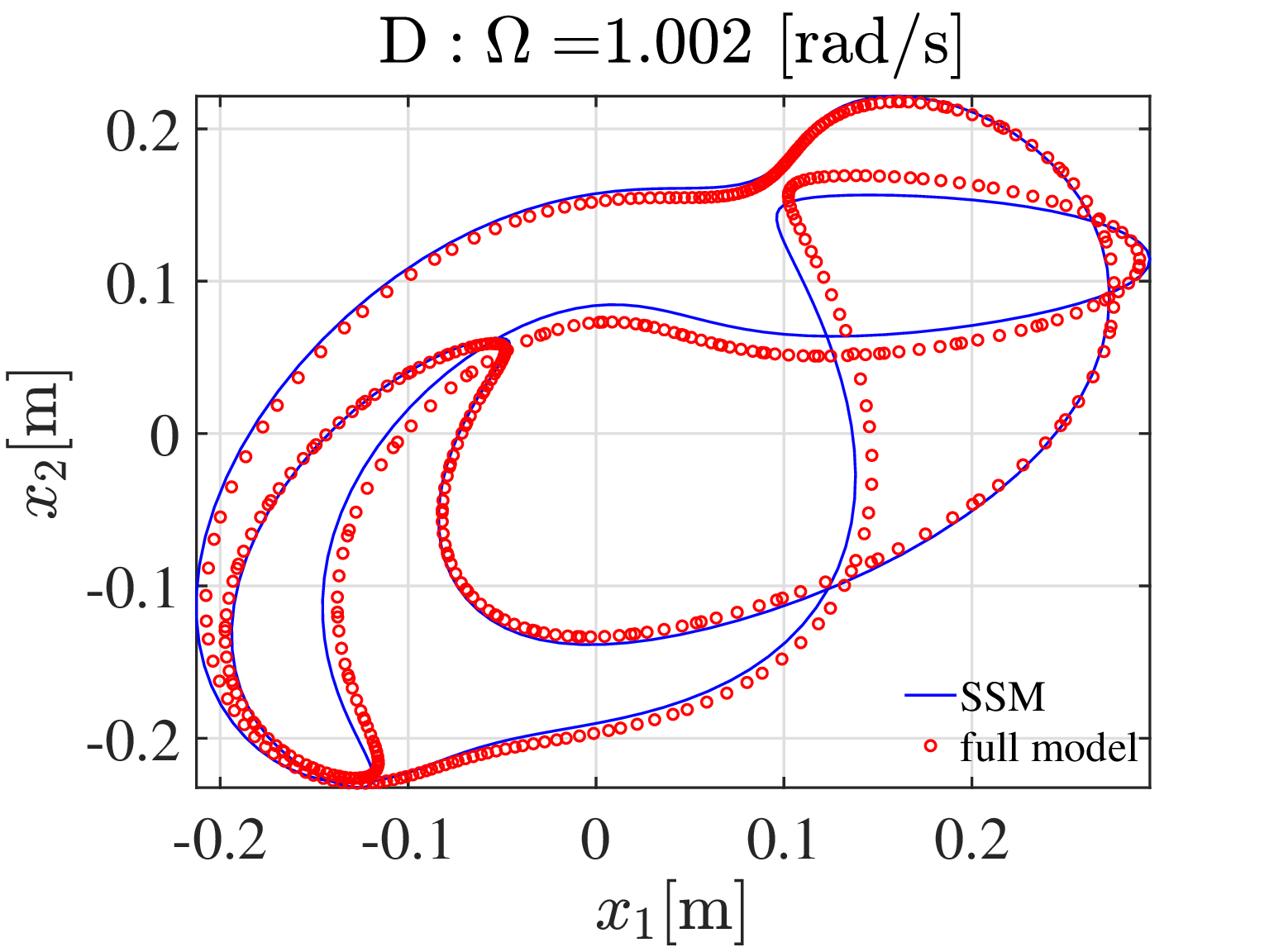}
    \includegraphics[width=0.45\textwidth]{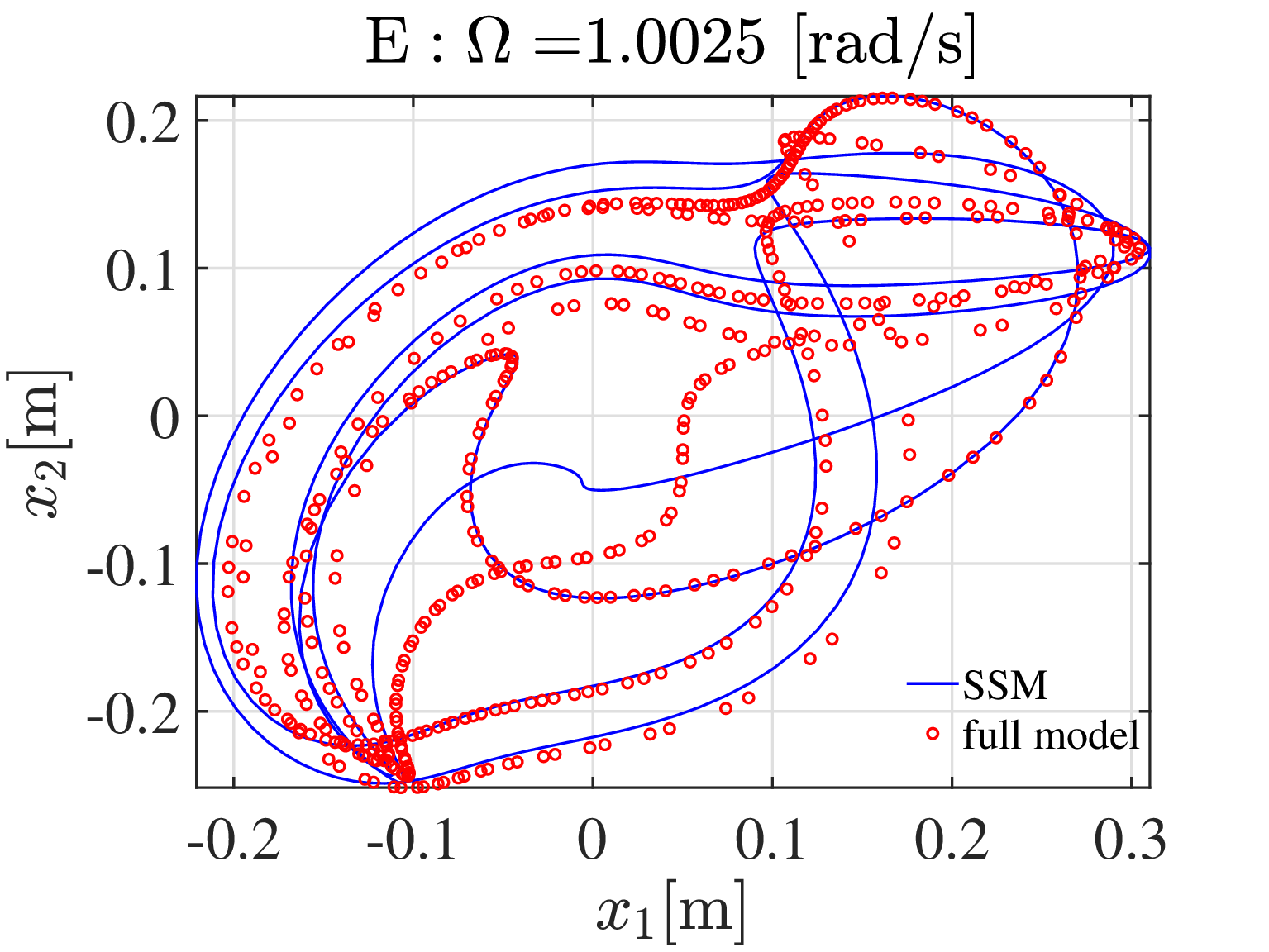}
    \caption{\small Intersections of the period-$2\pi/\Omega$ map of the quasi-periodic orbits D (upper panel) and E (lower panel) of the two coupled nonlinear oscillators with $\Omega$ = 1.002 and 1.0025 (cf.~Fig.~\ref{fig:L3_x2}). The blue lines denote the SSM-based predictions, while the red cycles denote reference results of the full model obtained using numerical integration.}
    \label{fig:verify_L2_stable}
\end{figure}

Here, we verify Period2 and Period4 quasi-periodic orbits in Fig.~\ref{fig:L3_x2}. In particular, we take points D and E in Fig.~\ref{fig:L3_x2} as representatives of Period2 and Period4 quasi-periodic orbits. We compute their Poincar{\'e} intersections associated with period-$2\pi/\Omega$-map using SSM-based prediction and the also numerical integration of the original system. Indeed, since quasi-periodic orbits D and E are stable, we can apply forward simulation of the original system and extract them in steady state. The obtained results are plotted in Fig.~\ref{fig:verify_L2_stable}. We observe that the results from SSM-based prediction match well with the reference solutions, especially for the quasi-periodic orbit D.

\begin{figure}[!ht]
    \centering
    \includegraphics[width=0.45\textwidth]{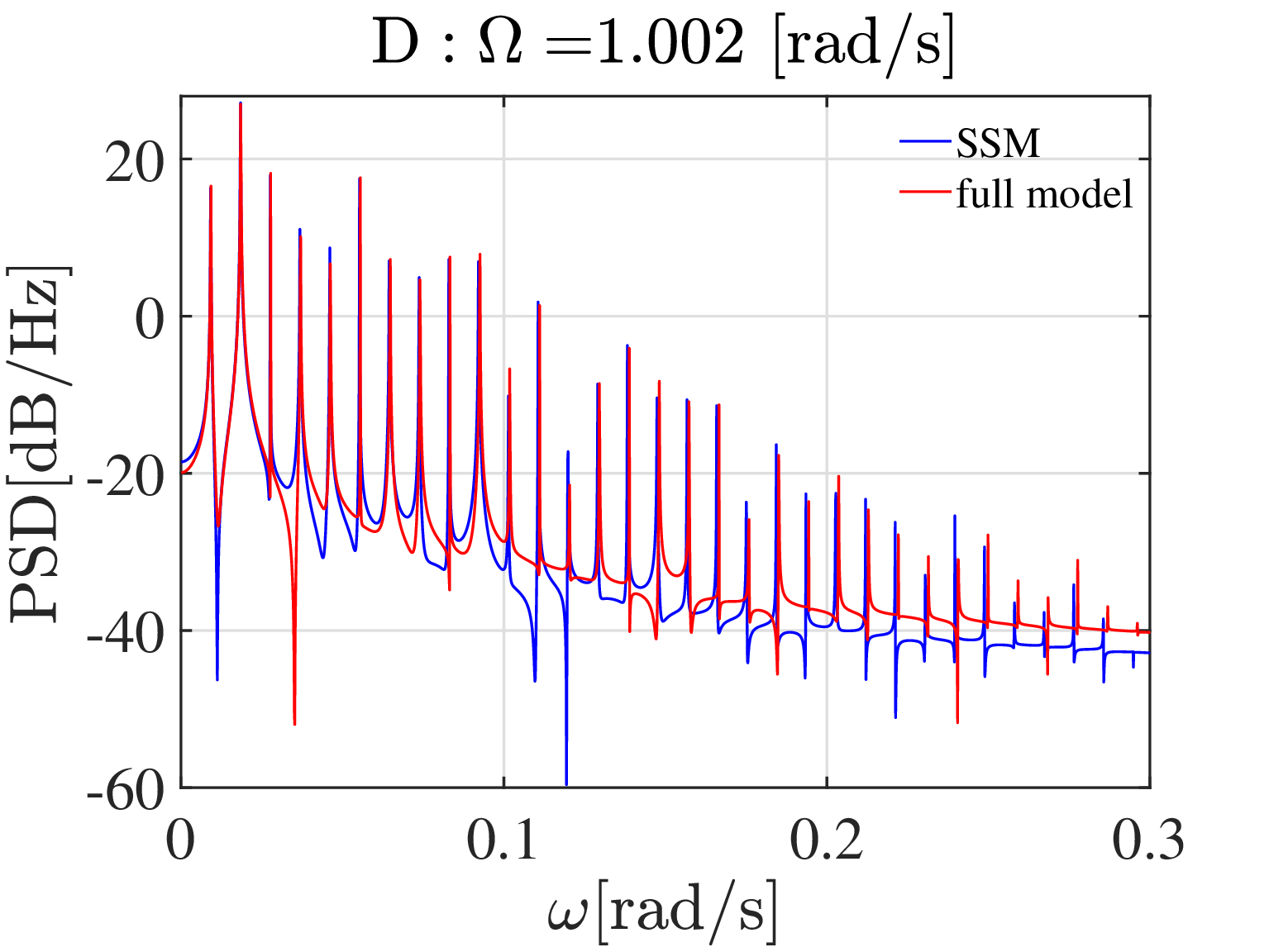}
    \includegraphics[width=0.45\textwidth]{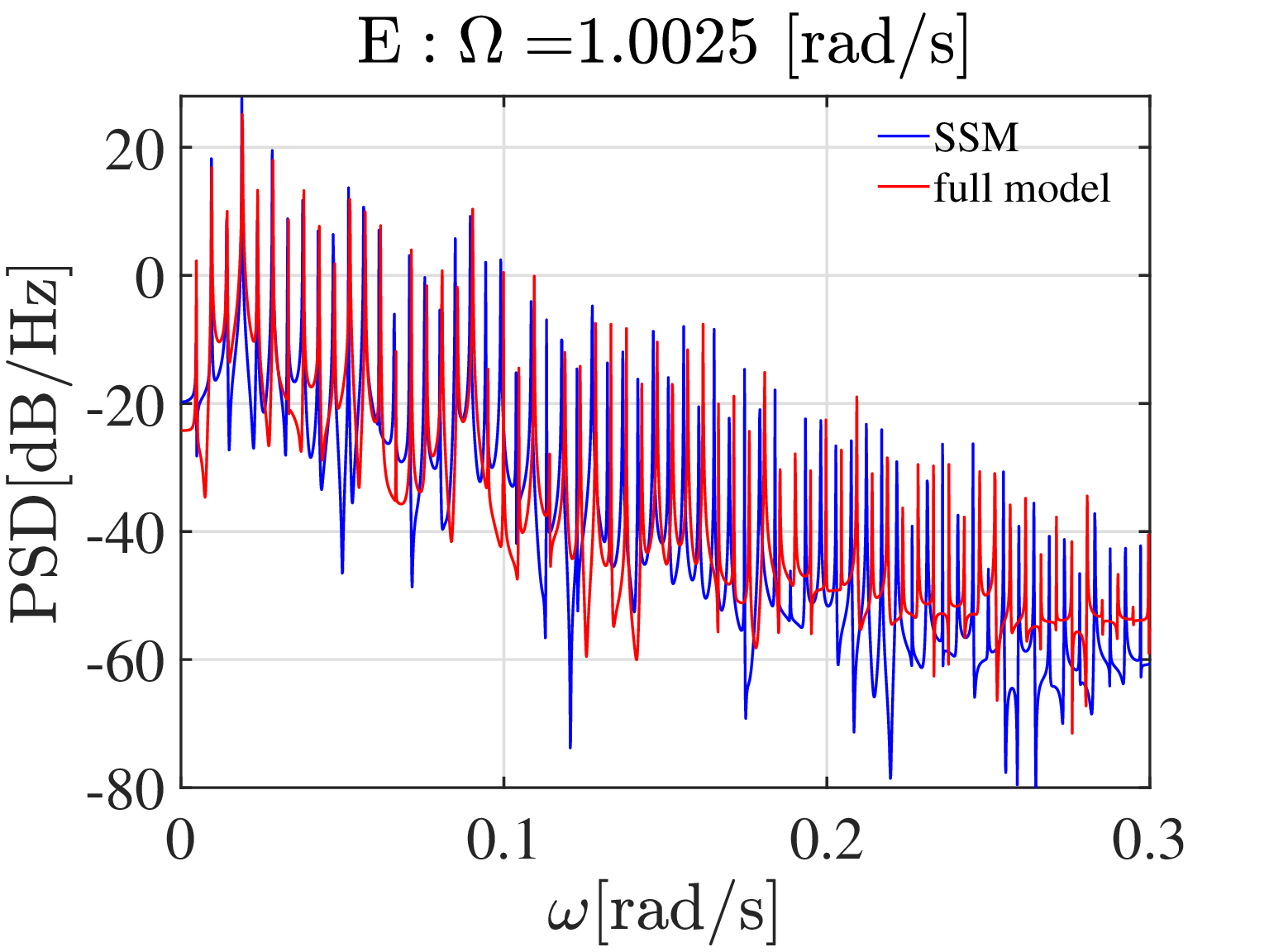}
    \caption{\small The power spectral density (PSD) of signals induced by the intersections in Fig.~\ref{fig:verify_L2_stable}. The blue/red lines denote the PSD of the results of SSM-based prediction and the reference solutions of the original model.}
    \label{fig:PSD-qusia}
\end{figure}

We further extract the principal frequency components of the intersected trajectories shown in Fig.~\ref{fig:verify_L2_stable}. Specifically, we perform power spectral density (PSD) analysis for these signals and present the obtained results in Fig.~\ref{fig:PSD-qusia}, from which we see that the SSM-based predictions match well with that of the reference solution of the original model. As expected, we see that the Period4 motion has more frequency components than that of the Period2 motion. Indeed, as a cascade of period-doubling bifurcations appears, more frequency components will be observed, and the PSD plot will behave as the one in the lower panel of Fig.~\ref{fig:chaos_x2} in the chaotic region.

\section{Supplementary materials for the shallow curved beam}
\label{sec:app-beam}

We first present the expressions for the constant matrices and vectors in \eqref{eq:system 2} as below
\begin{gather}
M_{ij} =\int_0^1{\varPhi _i ( \xi  ) \varPhi _j ( \xi  ) d\xi},\quad  K_{ij}=\int_0^1{\varPhi _i ( \xi  ) \varPhi _j^{''''} ( \xi  ) d\xi},\nonumber\\ 
C_{ij} =\int_0^1{2\bar{\mu}\varPhi _i ( \xi  ) \varPhi _j ( \xi  ) d\xi},\quad
B_{ij}=\int_0^1{\varPhi _i ( \xi  ) \varPhi _j^{''} ( \xi  ) d\xi},\nonumber\\ 
f_i =\int_0^1{\varPhi _i ( \xi  ) \bar{F}d\xi},\quad 
d_i =\int_0^1{\varPhi _i ( \xi  ) \varphi _{0}^{''}d\xi},\nonumber\\  H_{ij}=\kappa \int_0^1{\varPhi _i ( \xi  ) \varphi _{0}^{''}d\xi}\cdot \int_0^1{\varPhi _j ( \xi  ) \varphi _{0}^{''}d\xi}.\label{eq:matrix}
\end{gather}

\begin{figure}[!ht]
    \centering
   \includegraphics[width=0.45\textwidth]{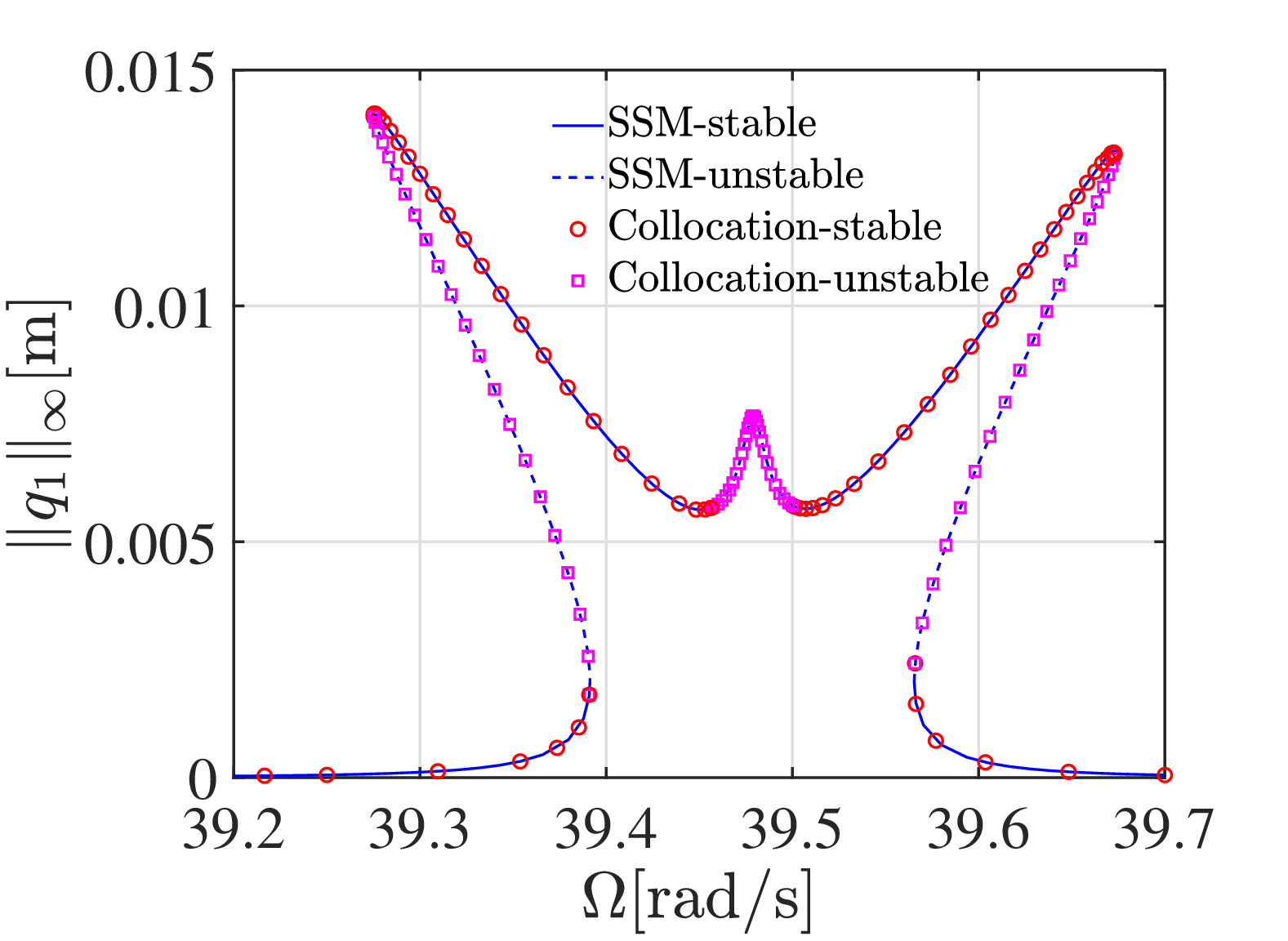}
   \includegraphics[width=0.45\textwidth]{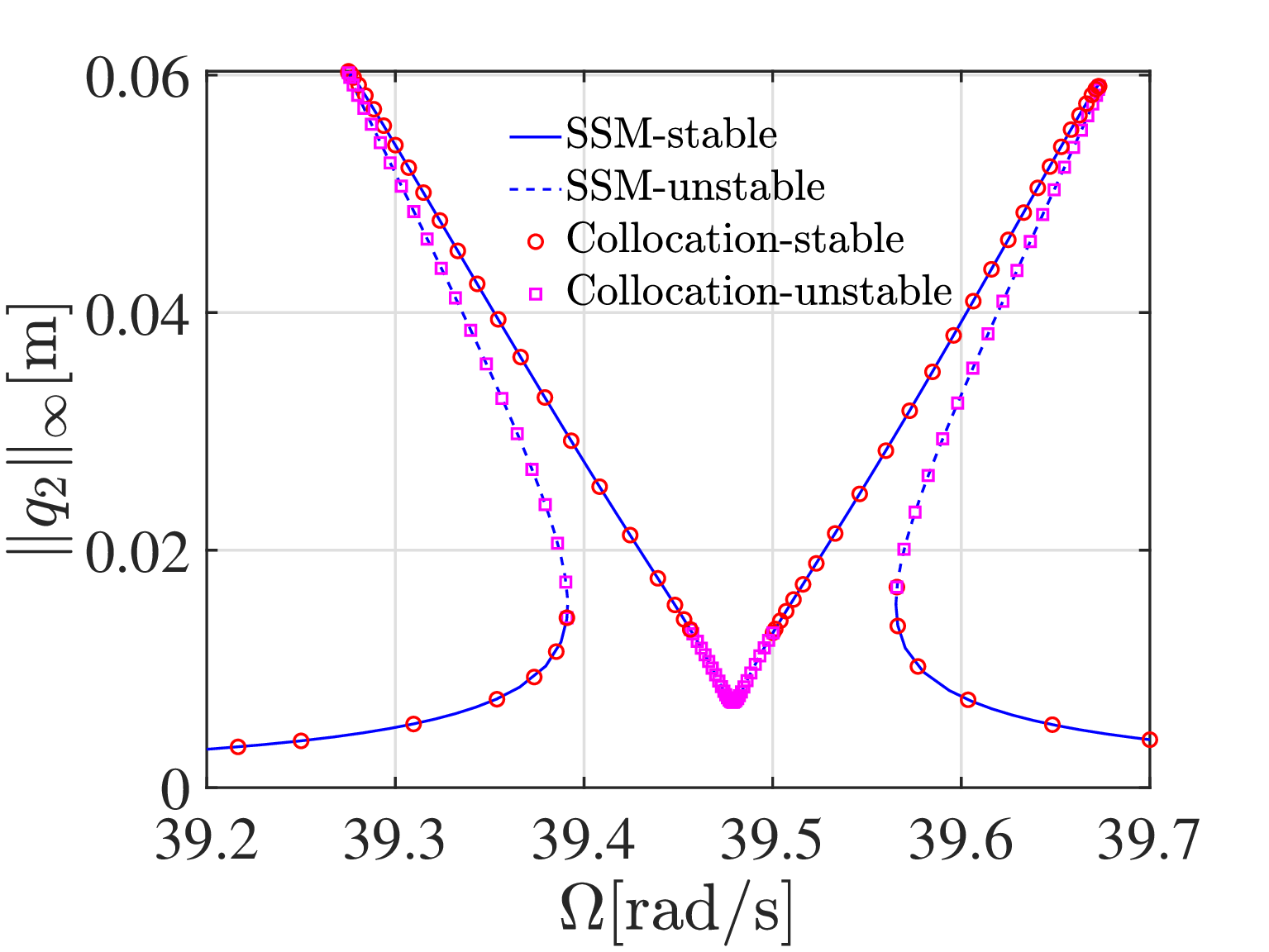} 
    \caption{\small FRCs of periodic orbits of the shallow curved beam system \eqref{eq:system 2} with $\Omega \in$ [39.2,39.7]. The solid/dashed lines denote the amplitudes of stable/unstable periodic solutions obtained by SSM analysis. The circles/squares denote the amplitudes of stable/unstable solutions obtained by \texttt{po}-toolbox of \textsc{COCO}, where periodic solutions are solved via collocation methods.}
    \label{fig:verify PO_q1q2}
\end{figure}

Next, we provide verification for the FRC-PO shown in Fig.~\ref{fig:ep_q1_q2} of the curved beam. We use the \texttt{po}-toolbox of \textsc{COCO}~\cite{COCO,dankowicz2013recipes,ahsan2022methods} to extract the FRC of periodic orbits of the full system. In the \texttt{po}-toolbox, a periodic orbit is formulated as a two-point boundary-value problem with periodicity boundary conditions, and the problem is solved using the collocation method. As seen in Fig.~\ref{fig:verify PO_q1q2}, the FRCs obtained from SSM-based predictions match well with that of the reference solutions. Here, the computational times of the FRC-PO for the SSM-based predictions and the collocation method are about 20 seconds and 9 minutes, respectively, demonstrating the speed-up gain of reductions via SSMs.

\begin{figure}[!ht]
    \centering
   \includegraphics[width=0.45\textwidth]{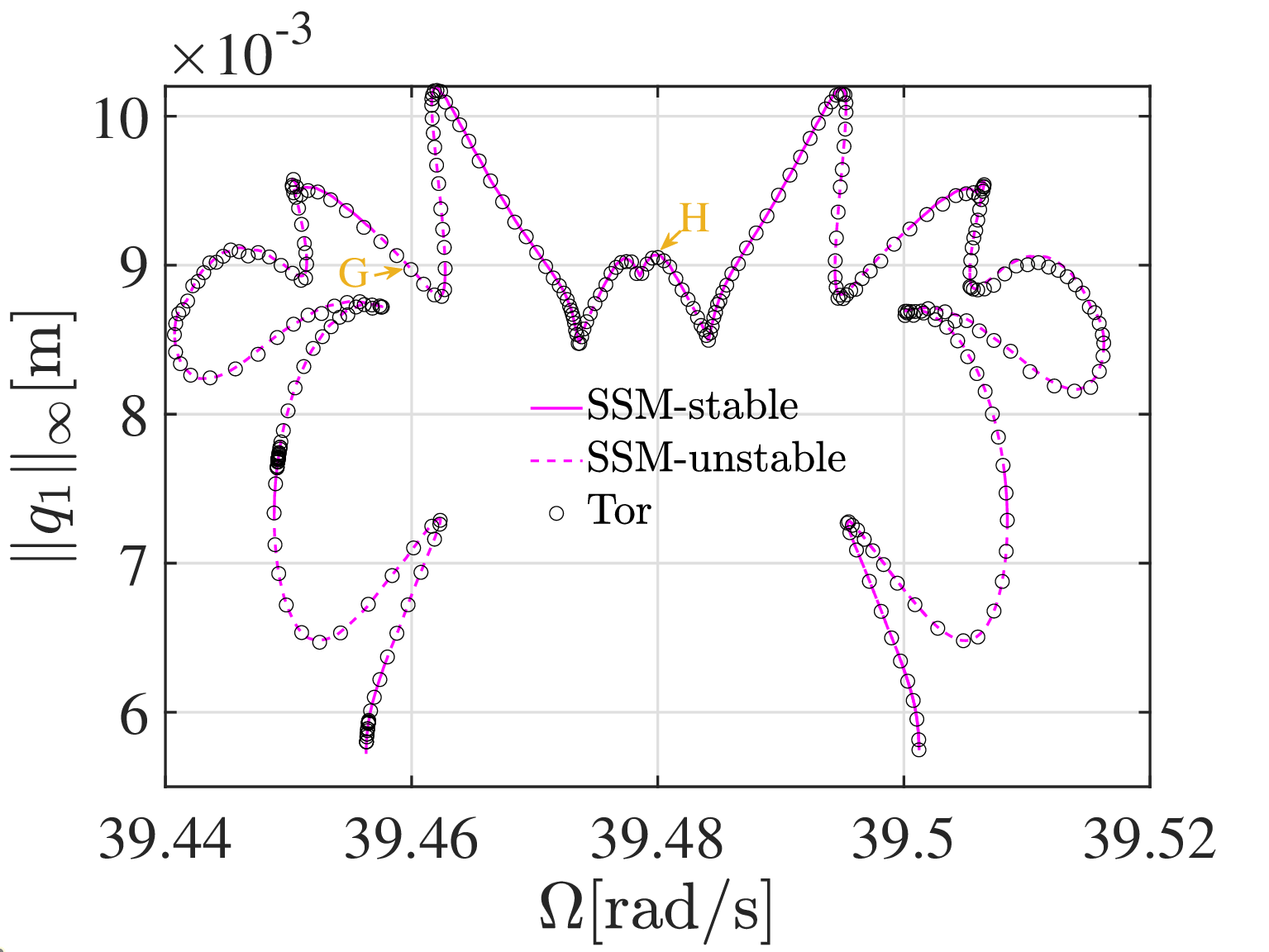}
   \includegraphics[width=0.45\textwidth]{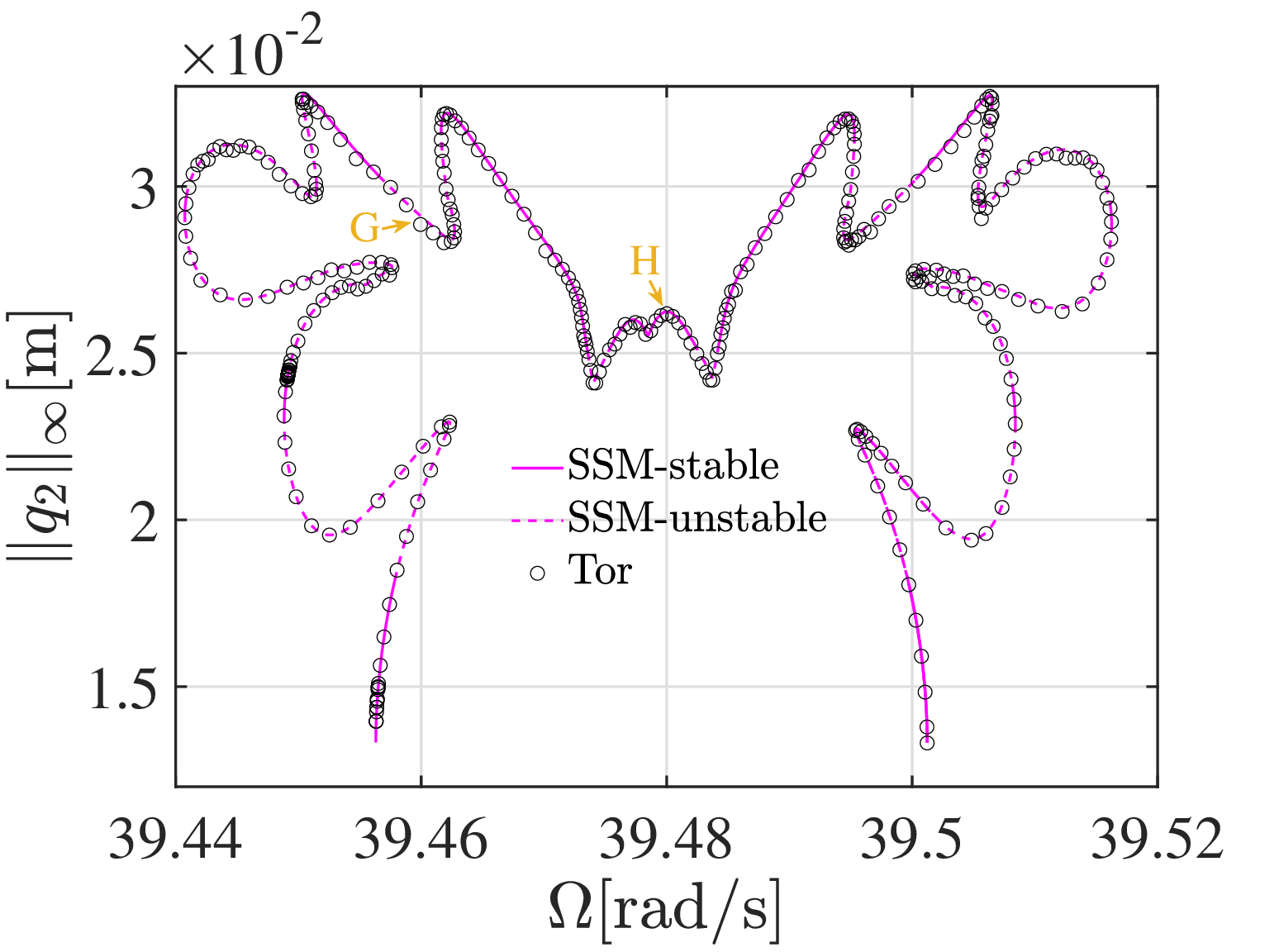} 
    \caption{\small FRCs of quasi-periodic orbits of the shallow curved beam system \eqref{eq:system 2} with $\Omega \in[39.44,39.52]$. The magenta solid/dashed lines denote stable/unstable quasi-periodic solutions obtained by SSM analysis (cf.~Fig.~\ref{fig:HB2po_q1_q2}). The circles denote quasi-periodic solutions obtained by \texttt{Tor} toolbox.}
    \label{fig:verify QO_q1q2}
\end{figure}

\begin{figure*}[!h]
    \centering
   \includegraphics[width=0.42\textwidth]{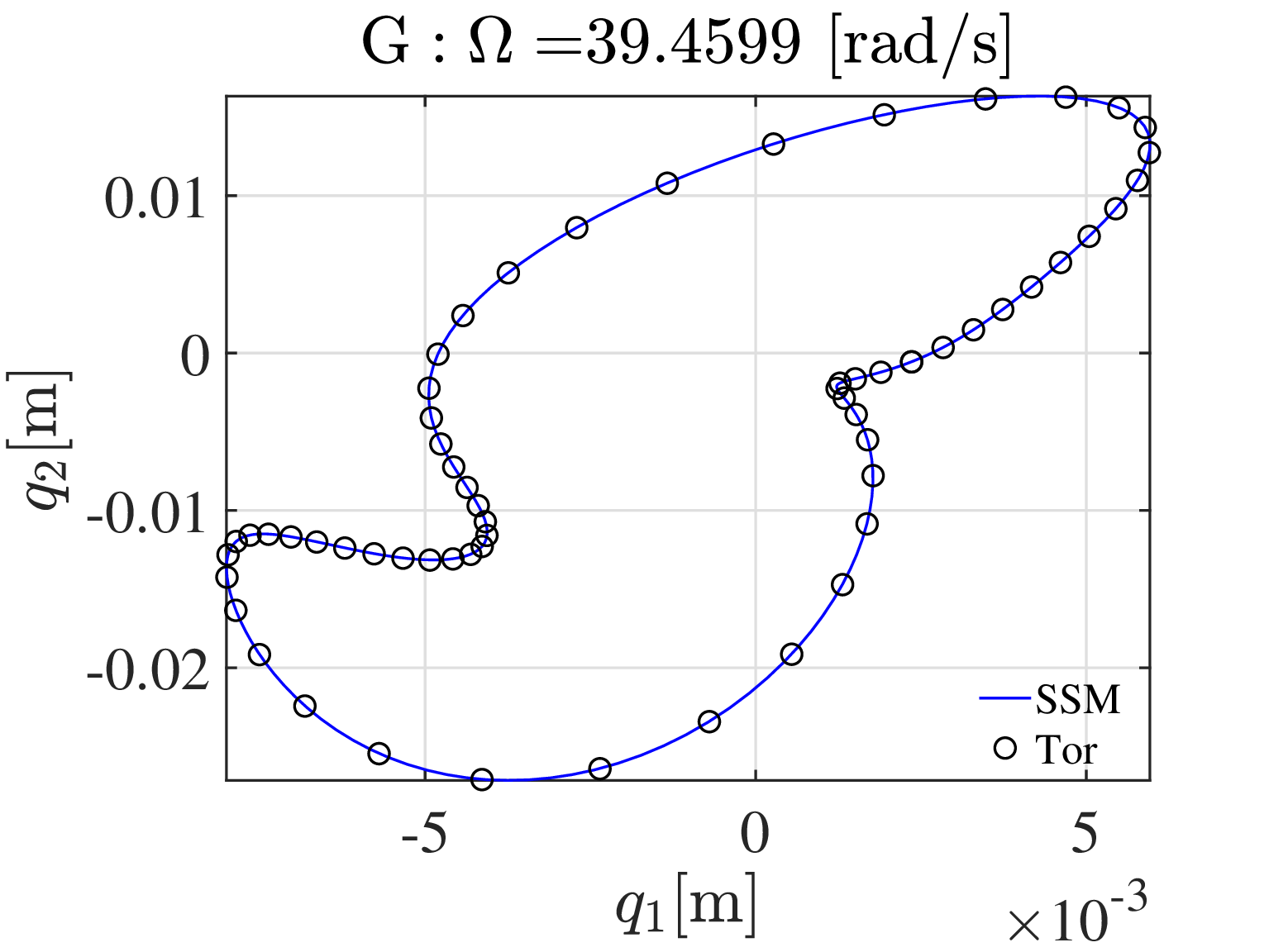}
   \includegraphics[width=0.42\textwidth]{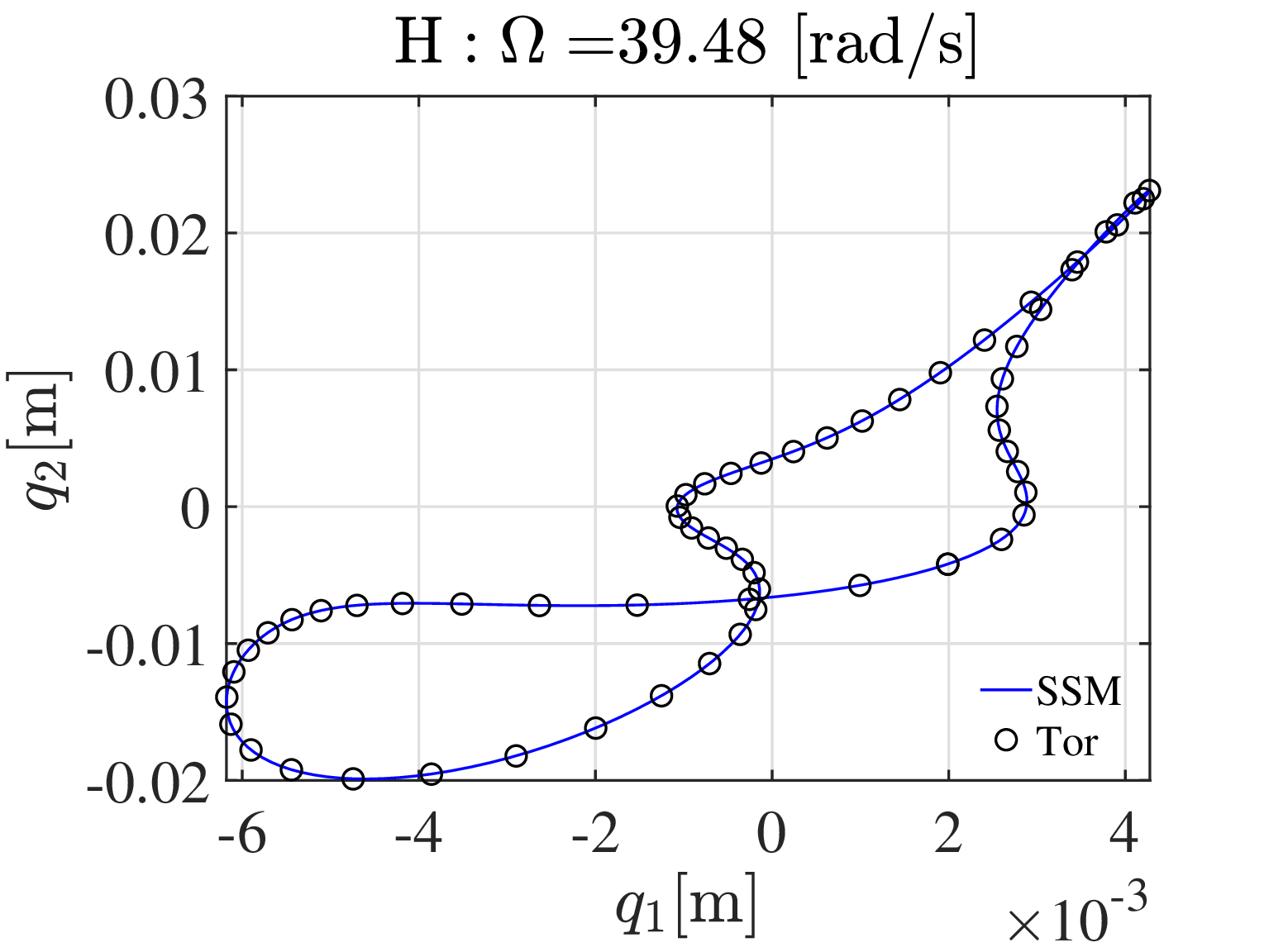} 
    \caption{\small Intersections of the period-$2\pi/\Omega$ map of the quasi-periodic orbits G (left panel) and H (right panel) of the shallow curved beam with $\Omega$ = 39.4599 and 39.48 (cf.~Fig.~\ref{fig:verify QO_q1q2}). The blue lines denote the SSM-based predictions while the black cycles denote reference results of the full model obtained using \texttt{Tor} toolbox.}
    \label{fig:verify Tor_q1q2}
\end{figure*}

We also provide verification for the FRC-QO shown in Fig.~\ref{fig:HB2po_q1_q2} of the curved beam. We use the \texttt{Tor} toolbox~\cite{li2020tor} of \textsc{COCO} to extract the FRC of quasi-periodic orbits of the full system. In the \texttt{Tor} toolbox, invariant tori on which quasi-periodic orbits stay are computed by solving some partial differential equations. We refer to~\cite{li2020tor,liNonlinearAnalysisForced2022c} for more details about the solution methods. As seen in Fig.~\ref{fig:verify QO_q1q2}, the FRC-QO obtained by \texttt{Tor} toolbox matches well with the one by SSM analysis. The intersections of the period-$2\pi/\Omega$ map of the quasi-periodic orbits G and H on the FRC-QO are shown in Fig.~\ref{fig:verify Tor_q1q2}, from which we see that the SSM-based predictions have high accuracy. These again demonstrate the effectiveness of the SSM-based ROM.~\textcolor{black}{Here, the computational times of the FRC-QO for the SSM-based predictions and the collocation method are about 10 minutes and 3 days, respectively, which further illustrates the speed-up gain of reductions via SSMs.}

\begin{figure*}[!ht]
    \centering
   \includegraphics[width=0.42\textwidth]{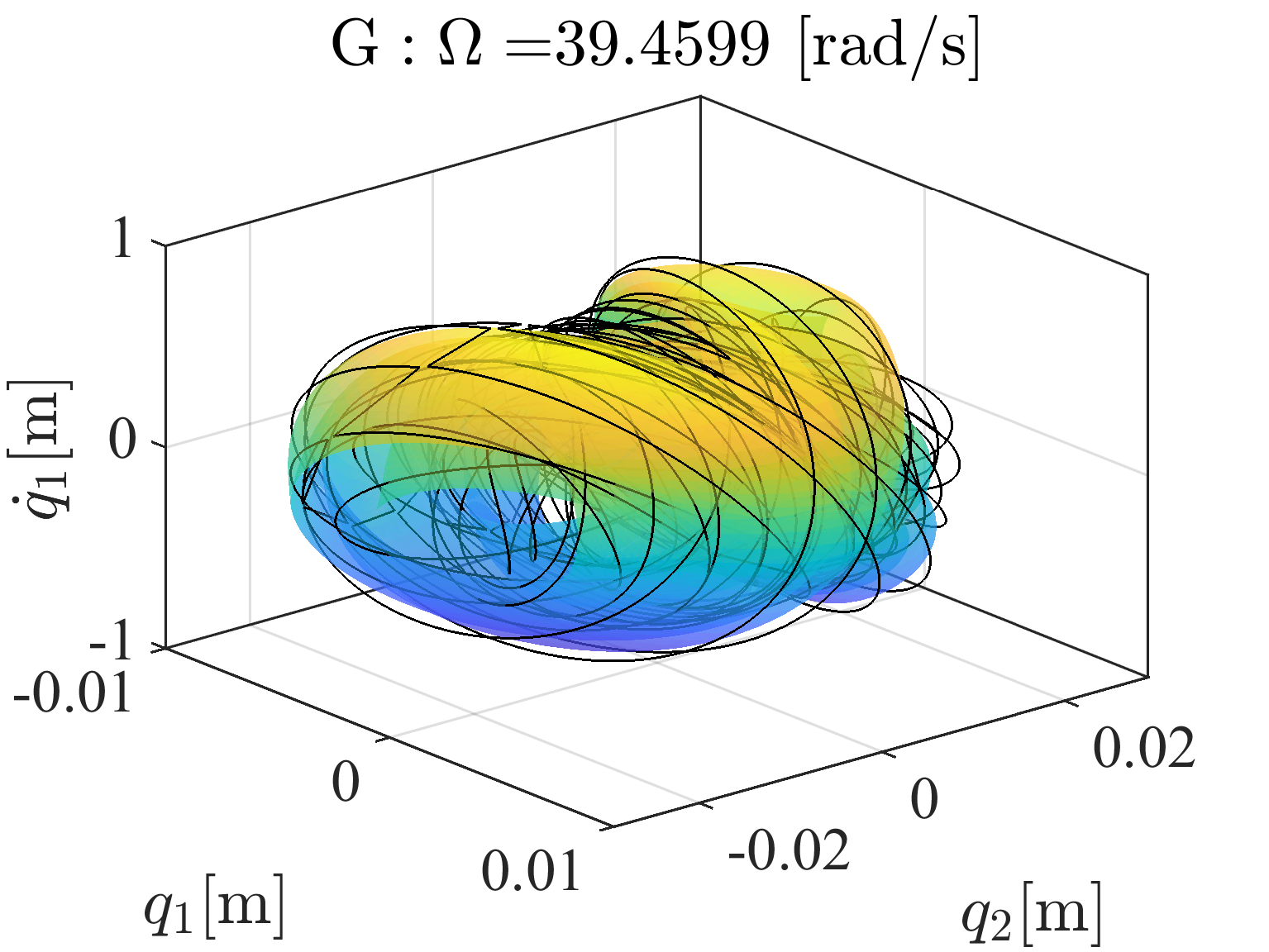}
   \includegraphics[width=0.42\textwidth]{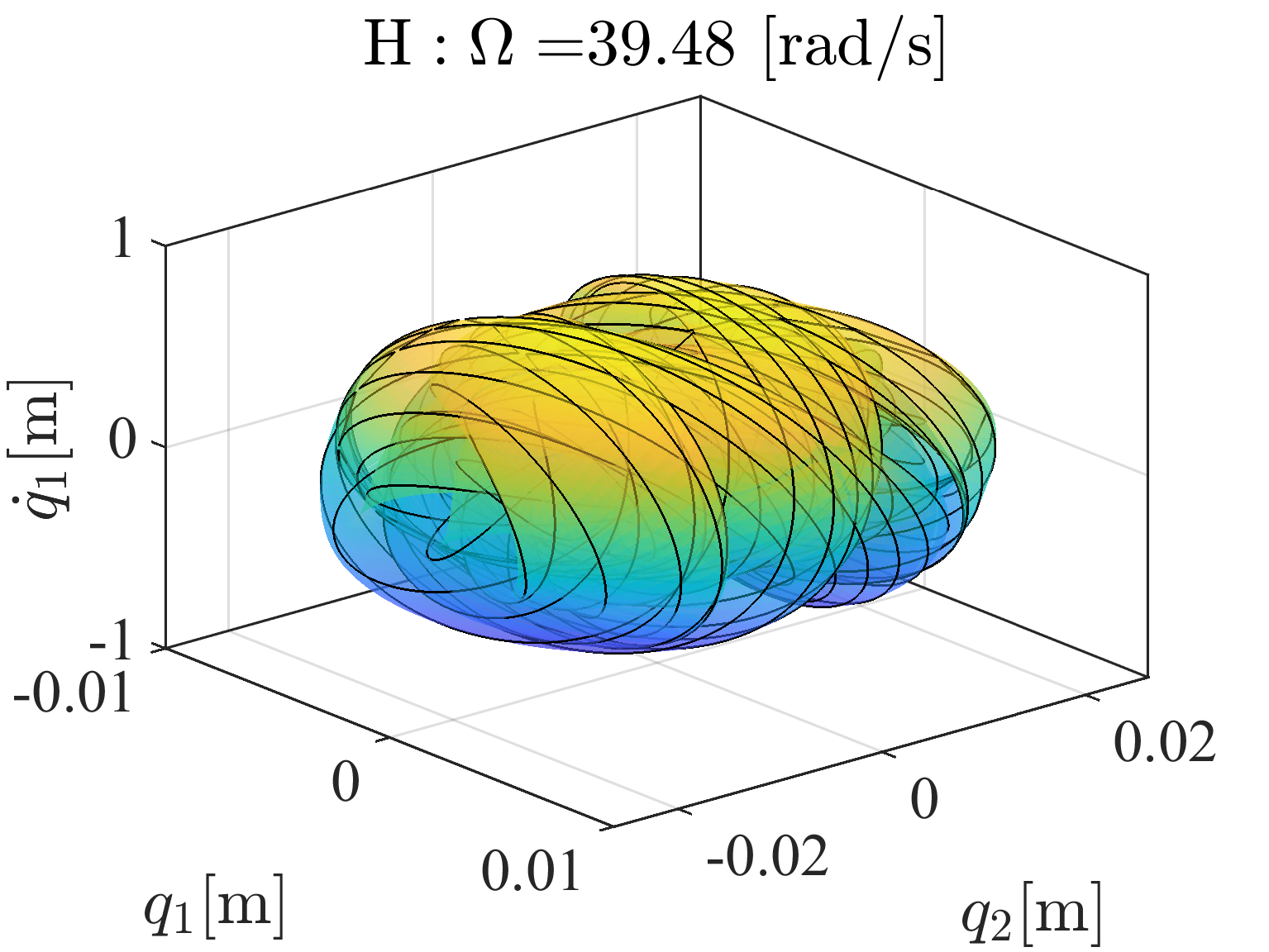} 
    \caption{\small Tori G and H of the shallow curved beam using SSM analysis (surface plot) and the trajectories by forward simulation of the full system~\eqref{eq:system 2} with initial states on the computed tori (black lines).}
    \label{fig:verify stability_q1q2}
\end{figure*}

We note that the current release of \texttt{Tor} toolbox does not support the stability analysis of quasi-periodic orbits. We use numerical integration to validate the stability types predicted by the SSM-based ROM. In particular, we take G and H in Fig.~\ref{fig:verify QO_q1q2} as representative of samples of unstable and stable quasi-periodic orbits (cf.~Fig.~\ref{fig:HB2po_q1_q2}). For each of these quasi-periodic orbits, we perform numerical integration of the full model with an initial point on the associated torus and monitor whether the trajectory stays on the invariant torus or not. If the quasi-periodic orbit is stable, the invariant torus is attracting, and then the trajectory will stay on it. Indeed, as seen in the right panel of Fig.~\ref{fig:verify stability_q1q2}, the trajectory with an initial condition on the invariant torus H stays on the torus, which confirms that the torus H is stable. In contrast, we observe from the {left} panel of Fig.~\ref{fig:verify stability_q1q2} that the trajectory with an initial condition on the unstable invariant torus G deviates from the torus, which is consistent with the prediction of SSM analysis.

\begin{figure*}[!ht]
    \centering
   \includegraphics[width=0.45\textwidth]{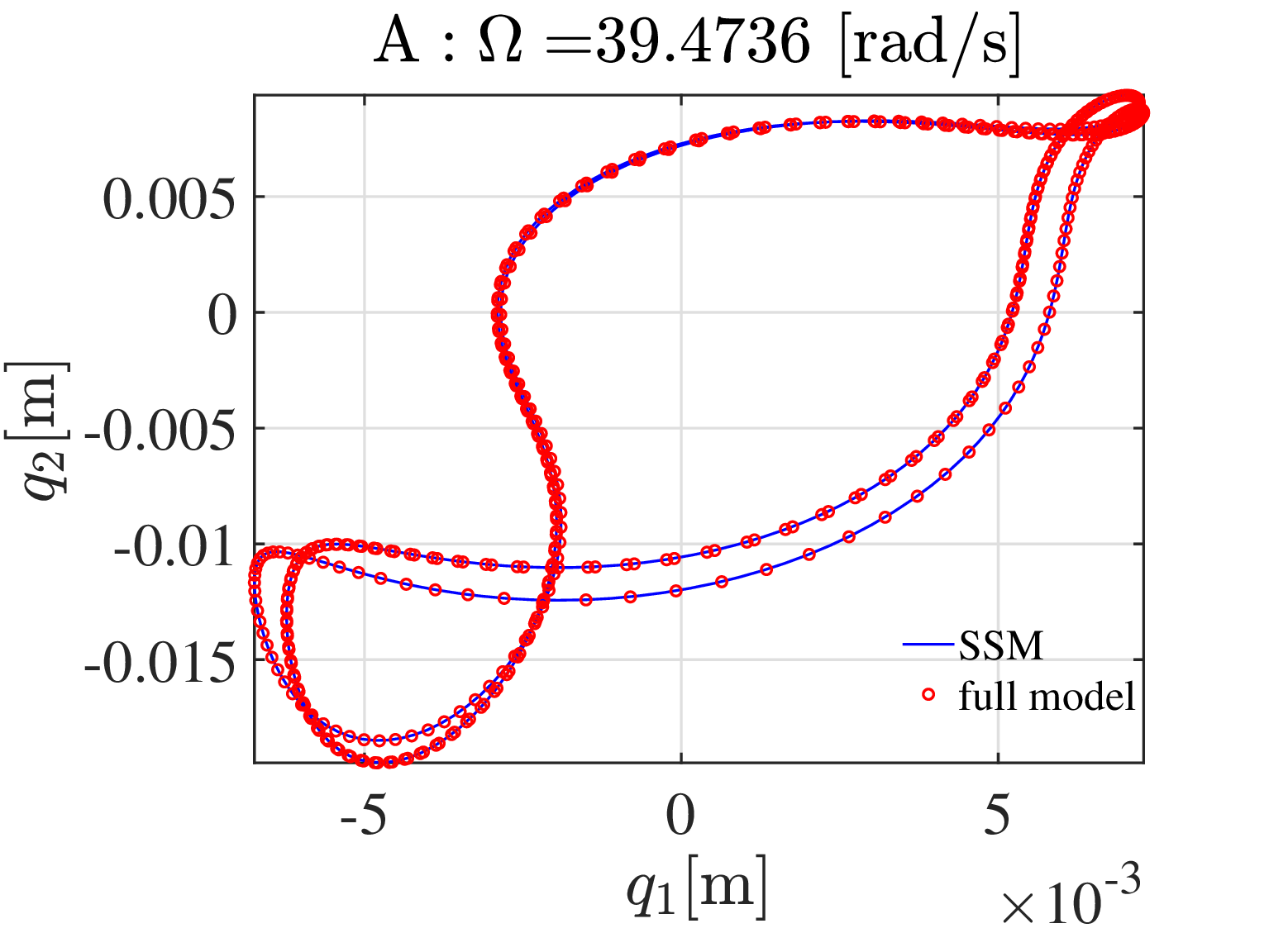}
   \includegraphics[width=0.45\textwidth]{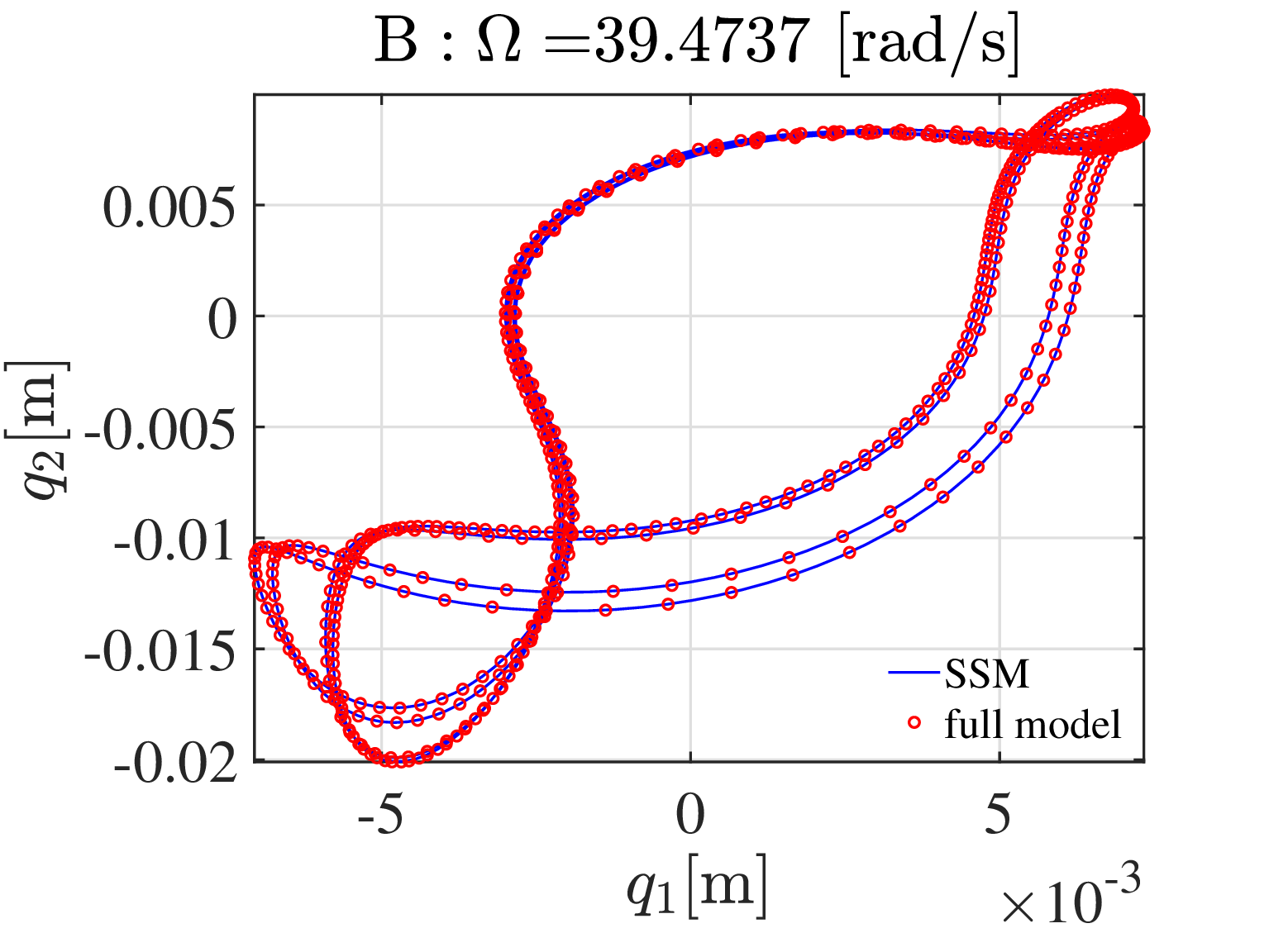} \\
   \includegraphics[width=0.45\textwidth]{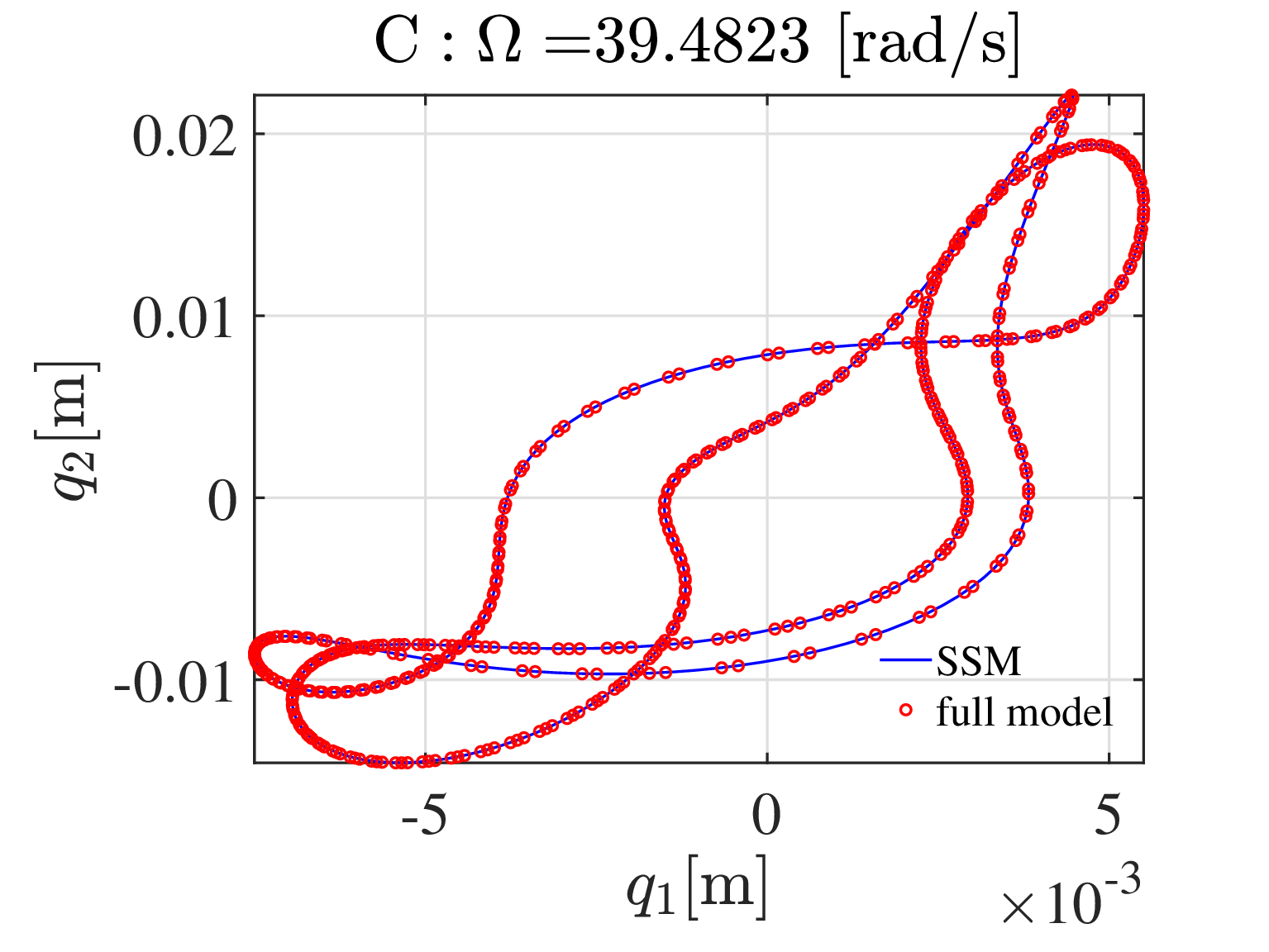}
   \includegraphics[width=0.45\textwidth]{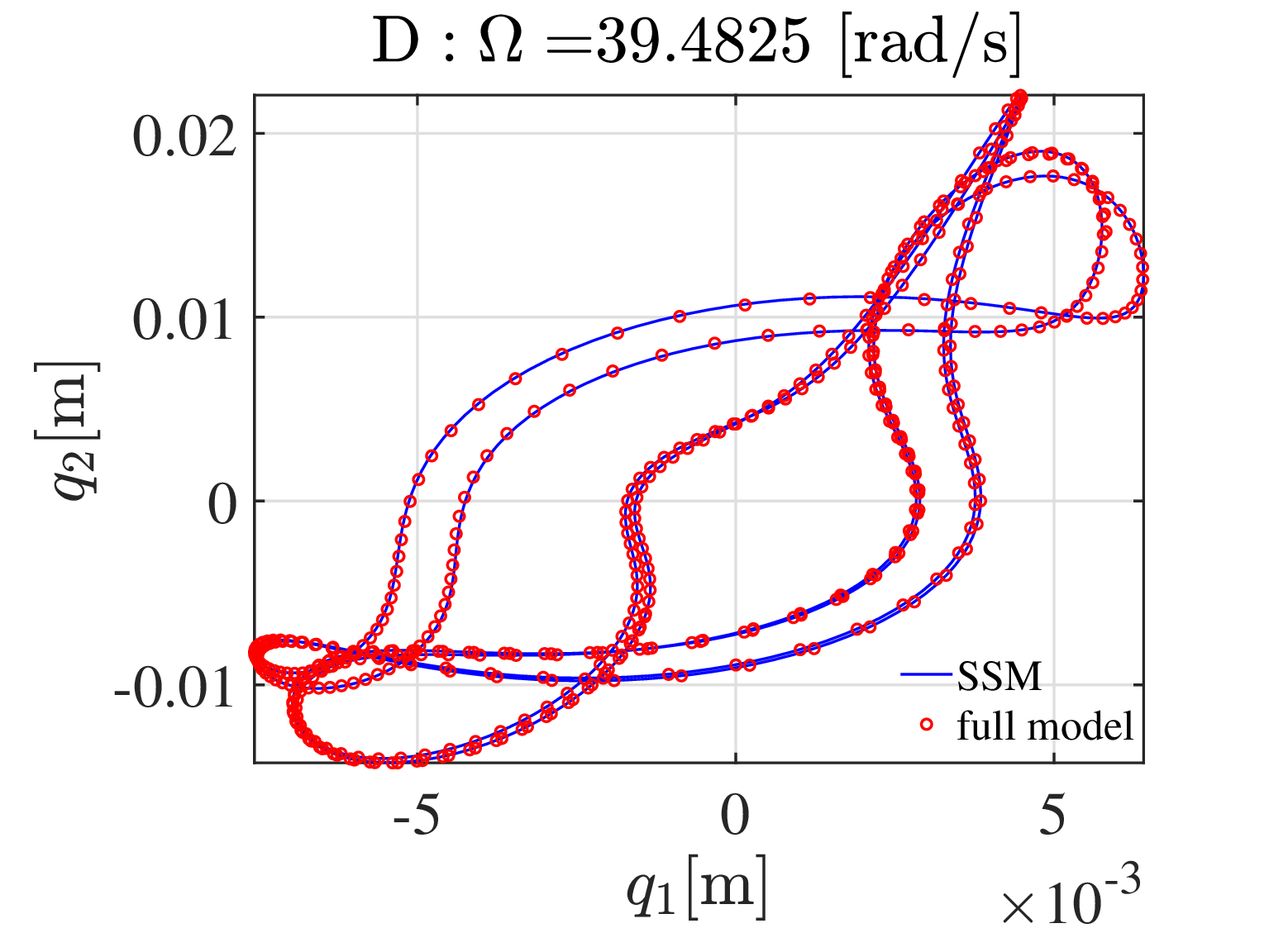}
    \caption{\small Intersections of the period-$2\pi/\Omega$ map of the quasi-periodic orbits A-D in Fig.~\ref{fig:isola_q2&L3}. The blue lines denote the results obtained by SSM analysis. The red cycles denote the results of the full system obtained from numerical integration.}
    \label{fig:verify L2&3 of system2}
\end{figure*}

Now we use numerical integration of the full system to verify the Period2 and Period4 quasi-periodic orbits in Fig.~\ref{fig:isola_q2&L3}. Specifically, we consider four stable limit cycles, namely A-D in Fig.~\ref{fig:isola_q2&L3} as the representative of Period2 (points A and C) and Period4 (points B and D) quasi-periodic orbits. The intersections of these quasi-periodic orbits and associated invariant tori with the period-$2\pi/\Omega$ map are shown in Fig.~\ref{fig:verify L2&3 of system2}, from which we see that the SSM-based predictions match excellently with that of the reference solutions of the full system.

\begin{figure}
    \centering
    \includegraphics[width=0.9\linewidth]{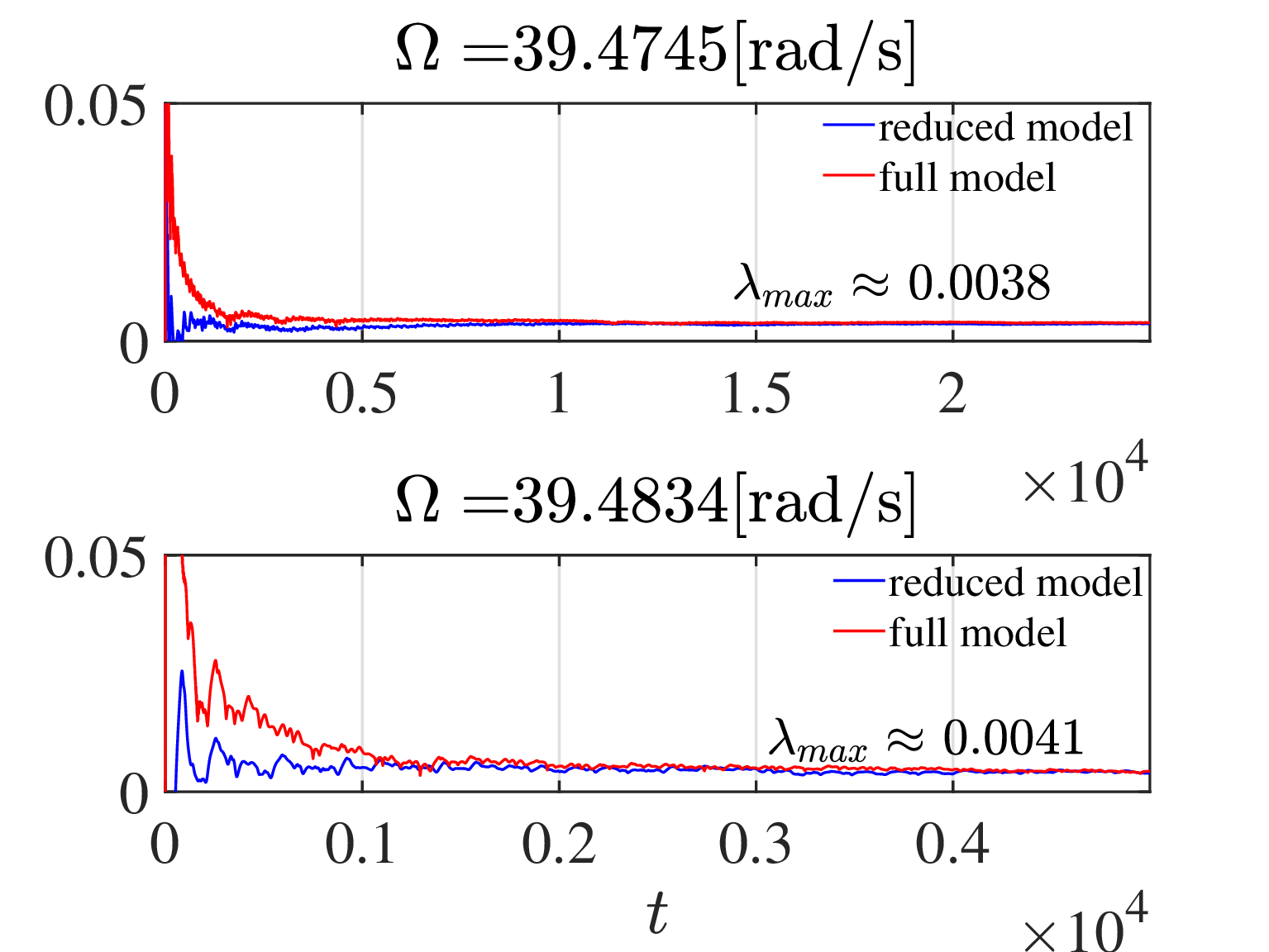}
    \caption{{Time history plots of maximum Lyapunov exponent of the two chaotic attractors in the right panel of Fig.~\ref{fig:chaos&isola}}.}
    \label{fig:lyapuv-ex2-chaos}
\end{figure}

{Finally, we present the maximum Lyapunov exponents of the two strange attractors shown in the right panel of Fig.~\ref{fig:chaos&isola}. We follow the same methods that produce the right panel of Fig.~\ref{fig:chaos_x2} to compute the maximum Lyapunov exponents of the strange attractors. In particular, for each of these strange attractors, we compute the maximum Lyapunov exponent using both the SSM-based ROM and the full model. The obtained results are shown in Fig.~\ref{fig:lyapuv-ex2-chaos}, from which we observe that the SSM-based predictions for the Lyapunov exponents match well with that of the reference solutions of the full system.}

\section{Supplementary analysis for the shallow shell}
\label{sec:app-shell}

\begin{figure*}[!ht]
    \centering
   \includegraphics[width=0.45\textwidth]{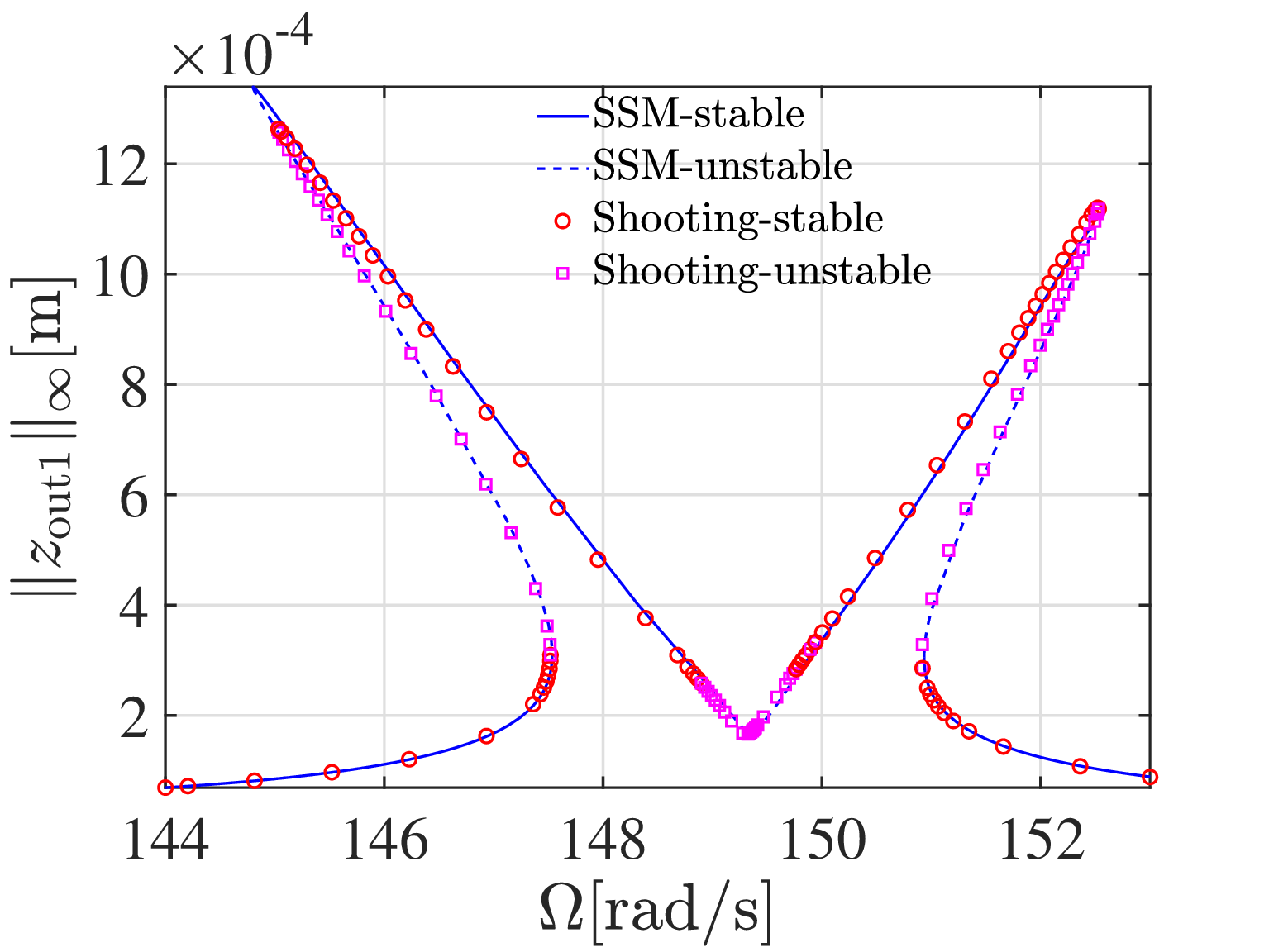}
   \includegraphics[width=0.45\textwidth]{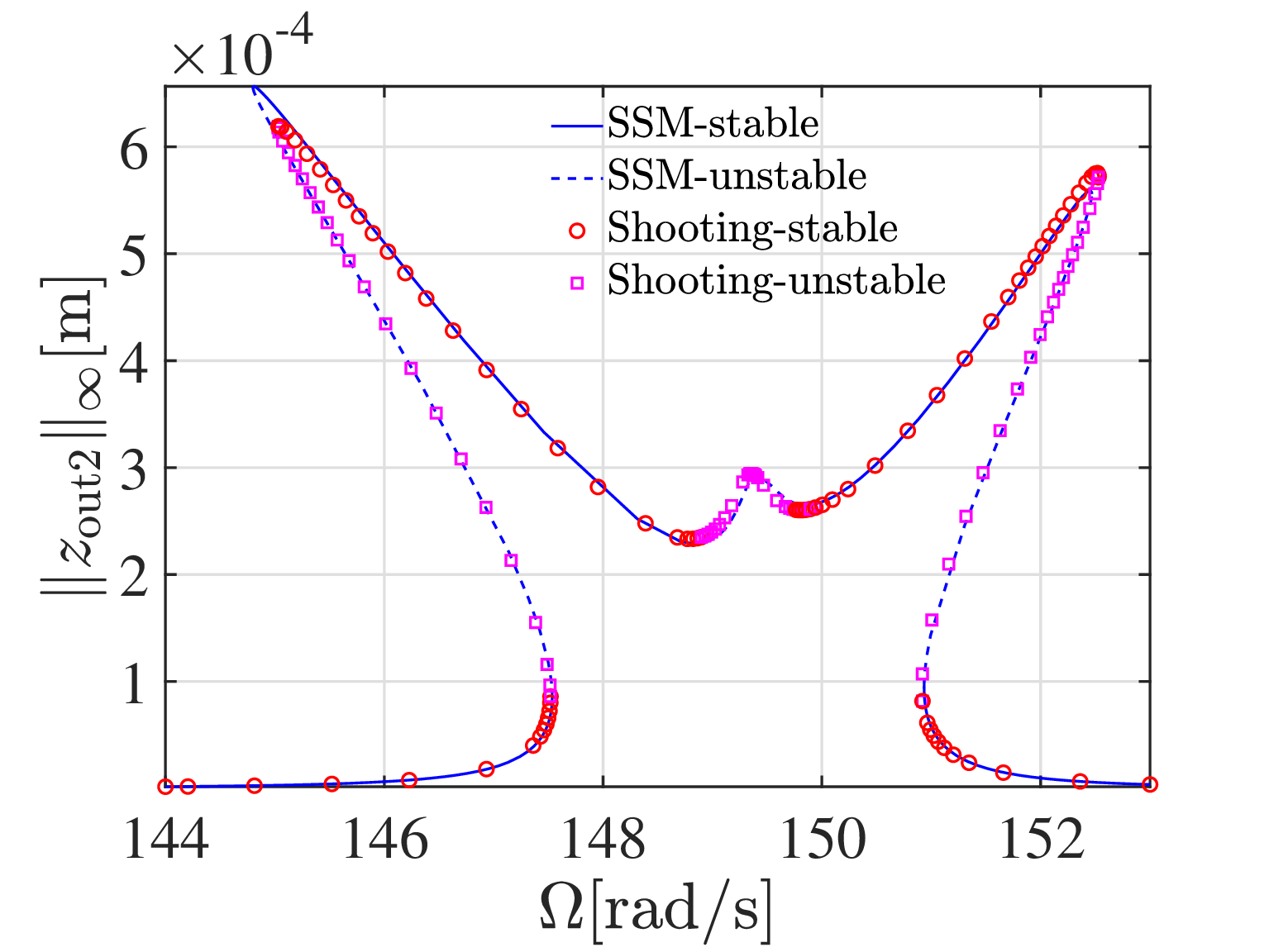} 
    \caption{\small FRCs of periodic orbits of the shallow curved shell system \eqref{eq:system 3} with $\Omega \in$ [144, 153]. The solid/dashed lines denote the amplitudes of stable/unstable periodic solutions obtained by SSM analysis. The circles/squares denote the amplitudes of stable/unstable solutions obtained by \texttt{shooting}-toolbox of \textsc{COCO}.}
    \label{fig:verify PO_z1z2}
\end{figure*}

We first extract the reference solutions of periodic orbits of the shallow shell under the variations in $\Omega$. Here, we consider an alternative method, namely the shooting method combined with parameter continuation (cf.~\cite{dankowicz2020multidimensional}), to extract the FRC-PO of the full nonlinear system. Specifically, the computation was performed using a COCO-based shooting toolbox~\cite{coco-shoot} with the Newmark integrator and the atlas algorithm of COCO. With 200 integration steps per excitation period and a maximum continuation step size 0.5, we obtain the FRC-PO of the full system shown in Fig. \ref{fig:verify PO_z1z2}. We observe from the figure that FRC-PO obtained from SSM-based predictions match well with that of the reference solutions via the shooting technique. Here, the computational time for SSM reduction is about one minute, while the time for the shooting method is about 2 days. Therefore, SSM reduction displays a significant speed-up gain relative to the shooting method in the above computations.

We now move to verifying the quasi-periodic orbits of the shallow shell system. Given this system has more than 1,300 DOFs, the \texttt{Tor} toolbox is inapplicable because it solves invariant torus using collocation methods and hence has an extremely high demand for memory cost. Here, we simply use numerical integration of the full system to validate the predicted quasi-periodic orbits. In particular, we take a point on a stable torus predicted by SSM reduction as the initial condition of the numerical integration and then perform a long time forward simulation such that it approaches a steady state. We then compare the quasi-periodic attractors obtained from the SSM-based predictions and the numerical integration. 

\begin{figure*}[!ht]
    \centering
    \includegraphics[width=0.38\textwidth]{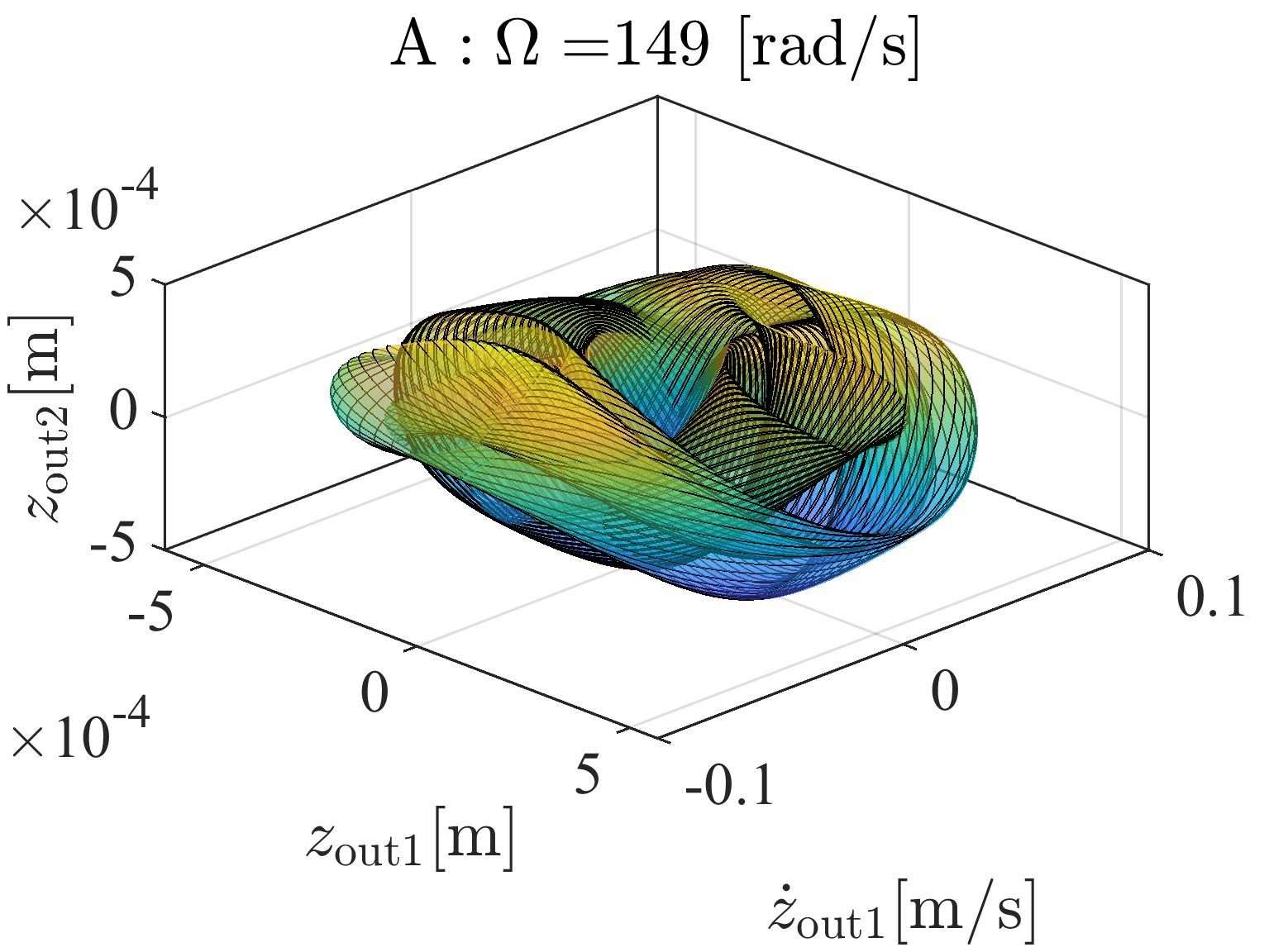}
    \includegraphics[width=0.38\textwidth]{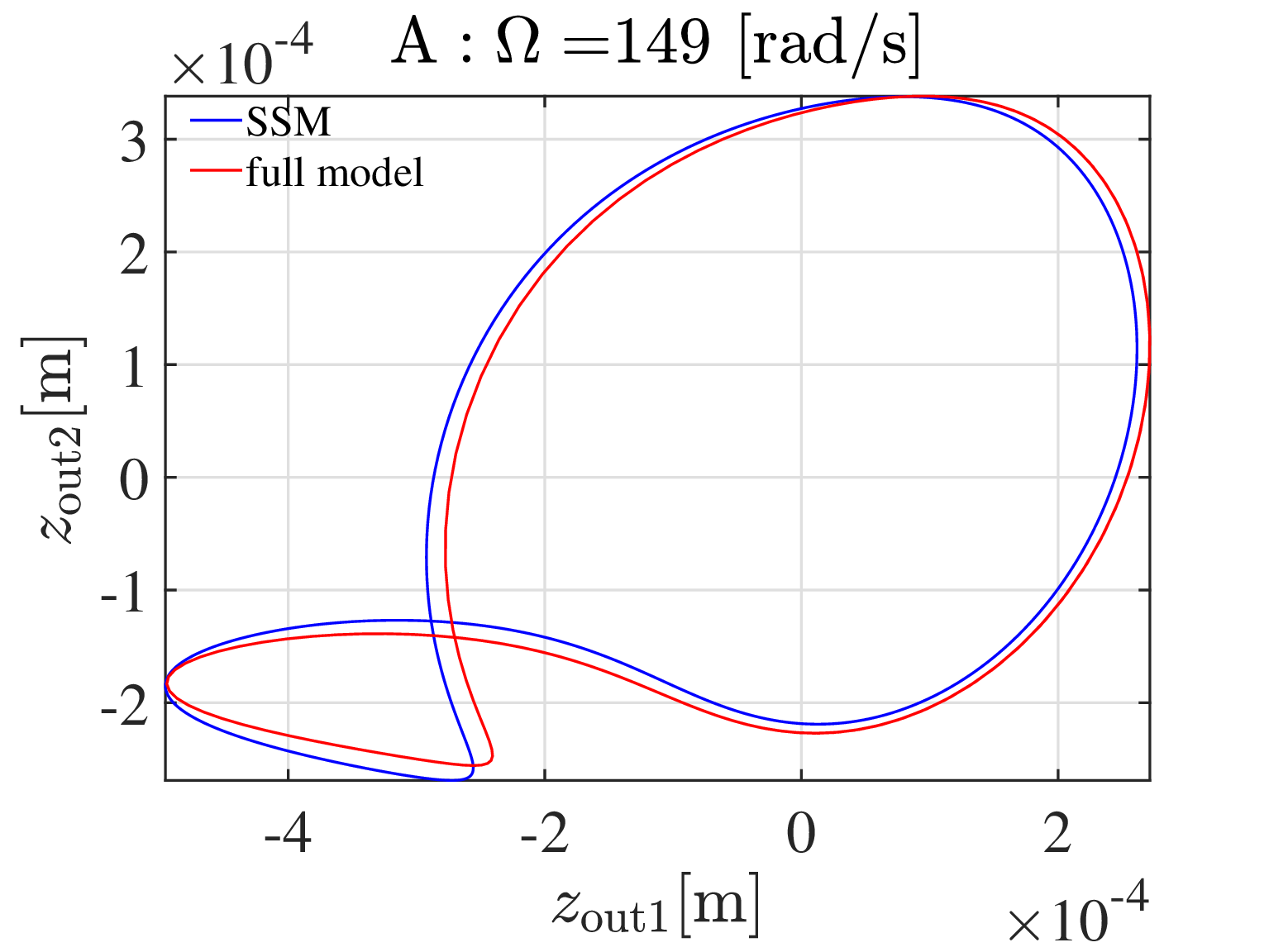}\\
   \includegraphics[width=0.38\textwidth]{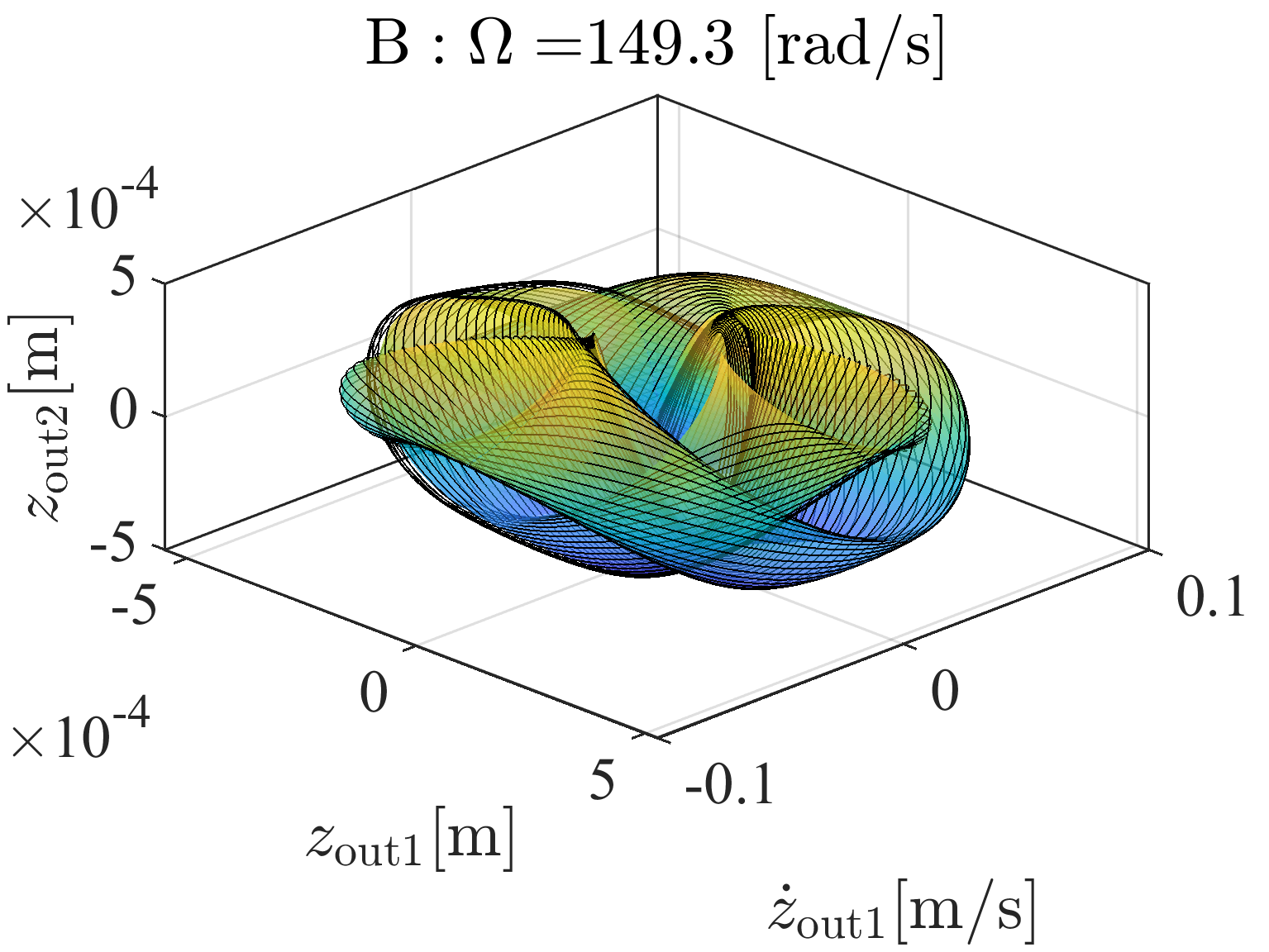} 
   \includegraphics[width=0.38\textwidth]{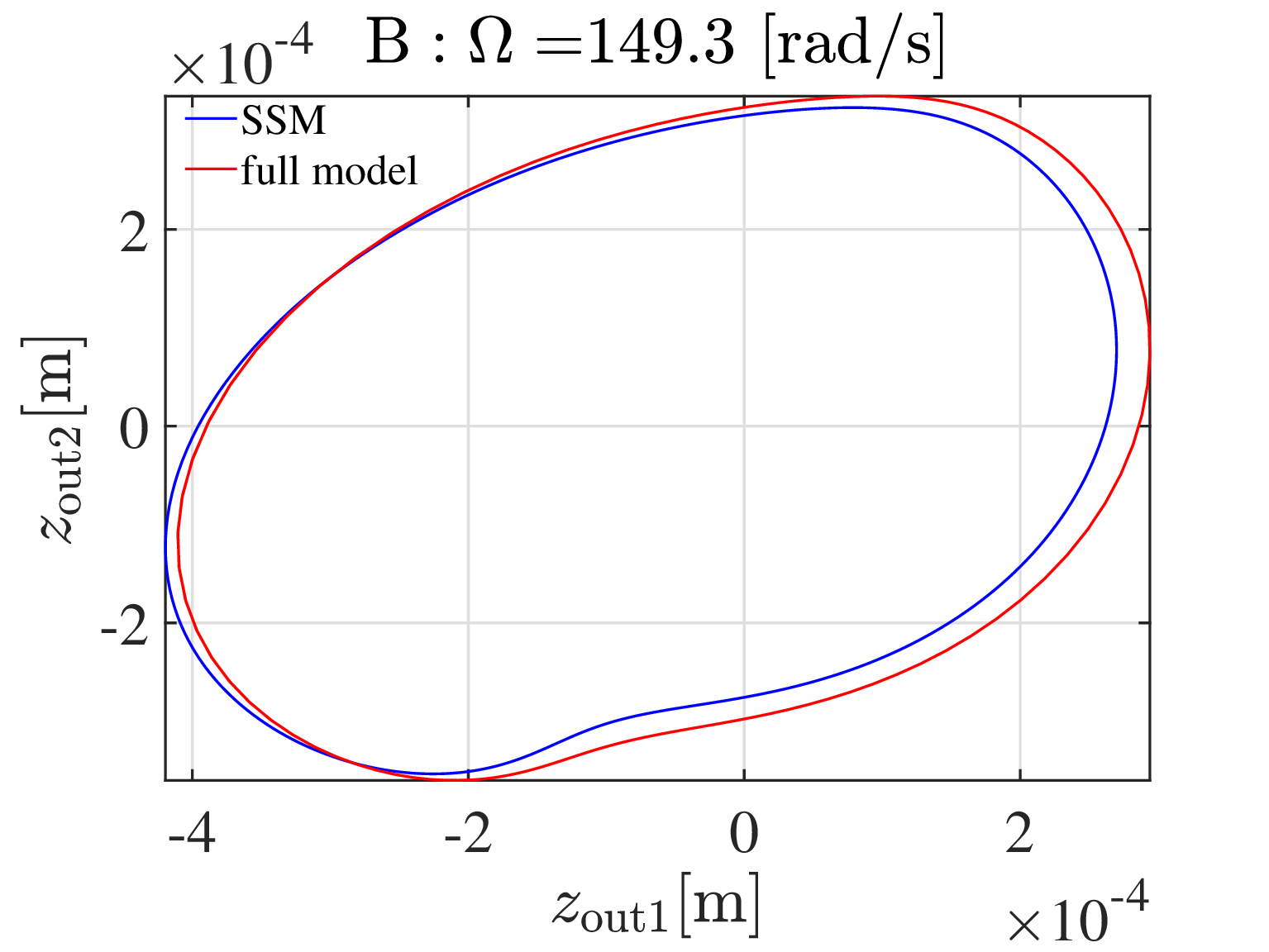}\\
     \includegraphics[width=0.38\textwidth]{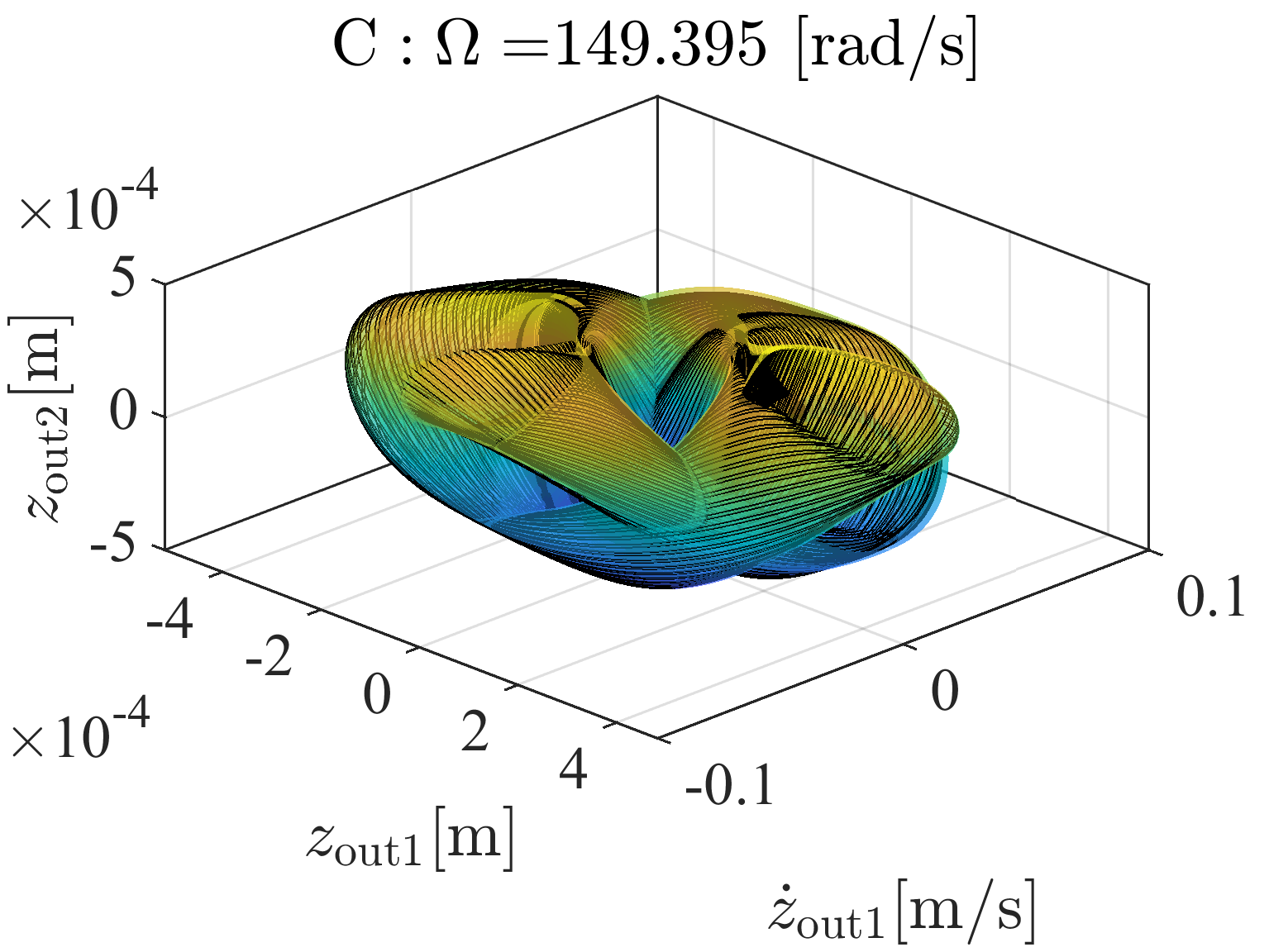}
   \includegraphics[width=0.38\textwidth]{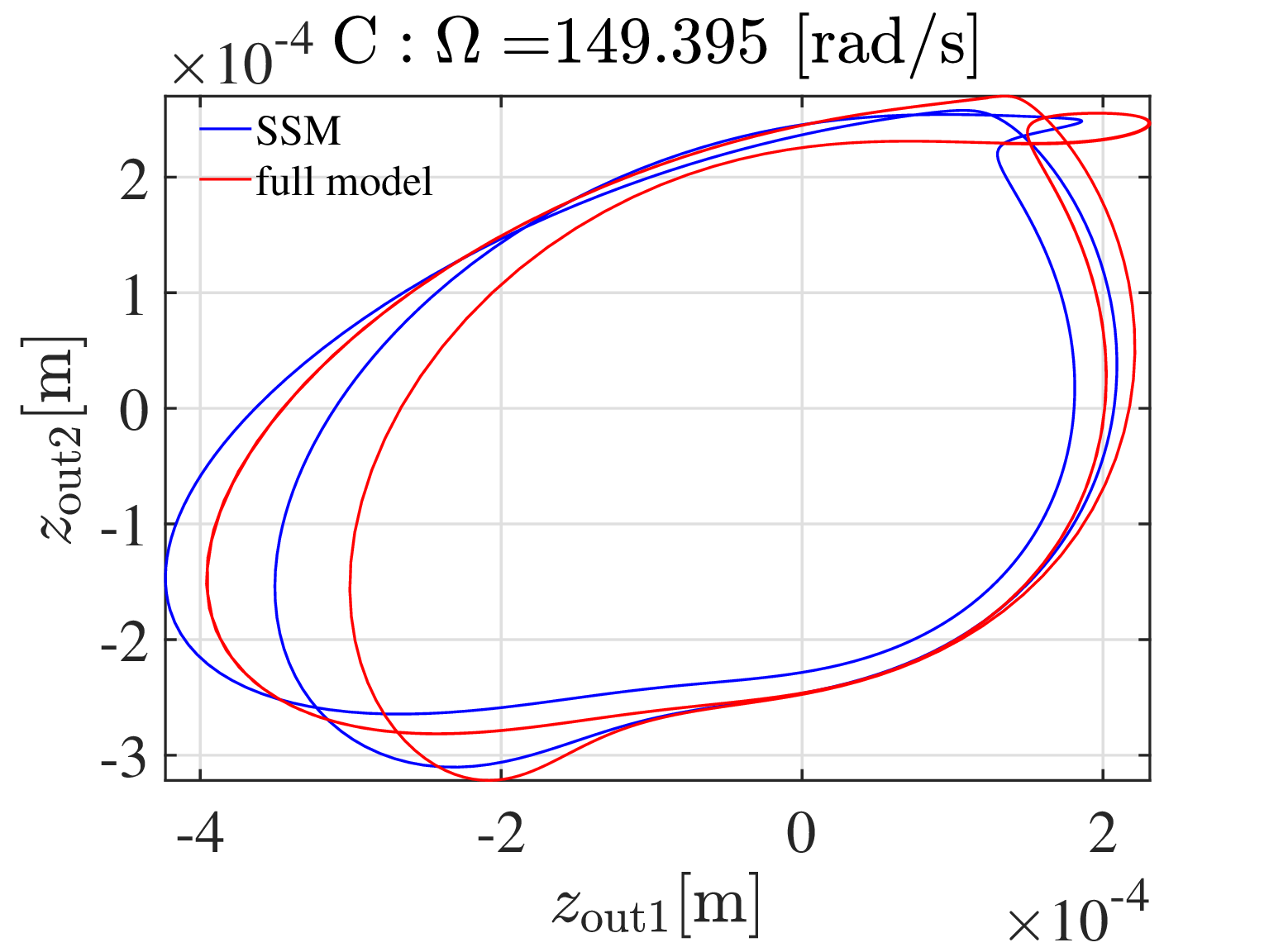}\\
   \includegraphics[width=0.38\textwidth]{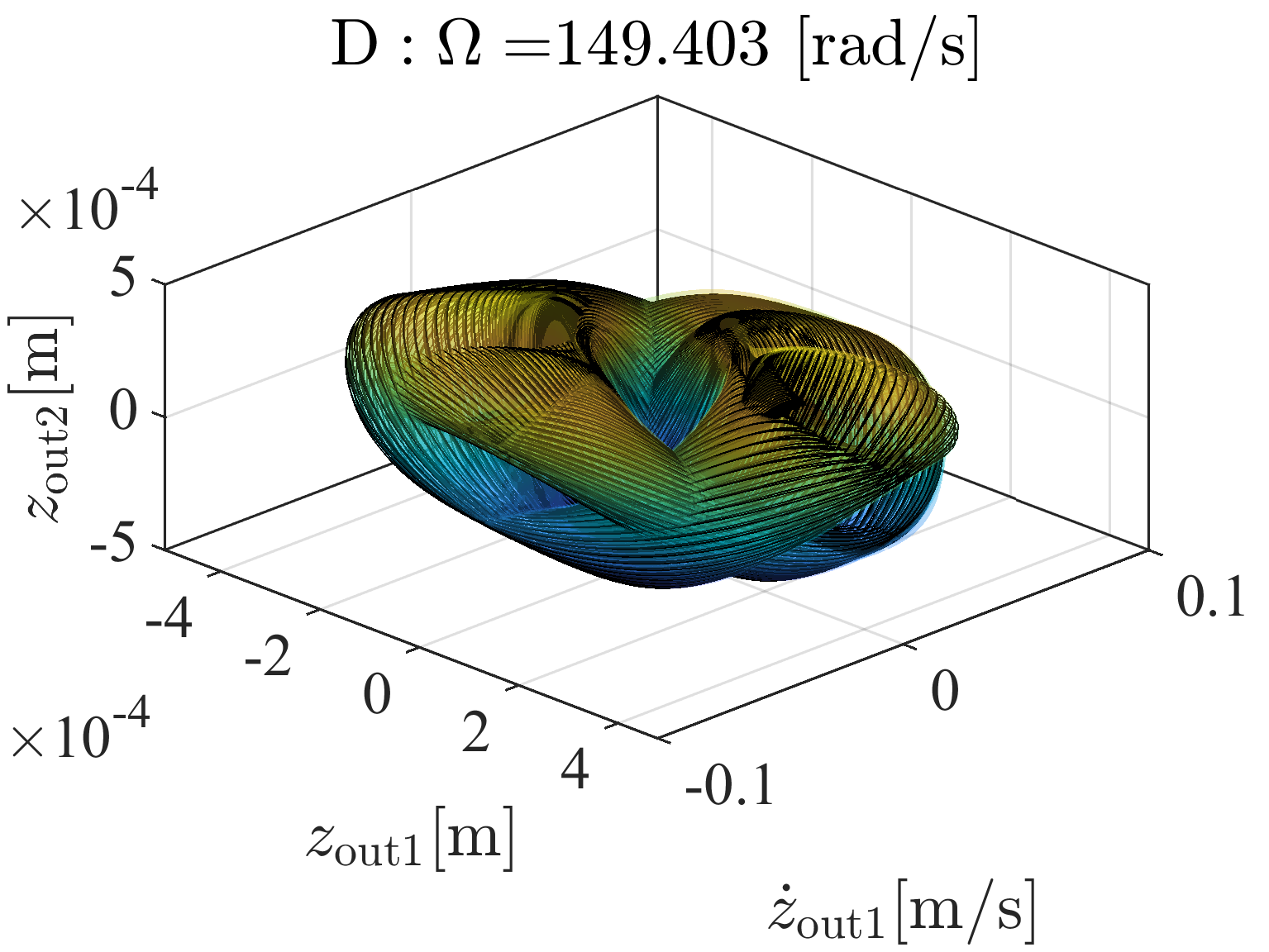}
   \includegraphics[width=0.38\textwidth]{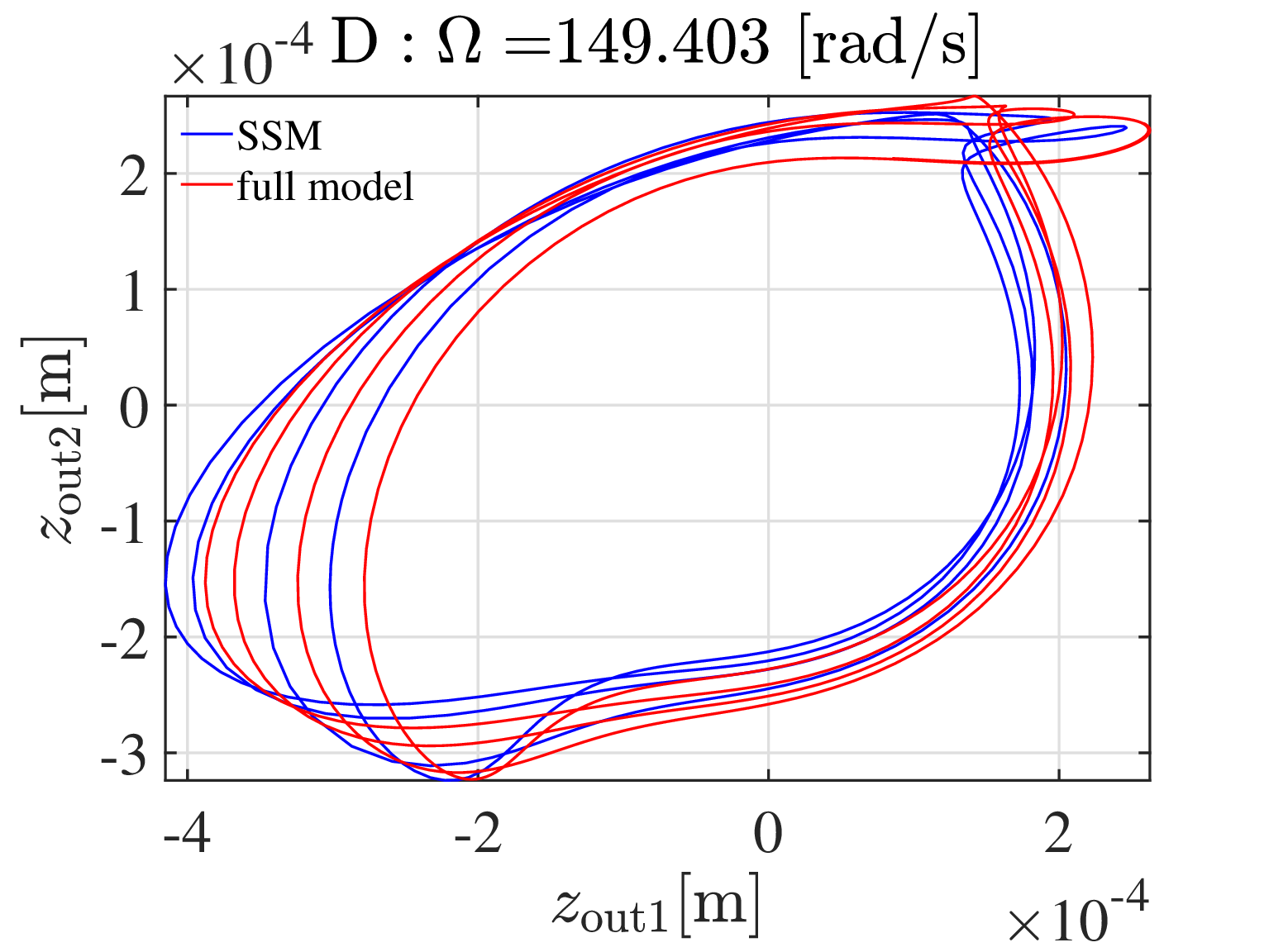}
    \caption{\small (Left panels) Tori of the shallow shell structure using SSM analysis (surface plot) and the trajectories by forward simulation of the full system with initial states on the computed tori (black lines). (Right panels) The intersections of these quasi-periodic attractors with the period-$2\pi/\Omega$ map were obtained from SSM analysis (blue lines) and direct numerical integration of the full systems (red lines). Here, the torus A and B in the upper two panels correspond to points A and B in the FRC-QO shown in Fig.~\ref{fig:HB2po_z1_z2}, and the torus C and D in the lower two panels corresponds to points C and D in the FRC-QO shown in Fig.~\ref{fig:PD2po_z1}.}
    \label{fig:verify QO-z1z2}
\end{figure*}

Here, we take points A and B ($\Omega_{\rm{A}} = 149, \Omega_{\rm{B}} = 149.3$) on the FRC-QO in Fig.~\ref{fig:HB2po_z1_z2} to conduct the verification because these two quasi-periodic orbits are stable. As for the numerical integration, we apply the generalized-$\alpha$ method in the forward simulation. We use 500 integration steps per excitation period and perform simulations over 4500 excitation cycles to ensure that the steady state has arrived. The obtained quasi-periodic attractors, along with their intersections with the period-$2\pi/\Omega$ map, are presented in the upper two panels of Fig.~\ref{fig:verify QO-z1z2}, from which we see that the SSM-based predictions match well with that of the reference solutions of the full system. This again demonstrates the accuracy of the SSM-based reduction. Moreover, the SSM-based reduction displays a significant speed-up gain relative to the forward simulations. Indeed, each forward simulation took about 5 days, yet the computational time for extracting the whole FRC of quasi-periodic orbits is less than 4 minutes.

Finally, we verify the point C with $\Omega=149.395$ on the FRC of quasi-periodic orbits with doubled period and the point D with $\Omega=149.403$ on the FRC of quasi-periodic orbits with quadrupled period in Fig.~\ref{fig:PD2po_z1}. We take these quasi-periodic orbits as representatives of the Period2 and Period4 quasi-periodic orbits and again apply the numerical integration of the full system to validate the SSM-based prediction. We present these two quasi-periodic attractors along with their intersections with the period-$2\pi/\Omega$ map in the lower two panels of Fig.~\ref{fig:verify QO-z1z2}. Indeed, the full system admits a quasi-periodic attractor of the doubled (quadrupled) period at $\Omega=149.395$ ($\Omega=149.403$). Moreover, we observe that the SSM-based prediction matches well overall with the reference solution of the full system. The observed discrepancies can be reduced when higher-order non-autonomous contributions are accounted~\cite{thurnher2024nonautonomous}.

\bibliographystyle{ieeetr}       
\bibliography{ref}   

\end{sloppypar}
\end{document}